%% file: nonlocal_long250416jcap.tex
\documentclass[a4paper,upright,11pt,english]{article}
\pdfoutput=1
\usepackage[vmargin = 2.5cm, hmargin = 2cm]{geometry}

\input epsf
\usepackage{babel} % passe en mode français
 \usepackage{amsmath} % package d’amsTex
\usepackage{mathtools}
\usepackage{amsthm} % redefinition des theoreme suivant amstex
\usepackage{amscd}
\usepackage{amscd}
\usepackage{amsbsy}
\usepackage{amssymb} % symboles amstex
\usepackage{mathrsfs} % fontes amstex
\usepackage{latexsym}
\usepackage{booktabs}
\usepackage{upgreek}
\usepackage{fancyhdr} % Ce package permet de définir des en-tête
\usepackage{graphicx} % permet d’inclure des graphiques
\usepackage[utf8]{inputenc} % permet l’encodage en UTF8
\usepackage{eurosym} % ce package définit le symbole euro.
\usepackage{color} % permet de changer les couleurs
\usepackage{multicol} % permet l’édition sur plusieurs colonnes
\usepackage{hyperref} % liens interne et externe
\usepackage{bbm}
\usepackage{amsthm}
\usepackage[usenames,dvipsnames]{pstricks}
\usepackage{epsfig}
\usepackage{pst-grad} % For gradients
\usepackage{pst-plot} % For axes
\usepackage[babel=true]{csquotes}

\usepackage{float}
\usepackage{cite}
\usepackage{appendix}

 \usepackage[usenames,dvipsnames]{pstricks}
 \usepackage{epsfig}
 \usepackage{pst-grad} % For gradients
 \usepackage{pst-plot} % For axes
 
\usepackage{pstricks}

\definecolor{Blue}{rgb}{0.15,0.15,0.70}
\definecolor{6dFcolor}{RGB}{0,0,255}
\definecolor{SDSScolor}{RGB}{160,160,160}
\definecolor{BOSSLOWZcolor}{RGB}{153, 102, 51}
\definecolor{BOSSCMASScolor}{RGB}{127.5, 0, 127.5}
\definecolor{WiggleZcolor}{RGB}{255,127,0}

\hypersetup{
colorlinks=true,
citecolor=Blue,
linkcolor=Blue,
urlcolor=Blue
}

\include{mydefs}

\theoremstyle{definition}

\theoremstyle{remark}

\begin{document} 

\begin{titlepage}

\vspace*{2cm}

\centerline{\Large \bf Non-local gravity and comparison with observational datasets.}

\vspace{5mm}

\centerline{\Large \bf 
II. Updated results and Bayesian model comparison with $\Lambda$CDM\footnote{Based on observations obtained with Planck (http://www.esa.int/Planck), an ESA science mission with instruments and contributions directly funded by ESA Member States, NASA, and Canada.}
}

%\vspace{5mm}
%
%\centerline{\Large \bf  }

\vskip 0.4cm
\vskip 0.7cm
\centerline{\large Yves Dirian$^a$, Stefano Foffa$^a$, Martin Kunz$^{a,b}$, Michele Maggiore$^a$ and Valeria Pettorino$^c$}
\vspace{3mm}
\centerline{\em $^a$D\'epartement de Physique Th\'eorique and Center for Astroparticle Physics,}  
\centerline{\em Universit\'e de Gen\`eve, 24 quai Ansermet, CH--1211 Gen\`eve 4, Switzerland}
\vspace{3mm}
\centerline{\em $^b$African Institute for Mathematical Sciences, 6 Melrose Road, Muizenberg, 7945, South Africa}
\vspace{3mm}
\centerline{\em $^c$Institut f\"{u}r Theoretische Physik, Universit\"{a}t Heidelberg,
Philosophenweg 16, D-69120 Heidelberg}

\vskip 1.9cm

\begin{abstract}

We present a comprehensive and updated  comparison   with cosmological observations of two non-local modifications of gravity previously introduced by our group, the so called RR and RT models. We implement the background evolution and the cosmological perturbations of the models in a modified Boltzmann code, using CLASS. We  then  test the non-local models against the {\em Planck} 2015 TT, TE, EE and Cosmic Microwave Background (CMB) lensing data,  isotropic and anisotropic Baryonic Acoustic Oscillations (BAO)  data, JLA supernovae,  $H_0$ measurements and growth rate data, and we perform Bayesian parameter estimation.  We then compare the RR, RT and $\Lambda$CDM models, using the Savage-Dickey  method. We find that the RT model and $\Lambda$CDM perform equally well, while    the performance of the RR model with respect to $\Lambda$CDM  depends on 
whether or not we include a prior on $H_0$ based on local measurements.

\end{abstract}

\end{titlepage}

\newpage

\section{Introduction}

In the last two decades $\Lambda$CDM has gradually become the standard cosmological paradigm. Indeed, $\Lambda$CDM fits all cosmological observations with great precision, with a limited set of free parameters. Still, the presence of  the cosmological constant raises a number of conceptual issues such as the coincidence problem, and the related  fact that a cosmological constant  is not technically natural from the point of view of the stability under radiative corrections. 
Thus, much effort is being devoted to looking for alternatives,  see e.g. \cite{Bull:2015stt} for a recent review. Most importantly, the wealth of high-quality cosmological data obtained in recent years, and expected in the near future, allows us  to test and possibly rule out any alternative model that does not simply reduce to $\Lambda$CDM in some limit.

An idea which is particularly fascinating from the theoretical point of view is to try  to explain the acceleration of the Universe by modifying General Relativity (GR) at cosmological scales, without introducing a cosmological constant.
This idea started to become popular several years ago with the DGP model~\cite{Dvali:2000hr}, and  also underlies the intense  activity of the last few years on the dRGT  theory of massive gravity~\cite{deRham:2010ik,deRham:2010kj,Hassan:2011hr} (see  \cite{Hinterbichler:2011tt,deRham:2014zqa} for  reviews),
as well as on the Hassan-Rosen  bigravity theory~\cite{Hassan:2011zd} (see  \cite{Schmidt-May:2015vnx} for a recent review). These studies  have revealed that, even if one succeeds in the  challenging task of  constructing  a theory of modified gravity  which is theoretically consistent (e.g. ghost-free and without causality violations),  one is in general still far from obtaining an interesting cosmological model.
For instance, the DGP model has a
self-accelerating branch~\cite{Deffayet:2000uy,Deffayet:2001pu}, but    it
succumbed to a fatal ghost-like instability which is present  in this  branch~\cite{Luty:2003vm,Nicolis:2004qq,Gorbunov:2005zk,Charmousis:2006pn,Izumi:2006ca}. Ghost-free massive gravity has difficulties even in recovering a viable cosmological evolution  at the background level, since there are no isotropic flat FRW solutions~\cite{D'Amico:2011jj}.  On the other hand there are open  FRW solutions, which however suffer from  strong coupling and ghost-like instabilities \cite{DeFelice:2013awa}. 
Bigravity has  cosmological FRW solutions, that have been much studied recently (see e.g. \cite{Volkov:2011an,Comelli:2011zm,vonStrauss:2011mq,Akrami:2012vf,Tamanini:2013xia,Fasiello:2013woa,Akrami:2013ffa,Konnig:2013gxa,Comelli:2014bqa,Solomon:2014dua,DeFelice:2014nja,Konnig:2014dna})
and fall into two branches. One is a so-called 
infinite branch, where at the background level there is a viable self-accelerating solution; however, at the level of  cosmological perturbations it suffers from  fatal ghost-like instabilities both in the scalar and in the tensor sectors
\cite{Lagos:2014lca,Cusin:2014psa,Konnig:2015lfa}. The finite branch contains instead a scalar instability at early time
\cite{Comelli:2012db,Konnig:2014xva,Lagos:2014lca}, which invalidates the use of linear perturbation theory and the predictivity of the theory. Tuning the Planck mass  associated to the second metric, this instability can 
be pushed back to unobservably early times~\cite{Akrami:2015qga}. However, in this limit the theory becomes indistinguishable from  $\Lambda$CDM with a  graviton mass which goes to zero. It is therefore  of limited interest,  for the purpose of testing GR and $\Lambda$CDM against a competitor theory.  Besides these two branches, in bigravity there is also  a `non-dynamical' branch, which also has a number of pathologies~\cite{Schmidt-May:2015vnx}.

The introduction of non-local terms opens up new possibilities for building models that modify gravity in the infrared (IR), since in the IR operators such as the inverse d'Alembertian, $\iBox$, become relevant. Even if locality is one of the basic principles of quantum field theory, at an effective level non-local terms unavoidably arise. Indeed, non-localities emerge  already  at a  classical level, when one
integrates out some fast degree of freedom to obtain an effective theory for the slow degrees of freedom, while at the quantum level loops involving light or massless particles generate non-local terms in the quantum
effective action. One can therefore begin with a purely phenomenological attitude, looking for non-local terms that, added to the standard Einstein equations without cosmological constant, produce a viable cosmological evolution. Of course,  for  this program to be successful, one must eventually also be able to derive such non-localities from a fundamental local theory, displaying the mechanism that generates them. However it is quite natural, and indeed probably unavoidable, to start with a purely phenomenological attitude,  since identifying first non-local terms that ``work",  i.e. give a viable cosmology,  might give crucial hints on the underlying mechanism that generates them.

In this spirit, Deser and Woodard \cite{Deser:2007jk,Deser:2013uya,Woodard:2014iga} proposed a non-local modification of gravity constructed adding to the Ricci scalar $R$ in the Einstein-Hilbert action a term of the form $Rf(\iBox R)$ (see also \cite{Barvinsky:2003kg,Barvinsky:2011hd,Barvinsky:2011rk} for related ideas). 
Observe that the function $f$ is dimensionless, so this model does not involve an explicit mass scale, contrary to the non-local models that we will discuss below.
There has been much activity in the literature  in identifying the form that the
function $f(\iBox R)$ should have in order to obtain a viable  background evolution, see e.g. \cite{Koivisto:2008xfa,Koivisto:2008dh,Capozziello:2008gu,Elizalde:2011su,Zhang:2011uv,Elizalde:2012ja,Park:2012cp,Bamba:2012ky}. The result, 
$f(X)=a_1[\tanh (a_2Y+a_3Y^2+a_4Y^3)-1]$, where $Y=X+a_5$, and $a_1,\ldots a_5$ are coefficients fitted to the observed expansion history, does not look very natural. More importantly, once the function $f(X)$ is fixed in this way, one can compute the cosmological perturbations of the model and compare with the data. The result is that this model is ruled out with great statistical significance,   at  7.8$\sigma$  from redshift space distortions, and at  5.9$\s$  from weak lensing~\cite{Dodelson:2013sma}.

From the above discussion it appears that constructing an IR modification of GR that is cosmologically viable (and still distinguishable from $\Lambda$CDM) is highly non-trivial. In the last couple of years, however, a different class of  non-local models has emerged, in which the non-local terms are associated to a mass scale $m$ (and which therefore are not of the Deser-Woodard type), and that appear to work remarkably well at the phenomenological level. The first model of this type was proposed 
in \cite{Maggiore:2013mea} (elaborating on previous works related to
the degravitation idea~\cite{ArkaniHamed:2002fu,Dvali:2006su,Dvali:2007kt}, as well as on attempts at writing massive gravity in non-local form~\cite{Porrati:2002cp,Jaccard:2013gla}), and is defined by
the non-local equation of motion
\be\label{RT}
\Gmn -\frac{m^2}{3}\(\gmn\iBox R\)^{\rm T}=8\pi G\,\Tmn\, .
\ee
The superscript T denotes the operation of taking the transverse part of a tensor (which is itself a non-local operation), $\Box$ is the covariant d'Alembertian computed with the curved-space metric $\gmn$, and its inverse
$\iBox$ is defined using the retarded Green's function, to ensure causality. The extraction of the transverse part ensures that energy-momentum conservation is  automatically satisfied. The factor $1/3$ provides a convenient normalization for the new mass parameter $m$. We will refer to this model as the ``RT" model, where R stands for the Ricci scalar and T for the extraction of the transverse part.

A closed form for the action corresponding to \eq{RT} is not known. This model is however  closely related to another non-local model, subsequently proposed in \cite{Maggiore:2014sia}, and defined by the action
\be\label{RR}
S=\frac{\mplr^2}{2}\int d^{4}x \sqrt{-g}\, 
\[R-\frac{1}{6} m^2R\frac{1}{\Box^2} R\]\, .
\ee
We will call this the ``RR" model. Computing the equations of motion derived from the action (\ref{RR}) and linearizing them over flat space  
one finds the same equations of motion as those obtained by linearizing \eq{RT} over flat space. However, at the full non-linear level, or linearizing over a background different from Minkowski, the RT and RR models are different. Thus, the RR model is in a sense the simplest non-local model of this class, at the level of the action, while the RT model can be considered as a sort of non-linear generalization of it. Of course, one can imagine different non-linear extensions of the RR model, so we just take the RR and RT models as two representative models of this class of non-local theories. It is however interesting to observe that,
at the   level of terms quadratic in the curvature, \eq{RR} is the most general viable non-local action of this class. Indeed, at the quadratic level in the curvature, a basis is given by $R^2$, $\Rmn\RMN$ and $C_{\mu\nu\rho\sigma} C^{\mu\nu\rho\sigma}$, where $C^{\mu\nu\rho\sigma}$ is the Weyl tensor. Thus, beside the term $m^2R\Box^{-2}R$, at the quadratic level the most general non-local term of this class also contains terms proportional to 
$m_2^2\Rmn \Box^{-2}\RMN$ and $m_3^2C_{\mu\nu\rho\sigma} \Box^{-2}C^{\mu\nu\rho\sigma}$, with some new mass parameters $m_2, m_3$. However, 
as recently shown in \cite{Cusin:2015rex}, a term of the form $m_2^2\Rmn \Box^{-2}\RMN$ does not provide a stable background evolution,\footnote{This is analogous to the fact that, in a Deser-Woodard type model,
a term of the form $\RMN f(\iBox\Rmn)$ shows instabilities \cite{Ferreira:2013tqn}.} while a term  $m_3^2C_{\mu\nu\rho\sigma} \Box^{-2}C^{\mu\nu\rho\sigma}$
 does not contribute to the background evolution in FRW, but at the level of perturbations induces instabilities in the tensor sector. Thus, at the level of terms in the action which are quadratic in the curvature, only the term $m^2R\Box^{-2}R$
is phenomenologically viable.  This already gives useful hints for searching the mechanism that could generate the required non-local terms. A recent discussion on how the non-local term of the RR and RT model could be generated in a fundamental local theory is given in \cite{Maggiore:2015rma,Maggiore:2016fbn}, where  it is observed that these non-local terms corresponds to a dynamical mass generation for the conformal mode of the metric.

The study of the RR and RT models shows that they have a very appealing phenomenology.
These models  have no vDVZ discontinuity, and no Vainshtein mechanism is needed to restore continuity 
with GR \cite{Maggiore:2013mea,Maggiore:2014sia,Kehagias:2014sda}. Therefore both models, with $m={\cal O}(H_0)$  as will be required by cosmology, pass without problems all solar system  tests.\footnote{See also app.~\ref{app:solar} below, where we will discuss a related  issue on the comparison with Lunar Laser Ranging, raised in \cite{Barreira:2014kra}.} 
In a cosmological setting,
at the background level they both 
dynamically generate  an effective  dark energy and have a realistic background FRW evolution, without the need of introducing a cosmological constant \cite{Maggiore:2013mea,Maggiore:2014sia,Foffa:2013vma}. The cosmological perturbations of these non-local models are well-behaved. This is true both  in the scalar sector, as shown 
in \cite{Dirian:2014ara}, and in the tensor sector, as we will show in this paper (see also \cite{Cusin:2015rex}). The study of the effect of these cosmological perturbations shows that  (contrary e.g. to the Deser-Woodard model), the predictions of both the RR and RT models are 
consistent with CMB, supernovae, BAO and structure formation 
data~\cite{Nesseris:2014mea,Dirian:2014ara,Barreira:2014kra}.\footnote{It should also be appreciated that a non-local model such as the RT or the RR model only introduces one new parameter $m$, which replaces the cosmological constant in $\Lambda$CDM. By comparison, bigravity replaces the cosmological constant by a set of 5 parameters $\beta_n$, $n=0,\ldots , 4$ and also introduces a new Planck mass associated to the second metric, and viable solutions are searched tuning this parameter space. Similarly,
in the Deser-Woodard model one tunes  a whole function $f(X)$.}
Further conceptual and phenomenological aspects of these models have been discussed in 
\cite{Modesto:2013jea,Foffa:2013sma,Conroy:2014eja,Cusin:2014zoa,Dirian:2014xoa,Mitsou:2015yfa,Barreira:2015fpa,Barreira:2015vra,Codello:2015pga,Cusin:2016nzi}.

It is also interesting to observe that, when the mass parameter $m$ is taken to be of order $H_0$,  non-local terms such as that in \eq{RR} modify the evolution near the present cosmological epoch, but 
are irrelevant at early times. These non-local models can therefore be supplemented with  any desired inflationary potential  in the early Universe, providing standard inflation at early time, independently of the fact that they generate dark energy at late time.  In particular this means that the typical  values for the amplitude and tilt $A_s$ and $n_s$ of the primordial scalar fluctuations can be obtained, in the non-local models, exactly as in $\Lambda$CDM, i.e. supplementing the model with an  inflationary model at high energies. Furthermore, 
as first discussed  in \cite{Maggiore:2015rma}
(see also \cite{Codello:2015pga}\cite{Cusin:2016nzi}\cite{Codello:2016neo}), for these non-local models there is a  very natural way of extending them so to  obtain Starobinsky inflation at high energies. For instance,  the action 
\be\label{6RRStar2}
S=\frac{\mplr^2}{2}\int d^4x\, \sqrt{-g}\, 
\[ R+\frac{1}{6M_{\rm S}^2} R\( 1- \frac{\Lambda_{\rm S}^4}{\Box^2}\) R\]\, ,
\ee
(where $M_{\rm S}\simeq 10^{13}$~GeV is the mass scale of the Starobinski model and $\Lambda_{\rm S}^4=M_{\rm S}^2m^2$) naturally interpolates between Starobinski inflation at early times and the RR model (\ref{RR}) at late times. Similarly, in  the RT model we can combine the non-local contribution in \eq{RT} with the contribution to the equations of motion coming from the $R^2$ term in the Starobinski model. As discussed in \cite{Maggiore:2015rma,Cusin:2016nzi} (see also \cite{Codello:2016neo}),
at early times the non-local term is irrelevant and we recover the evolution of  the  Starobinski inflationary model, while at late times the local $R^2$ term becomes irrelevant and we recover the evolution of the non-local models.

In this paper, improving and expanding the analysis that we have presented in ref.~\cite{Dirian:2014bma} (henceforth referred as Paper~I), 
we  implement the cosmological perturbations of the two non-local models into a Boltzmann code, and perform a detailed comparison with the data as well as Bayesian parameter estimation. We can then assess the performance of the two non-local models and see if they can `defy' $\Lambda$CDM, from the point of view of fitting the data. This is a level of comparison with the data, and with $\Lambda$CDM, that none of the modified gravity models discussed above have reached.

We improve on the analysis of Paper~I in several respects. We  use the  {\em Planck\,} 2015 data, which were not yet publicly available at the time that Paper~I appeared. We also include a more extended set of BAO data, including both  isotropic and anisotropic BAO  (while for  SNe we use, as in Paper~I, the JLA dataset) and we also add a comparison with  growth rate data. We then perform a Bayesian model comparison between the RR, RT and $\Lambda$CDM models, using the Savage-Dickey method, embedding them in a larger model with both the non-local term and the cosmological constant. Finally, we also show that tensor perturbations in these non-local models are well-behaved, and in fact basically indistinguishable from $\Lambda$CDM.

The paper is organized as follows. In section~\ref{sect:ObsConst} we discuss the datasets that we use and   give the results obtained from cosmological parameter extraction and goodness-of-fit for the non-local models, using CMB, BAO, type~Ia supernovae and $H_0$ measurements in different combinations, and we compare with the results obtained for $\Lambda$CDM with the same datasets. In sect.~\ref{sect:ModelComp} we perform a Bayesian comparison between the non-local models and $\Lambda$CDM, by computing the corresponding Bayes factors using the Savage-Dickey density ratio.
In sect.~\ref{sect:growth}, using the best-fit values of the cosmological parameters determined by CMB, BAO and supernovae, we compare the prediction of the non-local models with  structure formation and growth rate data.
Tensor perturbations are discussed in sect.~\ref{sect:tensor}. In app.~\ref{app:class} we provide technical details about the implementation of the nonlocal models into a Boltzmann code and exhibit explicitly the equations used, while in app.~\ref{app:solar}  we  discuss an issue on the comparison with Lunar Laser Ranging data. 
Our notation and conventions are the same as in Paper~I \cite{Dirian:2014bma} and in \cite{Dirian:2014ara}. 
Our modified Boltzmann code is publicly available on GitHub \cite{git_nonlocal}.

\section{Comparison with CMB, SNe and BAO data}\label{sect:ObsConst}

The equations of motion of the RR and RT non-local models, both at the background and linear scalar perturbations level over FRW, have been discussed at length in \cite{Dirian:2014ara}. For clarity we recall them briefly in app.~\ref{app:class}, in the format used in our numerical analysis. 
We have implemented these equations in CLASS \cite{Class}, and
constrained these models with observations using the Markov Chain Monte Carlo (MCMC) code Montepython \cite{MP}, originally interfaced with the CLASS Boltzmann code.  First of all this allows us to apply Bayesian parameter inference on the two non-local models. We can then compare the models to each other, as well as  to the concordance model $\Lambda$CDM, through Bayesian model selection, given the data.

\subsection{Datasets}

The main datasets considered throughout this paper are the same as the ones used in the \textit{Planck} 2015 papers, in particular in \cite{Planck_2015_CP} for constraining the base $\Lambda$CDM model, and in \cite{Planck_2015_DE} for constraining various phenomenological dark energy models. We adopt this choice so as to consider the most conservative data up to date, where the systematic uncertainties are mostly under control. For completeness we review them briefly in this section, referring the reader to the original papers for more detailed explanations. 

\paragraph{CMB.}

We consider the likelihoods given in the recent \textit{Planck} 2015 \cite{Planck_2015_1} measurements of the angular (cross-)power spectra of the CMB. This will update the analysis presented in Paper~I 
\cite{Dirian:2014bma}, where we used the \textit{Planck} 2013 nominal mission temperature data \cite{Planck_2013_1}.
In particular, we take the (full-mission) lowTEB data for low multipoles ($\ell \leq 29$) and the high-$\ell$ Plik  TT,TE,EE (cross-half-mission) ones for the high multipoles ($\ell > 29$) of the temperature and polarization auto- and cross- power spectra \cite{Planck_2015_DE, Planck_2015_wiki}. 

Furthermore, since the non-local models describe a  dynamical and clustering dark energy that emerges at late times\cite{Dirian:2014ara}, we also include the temperature$+$polarization (T$+$P) lensing data (where only the conservative multipole range $\ell =40-400$ is used), that provide CMB constraints on late-time cosmology, and more generically allow one to break degeneracies in the primary CMB anisotropies \cite{Planck_2015_Lkl,Planck_2015_lens}.

\vspace{5mm}
\noindent
In addition to CMB data, following the {\em Planck\,} analysis \cite{Planck_2015_CP, Planck_2015_DE}, we also include datasets from astrophysical measurements, so as to break further CMB degeneracies and reach tighter constraints on the parameter space. In particular, we consider the following datasets.

\paragraph{Type Ia supernovae.}

We consider the data from the SDSS-II/SNLS3 Joint Light-curve Analysis (JLA) \cite{JLA_2014} for  SN~Ia, using the complete (non-compressed) corresponding likelihoods.  Combining the latter with the ones obtained from the \textit{Planck} data allows one to put an independent constraint on the matter density fraction $\Omega_m$, and  breaks the CMB degeneracy in the 
$H_0-\Omega_m$ plane. This is particularly useful in constraining the non-local models, as already pointed out in Paper~I.

\paragraph{Baryon Acoustic Oscillations.}

As discussed in detail in \cite{Maggiore:2013mea,Maggiore:2014sia,Foffa:2013vma},
the RR and RT non-local models have a phantom dark energy equation of state, $ w_{\rm DE}(z) < -1$ for $z \geq 0$. Thus, the history of the growth of structures in the late universe is modified compared to a $\Lambda$CDM scenario. To further test this feature, it is useful to include different BAO scale measurements.
As datasets we consider the isotropic constraints provided by 6dFGS at $z_{\rm eff}=0.106$ \cite{Beutler_6dF_BAO_2012}, SDSS-MGS DR7 at $z_{\rm eff}=0.15$ \cite{Ross_SDSS_2014} and BOSS LOWZ at $z_{\rm eff}=0.32$  \cite{And_BOSS_2013}. These data provide measurements of the acoustic-scale distance ratio $D_{\rm V}(z_{\rm eff})/r_{\rm d}$. We also include the anisotropic constraints from CMASS at $z_{\rm eff}=0.57$ \cite{And_BOSS_2013}. Anisotropic constraints separate the clustering effects into their longitudinal and transverse component relative to the line-of-sight. This allows one to put separate constraints on  the ratios $H(z_{\rm eff})/r_{\rm d}$ and  $D_{\rm A}(z_{\rm eff})/r_{\rm d}$, where $D_{\rm A}$ is the angular diameter distance, and breaks the degeneracy in the $D_{\rm A}$-$H$ plane which arises when considering isotropic constraints. By comparison, in  Paper~I we only included the isotropic CMASS measurements of ref.~\cite{And_BOSS_2013}, and did not consider the point provided by \cite{Ross_SDSS_2014}. 

\paragraph{$H_0$ prior.}

Finally, as in Paper~I, we will perform several analysis including different combinations of these datasets. First we will constrain the models given the \textit{Planck} 2015 likelihoods only, second we will join to them the ones of JLA and BAO, and finally we will further add two different  priors on $H_0$ provided by different local observations. These values will be taken to be $H_0 = 70.6 \pm 3.3$ \cite{Efsta_H0_2013} and 
$H_0 = 73.8 \pm 2.4$ \cite{Riess:2011yx}.  In particular, we will  use the value  $H_0 = 73.8 \pm 2.4$ as an example of the impact of a high value of $H_0$. 
The most recent analysis of local measurements, which appeared after this work was finished, gives
$H_0 = 73.02 \pm 1.79$ \cite{Riess:2016jrr}.

\subsection{Parameter space and MCMC}

The  datasets discussed above will be used for  constraining the three statistical models that for definiteness we denote by $\mathcal{M}_\Lambda$, $\mathcal{M}_{\rm RT}$ and $\mathcal{M}_{\rm RR}$ and are associated to their respective cosmological models. For the consistency of our analysis, the initial conditions (inflationary scenario), ionization history and matter content of the universe will be chosen following the \textit{Planck} baseline \cite{Planck_2015_CP, Planck_2013_CP}. In each of the non-local models we have a parameter, $m^2$, that replaces the cosmological constant in $\Lambda$CDM, so the non-local models have the same number of parameters as
$\Lambda$CDM. Furthermore, in the spatially flat case that we are considering, in $\Lambda$CDM  the dark energy density fraction $\ola$ can be taken as a derived parameter, fixed in terms of the other parameters by the flatness condition. Similarly, in the non-local models $m^2$ can be taken as a derived parameter,  fixed again  by the flatness condition. Thus, not only the non-local models have the same number of parameters as $\Lambda$CDM, but in fact these can be chosen so that the independent parameters are exactly the same in the non-local models and in $\Lambda$CDM, which facilitates the comparison.\footnote{As discussed, e.g. in \cite{Maggiore:2013mea,Maggiore:2014sia,Foffa:2013vma}, these non-local models can be formally put in a local form introducing auxiliary fields such as   $U=-\iBox R$ and 
$S= -\iBox U$ (for the RR model, and similarly  for the RT model, which rather involves an auxiliary four-vector field). This in principle raises the question of whether the initial conditions for these auxiliary fields are further free parameters of the model.  At the conceptual level, it is important to realize that 
 the initial conditions of these fields are actually fixed within a given model, and in principle should follow  from the fundamental theory that generates  these effective non-localities. Therefore, they do not correspond to extra dynamical degrees of freedom (see in particular the discussion below eq.~3.6 in \cite{Foffa:2013vma} and refs. therein, as well as ref.~\cite{Foffa:2013sma}). From the practical point of view, this still leaves us with the issue of making an appropriate choice of initial conditions. For the RR model we will choose the initial conditions 
$U(t_*)=\dot{U}(t_*)=0$ and $S(t_*)=\dot{S}(t_*)=0$ at some initial time $t_*$ deep in RD (and similarly for the auxiliary fields of the RT model). As shown in sect.~4.1 of \cite{Foffa:2013vma}, the corresponding solution is an attractor in the space of solutions with different  initial conditions so, within a large attraction basin, our results are actually independent of  this choice. In other words, the homogeneous solutions associated to equation such as $\Box U=-R$ are all decaying modes. In fact, it is just this property that selected the RR and RT model with respect to other models, such as the one featuring a term $(\iBox\Gmn)^T$, originally considered in \cite{Jaccard:2013gla}, for which instead the homogeneous equations have growing modes.}

 For  the neutrino masses  will  use the same values used in  the \textit{Planck} 2015 baseline analysis \cite{Planck_2015_CP}, i.e. two massless and a massive neutrinos, with
$\sum_{\nu} m_{\nu}=0.06$~eV, and we are fixing the effective number of neutrino species to 
$N_{\rm eff}=3.046$. In the Conclusions, following \cite{DirainBarreira:inprep}, we will shortly comment  on the effect of the neutrino masses.

As independent cosmological parameters we take the Hubble parameter today
$H_0 = 100 h \, \rm{km} \, \rm{s}^{-1} \rm{Mpc}^{-1}$, the physical baryon and cold dark matter density fractions today $\omega_b = \Omega_b h^2$ and $\omega_c = \Omega_c h^2$, respectively, the amplitude  $A_s$ of primordial scalar perturbations, the spectral tilt $n_s$  and  the reionization optical depth  $\tau_{\rm re}$, so we have a $6$-dimensional parameter space. We choose unbounded flat priors on all these parameters, except for $\tau_{\rm re}$ which is taken to be bounded from below at the value of $0.01$. The corresponding vector in parameter space is given by, 
\begin{align}
\theta = ( H_0, \omega_b, \omega_c, A_s, n_s, \tau_{\rm re} ) \, . \label{base_param}
\end{align}
This parameter space will be explored via a Metropolis-Hasting algorithm. Our construction of the MCMC proceeds in two steps. The first consists in constructing a chain using wide Gaussian proposal distributions, which for all three models is taken to be centered initially on the \textit{Planck} best fit values obtained for $\Lambda$CDM. The posterior distribution obtained then provides a new proposal distribution (covariance matrix) for a second run, where we restart from the best-fit points of the latter posterior such that the burn-in  is avoided.
In order to get reliable results and well shaped final distributions, we typically constructed two or three chains of $10^5$ trials each and adjusted the jumping factor so as to get an acceptance rate $0.2 < r < 0.4$. The convergence of the set of chains is checked through a Gelman-Rubin convergence diagnostic. The best-fit points for each distribution are then refined by starting from the best-fit values of the (final) global run, and constructing a ``cold" chain, i.e. a chain controlled by a constant temperature parameter $T$ introduced into the MCMC acceptance probability as,
\begin{align}
\alpha(\theta_n, \theta_{n+1}) = \mathrm{min} \bigg\{ 1 , \bigg( \frac{P(\theta_{n+1})}{P(\theta_{n})} \bigg)^{1/T} \bigg\} \, ,
\end{align}
where $\theta_n$ is the $n$-th point sampled in parameter space and $P(\theta_n)$ its likelihood. Therefore the lower $T$ is, the stronger the chain will converge towards the nearest maximum of the likelihood 
distribution.\footnote{Observe that nothing guarantees that the corresponding maximum reached is \textit{the} global one, but starting the evaluation from the best-fit point obtained from the global run provides already representative best-fit values, sufficient for the purpose of our study.} In our analysis we choose $T=50$. 

\subsection{Results}\label{sect:SDresults}

In Table~\ref{partable} we show the mean values  and the corresponding $\chi^2$ for $\Lambda$CDM, the RT model and the RR model, for different combinations of datasets. In the upper left table we only use the {\em Planck} CMB data  presented in the previous subsection, while in the upper right table we combine them with BAO and JLA supernovae. In the two lower tables we further add a prior on $H_0$, namely $H_0 = 70.6 \pm 3.3$ for the lower left table  and a high value, chosen for definiteness $H_0 = 73.8 \pm 2.4$, for the lower right table. Beside giving the mean values of our six fundamental parameters, we also give the corresponding derived values of $\sigma_8$ and of the effective redshift to reionization $z_{\rm re}$.

\begin{table}[t]
\centering
\resizebox{15cm}{!}{
\begin{tabular}{|l||c|c|c||c|c|c|} 
 \hline 
\multicolumn{1}{|l||}{ } & \multicolumn{3}{|c||}{Planck} & \multicolumn{3}{|c|}{BAO+Planck+JLA} \\ \hline
Param & $\Lambda$CDM  & RT &RR & $\Lambda$CDM  & RT &RR\\ \hline 
$100~\omega_{b }$  &$2.225_{-0.016}^{+0.016}$  &  $2.224_{-0.016}^{+0.016}$ &  $2.227_{-0.016}^{+0.016}$ 

& $2.228_{-0.015}^{+0.014}$ &$2.223_{-0.014}^{+0.014}$ &$2.213_{-0.014}^{+0.014}$ \\
$\omega_c$  & $0.1194_{-0.0015}^{+0.0014}$ &  $0.1195_{-0.0015}^{+0.0014}$ & $0.1191_{-0.0015}^{+0.0014}$  

& $0.119_{-0.0011}^{+0.0011}$  & $0.1197_{-0.00096}^{+0.0011}$ &  $0.121_{-0.001}^{+0.001}$ \\ 
$H_0$  &$67.5_{-0.66}^{+0.65}$&  $68.86_{-0.7}^{+0.69}$ & $71.51_{-0.84}^{+0.81}$

& $67.67_{-0.5}^{+0.47}$ &$68.76_{-0.51}^{+0.46}$ &$70.44_{-0.56}^{+0.56}$\\ 
$\ln (10^{10}A_{s })$& $3.064_{-0.025}^{+0.025}$&$3.057_{-0.026}^{+0.026}$ & $3.047_{-0.025}^{+0.026}$ 

& $3.066_{-0.026}^{+0.019}$&$3.056_{-0.023}^{+0.021}$ & $3.027_{-0.023}^{+0.027}$ \\ 
$n_{s}$  & $0.9647_{-0.0049}^{+0.0048}$&  $0.9643_{-0.005}^{+0.0049}$  &  $0.9649_{-0.0049}^{+0.0049}$

 & $0.9656_{-0.0043}^{+0.0041}$& $0.9637_{-0.0041}^{+0.0039}$ &  $0.9601_{-0.0039}^{+0.004}$ \\ 
$\tau_{\rm re}$  & $0.0653_{-0.014}^{+0.014}$  & $0.06221_{-0.014}^{+0.014}$& $0.05733_{-0.014}^{+0.014}$

 & $0.06678_{-0.013}^{+0.011}$ &  $0.0611_{-0.013}^{+0.011}$& $0.04516_{-0.012}^{+0.014}$ \\ 
$z_{\rm re}$  & $8.752_{-1.2}^{+1.4}$& $8.442_{-1.2}^{+1.5}$ &  $7.932_{-1.2}^{+1.5}$ 

& $8.893_{-1.2}^{+1.1}$ & $8.359_{-1.2}^{+1.2}$ & $6.707_{-1.2}^{+1.7}$\\ 
$\sigma_8$  & $0.8171_{-0.0089}^{+0.0089}$& $0.8283_{-0.0096}^{+0.0092}$ &  $0.8487_{-0.0096}^{+0.0097}$

&$0.817_{-0.0095}^{+0.0076}$ & $0.8283_{-0.0093}^{+0.0085}$ & $0.8443_{-0.0099}^{+0.01}$ \\   
\hline 
$\chi^2_{\rm min}$ &$12943.3$ &$12943.1$&$12941.7$ &$13631.0$  &$13631.6$ &$13637.0$\\
$\Delta \chi^2_{\rm min}$ &$1.6$ & $1.4$&$0$ &$0$  &$0.6$ &$6.0$\\
\hline
\end{tabular}
}
\resizebox{15cm}{!}{
\begin{tabular}{|l||c|c|c||c|c|c|} 
 \hline 
\multicolumn{1}{|l||}{ } & \multicolumn{3}{|c||}{BAO+Planck+JLA+$(\rm{H_0}=70.6$)} & \multicolumn{3}{|c|}{BAO+Planck+JLA+$(\rm{H_0}=73.8)$} \\ \hline
Param & $\Lambda$CDM  & RT &RR & $\Lambda$CDM  & RT &RR\\ \hline 
$100~\omega_{b }$  &$2.229_{-0.015}^{+0.014}$ & $2.223_{-0.014}^{+0.014}$ &  $2.215_{-0.014}^{+0.014}$

& $2.233_{-0.014}^{+0.014}$ &$2.226_{-0.014}^{+0.014}$ &$2.217_{-0.014}^{+0.014}$ \\
$\omega_c$  &  $0.1188_{-0.0011}^{+0.001}$& $0.1197_{-0.0011}^{+0.001}$ & $0.1208_{-0.001}^{+0.00099}$

& $0.1185_{-0.0011}^{+0.00097}$& $0.1194_{-0.001}^{+0.001}$ &$0.1207_{-0.00097}^{+0.00096}$ \\ 
$H_0$  & $67.75_{-0.47}^{+0.48}$& $68.75_{-0.48}^{+0.49}$&$70.57_{-0.56}^{+0.54}$

 & $67.93_{-0.43}^{+0.48}$ & $68.91_{-0.5}^{+0.49}$ & $70.65_{-0.54}^{+0.52}$ \\ 
$\log (10^{10}A_{s })$ & $3.069_{-0.024}^{+0.024}$& $3.056_{-0.022}^{+0.026}$&$3.03_{-0.021}^{+0.021}$

& $3.077_{-0.019}^{+0.026}$&  $3.061_{-0.022}^{+0.026}$ & $3.031_{-0.022}^{+0.018}$ \\ 
$n_{s}$  & $0.9662_{-0.0042}^{+0.0042}$& $0.9637_{-0.0042}^{+0.0041}$  & $0.9607_{-0.0041}^{+0.0039}$

& $0.9671_{-0.0041}^{+0.0041}$ &$0.9645_{-0.0041}^{+0.004}$ &$0.9611_{-0.004}^{+0.0038}$\\    
$\tau_{\rm re}$  &$0.06883_{-0.013}^{+0.012}$& $0.06099_{-0.011}^{+0.014}$ & $0.04701_{-0.011}^{+0.011}$

 & $0.07275_{-0.01}^{+0.014}$ &  $0.0641_{-0.012}^{+0.013}$& $0.04791_{-0.011}^{+0.01}$ \\  
$z_{\rm re}$  & $9.081_{-1.1}^{+1.2}$& $8.341_{-1}^{+1.4}$ & $6.922_{-1.1}^{+1.3}$ 

& $9.435_{-0.85}^{+1.3}$ & $8.636_{-1.1}^{+1.3}$ &  $7.02_{-1.2}^{+1.1}$\\  
$\sigma_8$  & $0.8179_{-0.0089}^{+0.0089}$& $0.8283_{-0.0089}^{+0.0095}$& $0.8452_{-0.0086}^{+0.0085}$

 & $0.8197_{-0.0075}^{+0.0096}$ &$0.8298_{-0.0086}^{+0.0095}$ & $0.8456_{-0.0088}^{+0.0081}$ \\  
\hline 
$\chi^2_{\rm min}$ &$13631.9$ &$13631.9$&$13637.0 $&$13637.5$  &$13636.1$ &$13638.9$\\
$\Delta \chi^2_{\rm min}$ &$0$ &$0$&$5.1$ &$1.4$  &$0$ &$2.8$\\
\hline
\end{tabular}
}
\caption{\label{partable} Parameter tables for $\Lambda$CDM and the non-local models.}
\end{table}

\begin{figure}[t]
\centering
\includegraphics[scale=0.8]{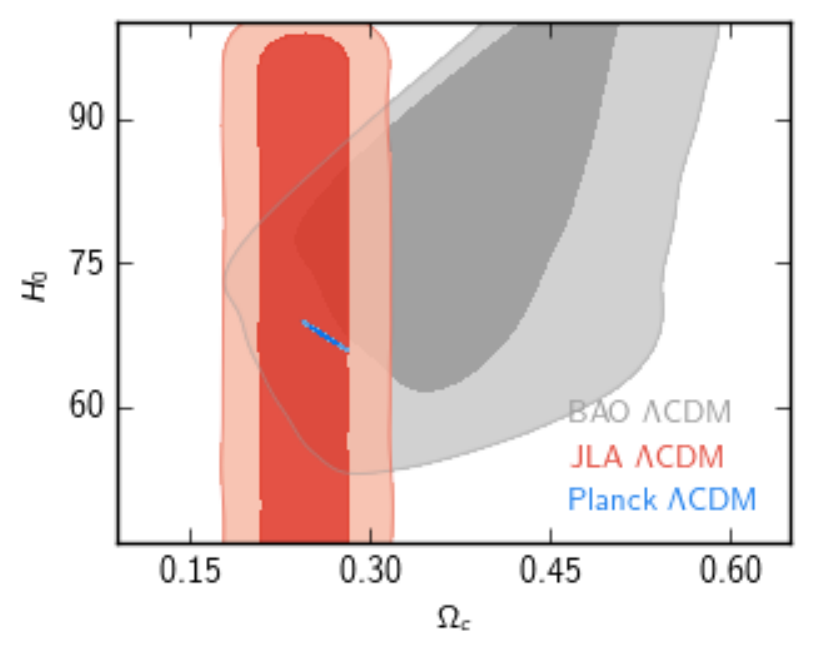}~~ \includegraphics[scale=0.8]{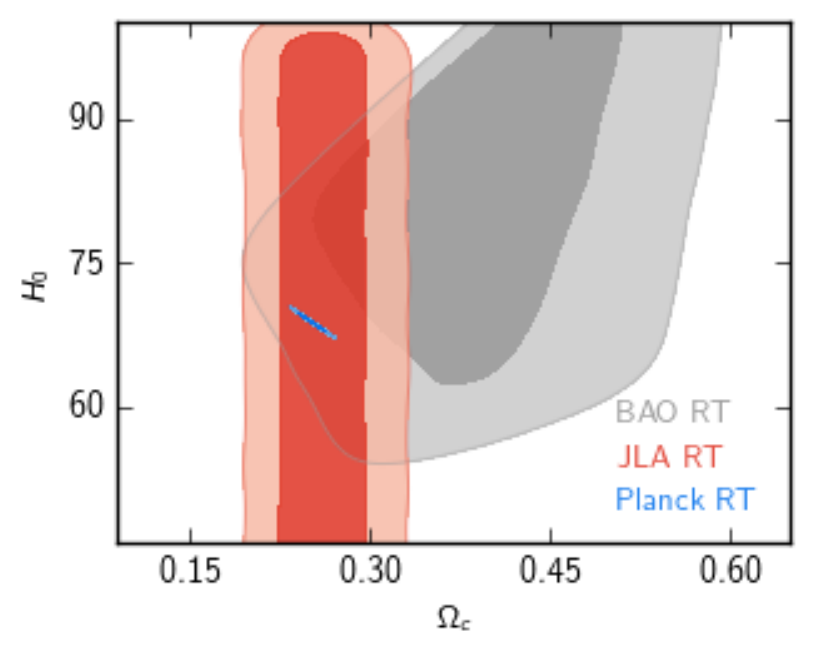} \\
\includegraphics[scale=0.8]{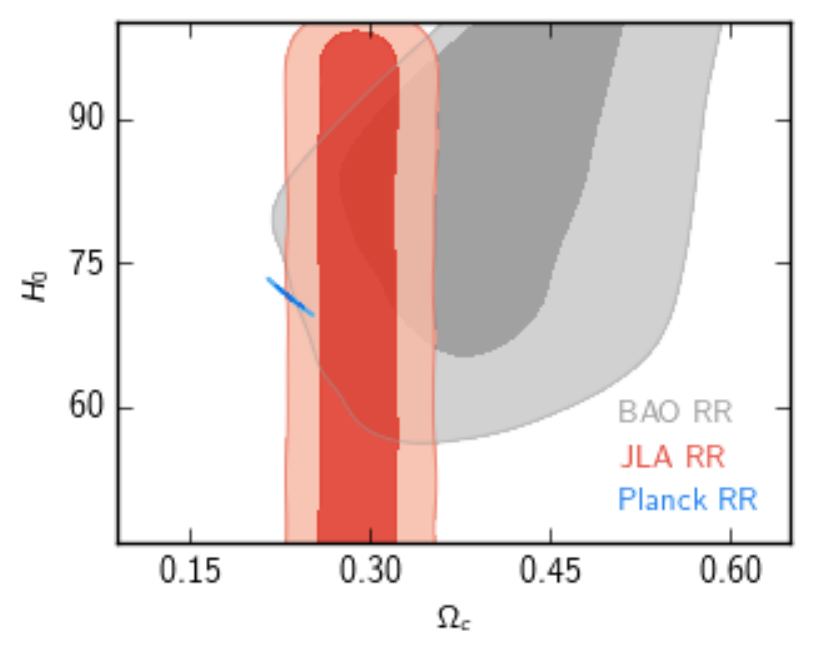} \\
\caption{\label{contours} The separate $1\sigma$ and $2\sigma$ contours for CMB, BAO and SNe in the plane
$(H_0,\Omega_c)$ for  the models and  $\Lambda$CDM, RT and RR.}
\end{figure}

In our case  the models to compare among each other have the same number of parameters.
Let us recall that, in this case, the  standard Akaike or Bayesian information criteria \cite{Trotta_BITS}  amount to comparing directly their respective goodness-of-fit, computing $|\Delta \chi^2_{ij}| = |\chi_{\mathrm{min},i}^2 - \chi_{\mathrm{min},j}^2|$ between models $\mathcal{M}_i$ and $\mathcal{M}_j$.
The larger the value of $|\Delta \chi^2_{ij}|$, the higher the evidence for the model with smaller goodness-of-fit. A difference $|\Delta \chi^2_{ij}| \leq  2$ implies  statistically equivalence between the two models compared, while $|\Delta \chi^2_{ij}| \gtrsim 2$ suggests ``weak evidence'', and $|\Delta \chi^2_{ij}|\gtrsim 6$ indicates ``strong evidence''.  
We will come back to the model comparison  in a fully Bayesian manner in Section~\ref{sect:ModelComp}, and for the moment we stick to the comparison of the $\chi^2$.

Comparing with Tables~1-3 of Paper~I, we see that the present results confirm and reinforce the trends already observed in Paper~I. 
In particular,  with CMB only, we see from the values of the $\chi^2$ that the three models perform in a very comparable manner, with the RR model being slightly better, but at a level which is not statistically significant.
However, once we add BAO and SNe, 
$\Lambda$CDM and the RT model remain statistically indistinguishable, while the RR model is significantly disfavored. This trend is strongly reinforced in the present analysis compared to Paper~I, since now 
$\chi^2_{\rm min, RR}-\chi^2_{{\rm min}, \Lambda{\rm CDM}}\simeq 6.0$, while in Paper~I we found $\chi^2_{\rm min, RR}-\chi^2_{{\rm min}, \Lambda{\rm CDM}}\simeq 3.2$. 
Just as in Paper~I, the origin of this result can be understood looking at the two-dimensional marginalized likelihoods in the $H_0-\Omega_c$ plane, shown in Fig.~\ref{contours} (to be compared with Fig.~1 of Paper~I). We see that for the RT model and for $\Lambda$CDM   the separate contours obtained from CMB, BAO and JLA supernovae agree very well, while for the RR model there is a slight tension between CMB and JLA. We emphasize, though, that even this `tension' is just at the $2\sigma$ level. In other words even the RR model, by itself, fits  the data at a fully acceptable level. However, in the comparison with the RT model and with $\Lambda$CDM, it clearly performs less well,  when no prior on $H_0$ is imposed (and assuming the Planck baseline value for the neutrino masses, see the Conclusions).
Observe that the shift in the BAO contour, with respect to Paper~I, is mostly due to the inclusion of the SDSS-MGS DR7 point at $z_{\rm eff}=0.15$. 

Concerning the mean values, we see from the table that the most significant difference is in $H_0$, with the non-local model (and particularly the RR model) favoring slightly higher values of $H_0$, compared to $\Lambda$CDM. This confirms again a result obtained in Paper~I. So, not surprisingly, adding as a prior the value of $H_0$ from local measurements goes in the direction of favoring the non-local models. Actually, with the value $H_0 = 70.6 \pm 3.3$ given in \cite{Efsta_H0_2013} the situation changes little, with $\Lambda$CDM and the RT model still statistically equivalent, and the RR model disfavored, while a higher value, such as $H_0 = 73.8 \pm 2.4$, would bring back the RR model to values of $\Delta\chi^2$ only weakly disfavored.
In all cases, $\Lambda$CDM and the RT model are statistically indistinguishable, with differences in favor of one or the other which depend on the  datasets and priors used.

From the above results we can also obtain the mean value for the derived parameter $m^2$ of the non-local models. Actually, it is convenient to use the dimensionless quantity
\be\label{defgammamH0}
\gamma =\frac{m^2}{9H_0^2}\, ,
\ee
that  enters in the study of the cosmological evolution~\cite{Maggiore:2013mea,Maggiore:2014sia}. Based on the values of the parameters given in Table~\ref{partable} we find, for $\gamma$, the values given in Table~\ref{tab:gamma}.

\begin{figure}[t]
\centering
\includegraphics[width=0.85\columnwidth]{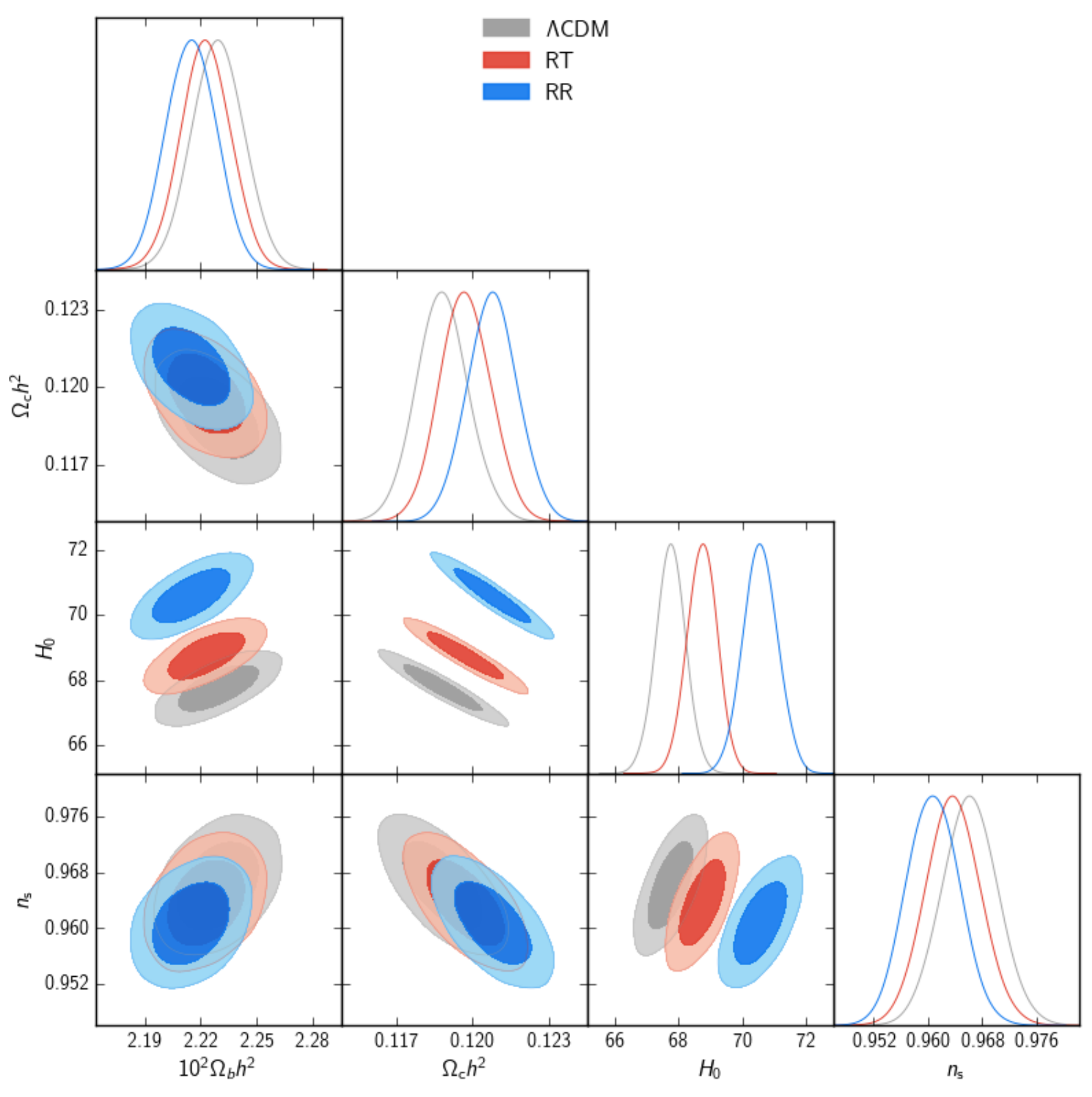} \\
\caption{\label{triangle} Triangle plot for Planck+BAO+JLA+$(H_0=70.6)$. }
\end{figure}

\begin{figure}[t]
\centering
\includegraphics[width=0.55\columnwidth]{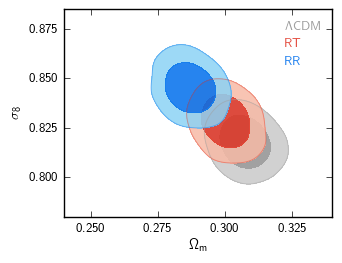}
\caption{\label{s8_Om} $\sigma_8 - \Omega_m$ contour plot for Planck+BAO+JLA+$(H_0=70.6)$.}
\end{figure}

\begin{figure}[th]
\centering
\includegraphics[width=0.7\columnwidth]{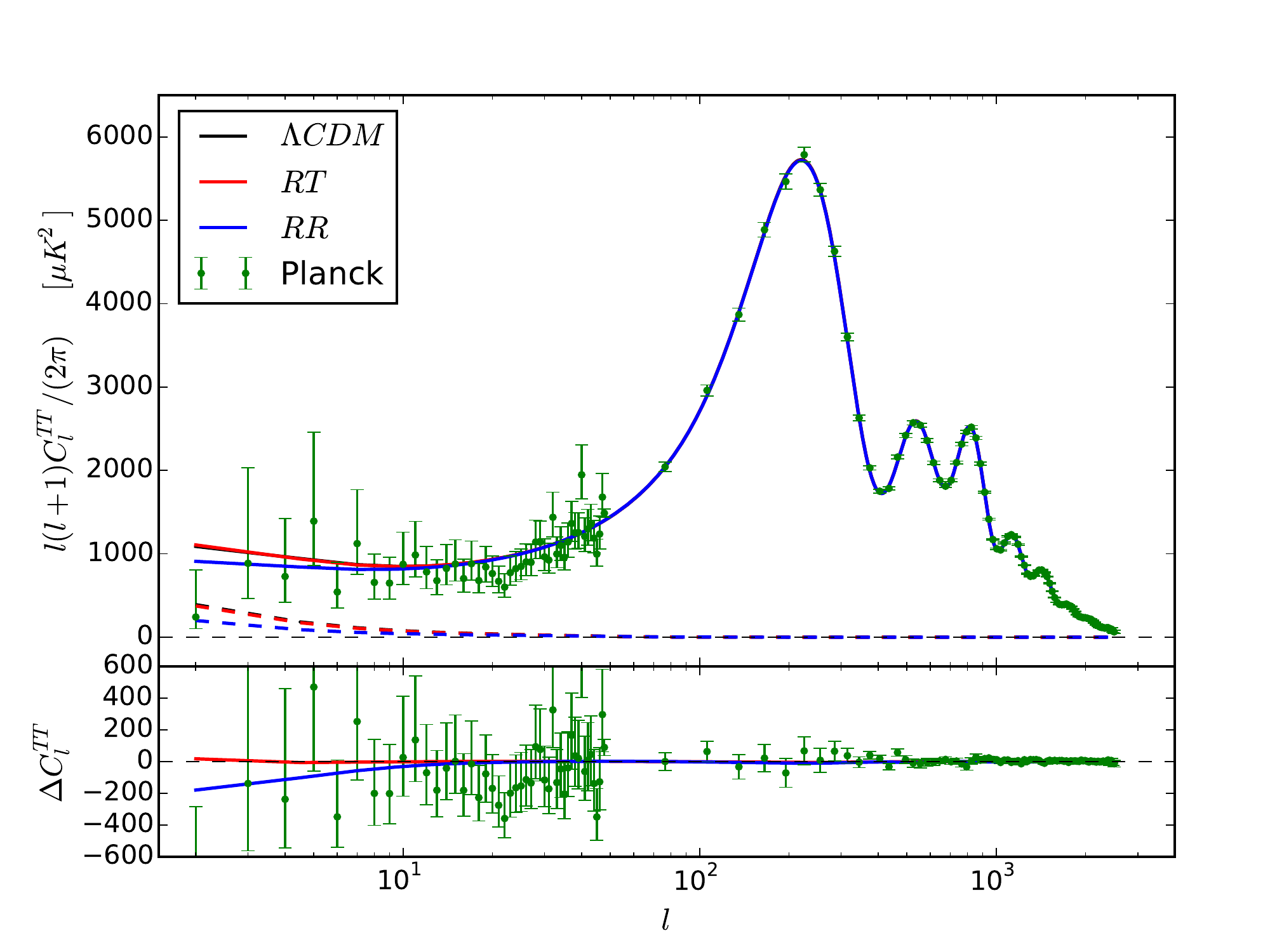} 
\caption{\label{ClTT} Upper plot: temperature power spectrum (thick), and the separate contribution from the late ISW contributions (dashed), for $\Lambda$CDM (black), RT (red) and RR (blue), using the best fit values of the parameters determined from BAO+JLA+Planck.
The black and red lines are indistinguishable on this scale. The lower plot shows the residuals for $\Lambda$CDM and difference of RT (red) and RR (blue) with respect to $\Lambda$CDM. Data points are from Planck 2015 \cite{Planck_2015_CP} (green bars). Error bars correspond to $\pm 1 \sigma$ uncertainty.
}
\end{figure}

\begin{figure}[th]
\centering
\includegraphics[width=0.7\columnwidth]{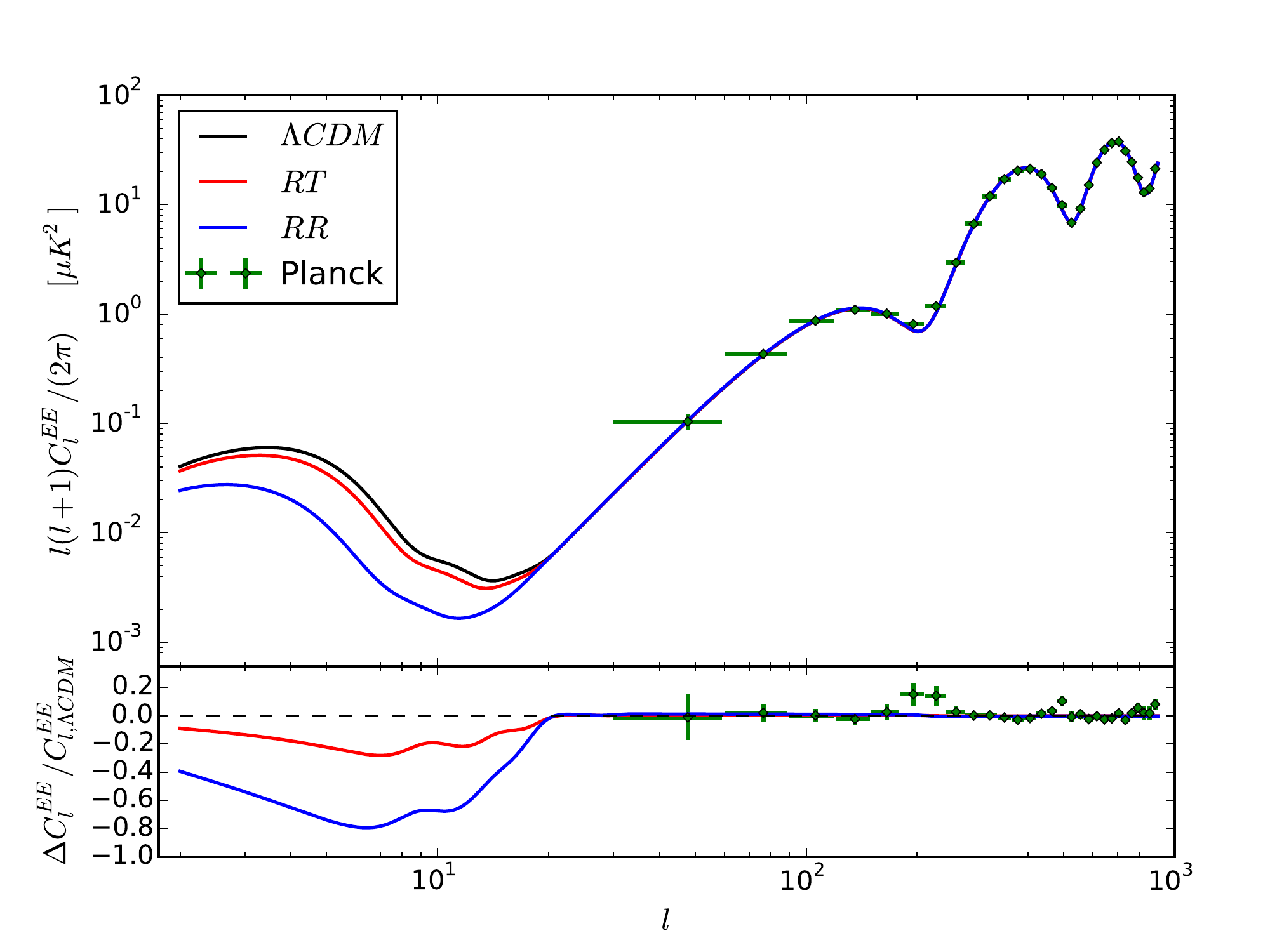}
\caption{Similar,  for the $EE$ spectra except that the separate late ISW contributions are not shown. The lower plot shows the \textit{relative} difference of RT (red) and RR (blue) with respect to $\Lambda$CDM.
\label{ClEE}}
\end{figure}

\begin{figure}[h]
\centering
\includegraphics[width=0.7\columnwidth]{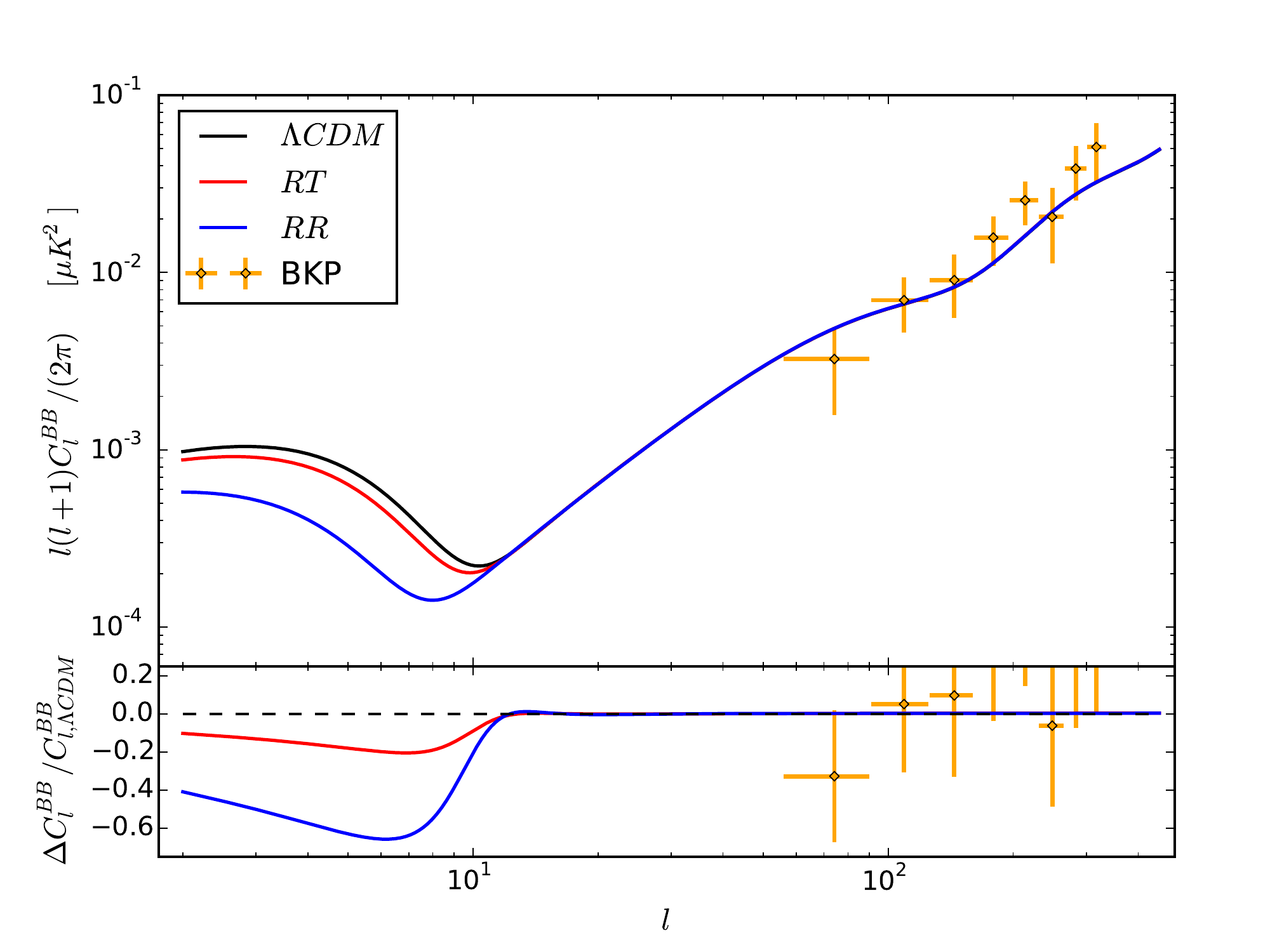}
\caption{As in Fig.~\ref{ClEE}
for the lensed $BB$ spectra the data points are from the joint BICEP2+Keck+Planck  \cite{Ade:2015tva} (orange crosses). 
\label{ClBB}}
\end{figure}

\begin{figure}[h]
\centering
\includegraphics[width=0.7\columnwidth]{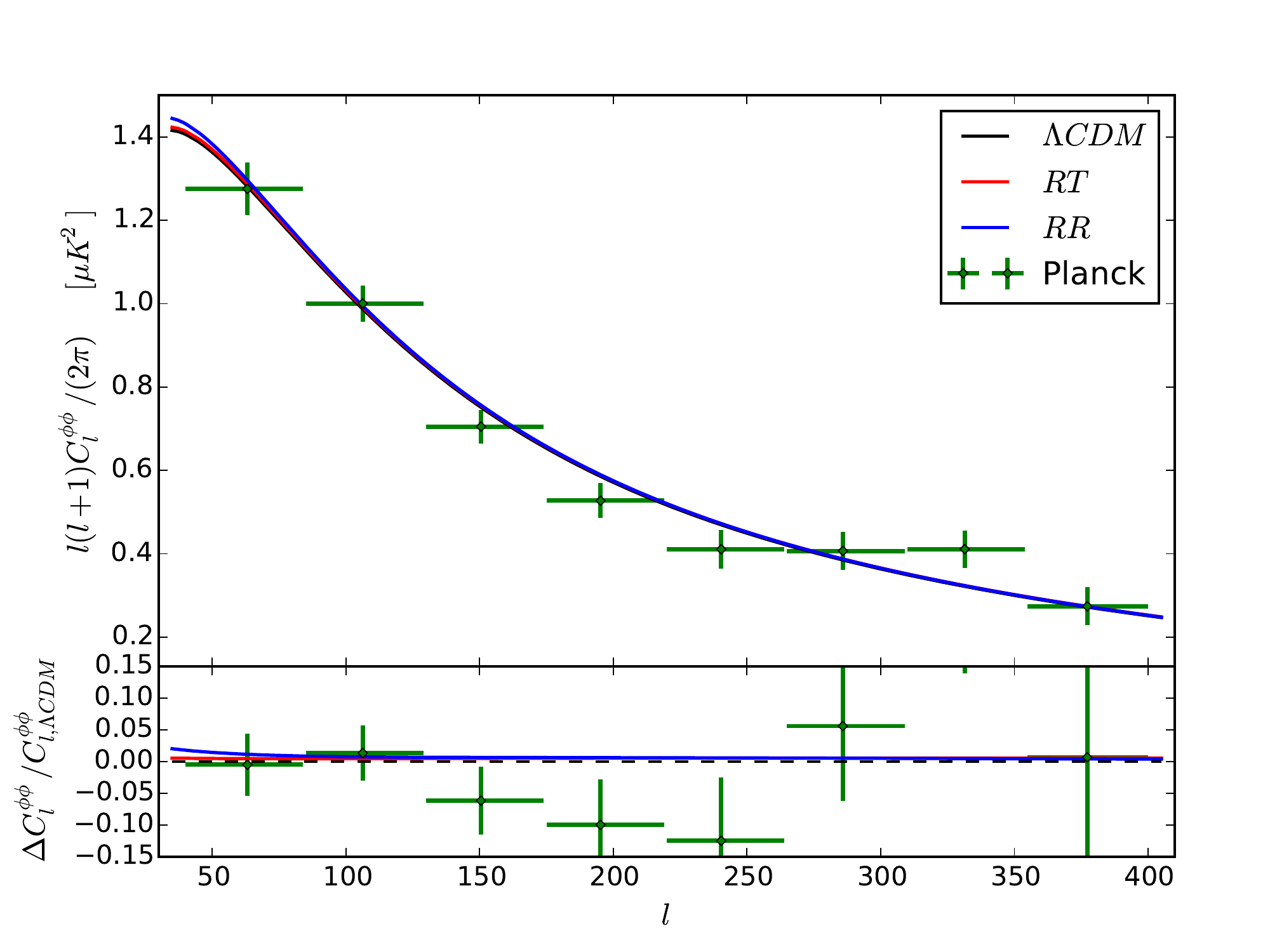}
\caption{The lensed $\phi \phi$ spectra (lensing potential).
\label{ClFF}}
\end{figure}

In  Fig.~\ref{triangle} we show the triangle plot for the case  Planck+BAO+JLA 
plus a prior $H_0=70.6$, while in Fig.~\ref{s8_Om} we plot the 2-dimensional likelihood in the plane $\sigma_8-\Omega_m$.
Finally, having also determined the best fit values of the cosmological parameters, it is interesting to display explicitly how the CMB data are fitted by the three models. We use for definiteness the best fit determined from BAO+JLA+Planck. In Figs.~\ref{ClTT}-\ref{ClFF} we show, for the three models, the fit to the temperature power spectrum, the $EE$ spectrum, the lensed $BB$ spectrum and the lensing potential, respectively.

\begin{table}[t]
\centering
\resizebox{8cm}{!}{
\begin{tabular}{|l|c|c|} 
 \hline 
Data &  RT & RR  \\ \hline 
Planck & $5.17(4) \times 10^{-2}$ & $9.35(7) \times 10^{-3}$\\
BAO+Planck+JLA & $5.15(4) \times 10^{-2}$ & $9.21(7) \times 10^{-3}$ \\
BAO+Planck+JLA+$(\rm{H_0}=70.6$)& $5.15(4) \times 10^{-2}$ & $9.22(7) \times 10^{-3}$\\
BAO+Planck+JLA+$(\rm{H_0}=73.8)$& $5.17(4) \times 10^{-2}$ & $9.24(7) \times 10^{-3}$\\
\hline
\end{tabular}}
\caption{Mean values for $\gamma=m^2/(9H_0^2)$.\label{tab:gamma}}
\end{table}

\clearpage

\section{Bayesian model comparison}\label{sect:ModelComp}

\subsection{The Bayes factor}\label{sect:SDtheory}

In the previous section, in order to compare the performance of the three models, we have used the differences 
$\Delta \chi^2_{ij}$. This method is not genuinely Bayesian,  and only leads to sensible results if a number of  assumptions are satisfied by the posterior distributions under interest (e.g. large number of data points, weak skewness, etc), so it is possible that the results drawn from it are biased.  A more accurate method for comparing models, which is fully Bayesian, is based on Bayes factors \cite{Trotta_ABS}.  In this section 
we use Bayes factors to compare the three models of interest among each other, given the  \textit{Planck}+BAO+JLA data. 
Starting from Bayes theorem, 
\begin{align}
 P(\theta | d, \mathcal{M}) = \frac{P(d | \theta, \mathcal{M}) P(\theta |\mathcal{M})}{P(d| \mathcal{M}) } \, ,  \label{eq:bayes}
\end{align}
which says that the posterior distribution, i.e. the probability  of the parameters $\theta$ given the data $d$ and the model $\mathcal{M}$, $P(\theta | d, \mathcal{M}) $, equals the product of the likelihood function $P(d | \theta, \mathcal{M})$ by the prior $P(\theta |\mathcal{M})$, divided by the evidence $P(d| \mathcal{M})$ (marginal likelihood). General Bayesian model comparison is based on the model probability $P(\mathcal{M}|d)$. This number is however difficult to interpret in absolute terms, and instead one considers relative model probabilities. Using again Bayes theorem we can express the relative model probability for two models $\mathcal{M}_i$ and $\mathcal{M}_j$ as
\begin{align}
\frac{P(\mathcal{M}_i|d)}{P(\mathcal{M}_j|d)} =  \frac{P(d| \mathcal{M}_i)}{P(d| \mathcal{M}_j)} \frac{P(\mathcal{M}_i)}{P(\mathcal{M}_j)}
\equiv B_{ij} \frac{P(\mathcal{M}_i)}{P(\mathcal{M}_j)} \, ,
\end{align}
i.e.\ up to the prior model probabilities $P(\mathcal{M})$ the relative model probabilities are just given by the Bayes factors $B_{ij}$ which
are defined as the ratio of the evidences computed within the two models $\mathcal{M}_i$ and $\mathcal{M}_j$.
The Bayes factors correspond to betting odds, their numerical value is conventionally translated into statements about the evidence of  model $i$ with respect to model $j$ using the Jeffreys' scale. Model $i$ is favored with respect to model $j$ if $B_{ij}>1$. For $1<B_{ij}<3$ the evidence is deemed `weak',  for $3<B_{ij}<20$ is `definite', for $20<B_{ij}<150$ is `strong', and for $B_{ij}>150$ is 
`very strong'.\footnote{Some subtleties may appear in the use of the Jeffreys' scale when comparing models with a different number of parameters \cite{Nesseris:2012cq}, which however is not our case.}

For multi-parameter models the computation of the respective evidences is in general numerically expensive. However when the models to compare with each other can be nested together, i.e. can be embedded in a larger model that reduces to one model in a limit, and to the other model in another limit, this task is made easier by the Savage-Dickey density ratio (SDDR) method. In our context this nesting is possible and the SDDR can be applied. 

\subsection{Model nesting and Savage-Dickey density ratio}

Our task is to compute the Bayes factors $B_{\Lambda i}$ constructed out of the evidence computed assuming the $\Lambda$CDM model, our null hypothesis, and one of the two non-local models of interest ($i = {\rm RT, RR}$), the alternative hypothesis, for a given set of data. Computing the evidence for a model is generally quite hard as it requires a high-dimensional integration. Marginalising Eq.\ (\ref{eq:bayes}) over $\theta$ we find immediately that
\begin{align}
P(d| \mathcal{M}) = \int d\theta P(d | \theta, \mathcal{M}) P(\theta |\mathcal{M}) \, .
\end{align}
If however two models are nested, it becomes  possible to use the SDDR instead to find directly the Bayes factor between the models. In order to be able to exploit the SDDR we consider the extended model constructed from each of the non-local models by adding a cosmological constant. For the model RT the equation of motion (\ref{RT}) is then modified into 
\be\label{RTLambda}
\Gmn -\frac{m^2}{3}\(\gmn\iBox R\)^{\rm T}+ g_{\mu \nu} \Lambda=8\pi G\,\Tmn\, ,
\ee
while for the model RR  the action (\ref{RR})  becomes 
\be\label{SRRLambda}
S=\frac{\mplr^2}{2}\int d^{4}x \sqrt{-g}\, 
\[R -2 \Lambda -\frac{1}{6} m^2R\frac{1}{\Box^2} R\]\, .
\ee
The model (\ref{RTLambda}) reduces to $\Lambda$CDM in the limit $m^2\ra 0$, and to the RT model in the limit $\Lambda\ra 0$, and similarly the model (\ref{SRRLambda}) reduces in these limits to $\Lambda$CDM and to the RR model, respectively. Of course, the models (\ref{RTLambda}) and
(\ref{SRRLambda}) could be interesting in their own right. However, we will only use them as a tool for comparing the RR and RT models to $\Lambda$CDM.

These two models for gravity allow us to construct two associated statistical models, denoted 
$\mathcal{M}_{\Lambda+{\rm RT}}$ and $\mathcal{M}_{\Lambda+{\rm RR}}$ respectively, which can be constrained by the data with the same method described in the previous section. The only difference with respect to the analysis performed previously is that now the total dark energy component is a  mixture of the one induced by the cosmological constant, $\Omega_{\Lambda}$, and the one induced by the non-local modification of gravity $\Omega_{X_i}$, which depends on the  the mass parameter $m$. These models  therefore have an extended parameter space that we take to be spanned by 
\begin{align}
\tilde{\theta} = ( H_0, \omega_b, A_s, n_s, \tau_{\rm re}, \Omega_{\Lambda}, \Omega_{X_i} ) \, .
\end{align}
Observe that now the physical dark matter density fraction $\omega_c$ is taken as a derived parameter, and instead we vary the density fractions of the two types of dark energy. We choose to proceed in this way in order to keep maximal control on the choice of prior for the latter, since this condition can be very important when using the SDDR method \cite{Wagen_SDDR} that we discuss now.

Consider the nested model $\mathcal{M}_{\Lambda + i}$ (where $i = \mathrm{RT~or~RR}$). Since this model is a nesting of $\mathcal{M}_{\Lambda}$ and $\mathcal{M}_{i}$, we assume (by continuity) that its likelihood function taken at $\Omega_\Lambda = 0$ reproduces the one of $\mathcal{M}_{i}$, and equivalently putting $\Omega_{X_i}=0$ reproduces the one of $\mathcal{M}_{\Lambda}$, 
\begin{align}
P\big(d | \bar{\theta}_\Lambda, \Omega_\Lambda = 0, \mathcal{M}_{\Lambda + i}\big) &=  P\big(d |  \bar{\theta}_\Lambda, \mathcal{M}_i  \big) \, , \\
P\big(d | \bar{\theta}_{X_i}, \Omega_{X_i}=0, \mathcal{M}_{\Lambda + i}\big)  &=  P\big(d | \bar{\theta}_{X_i}, \mathcal{M}_\Lambda  \big) \, , 
\end{align}
where $\bar{\theta}_{j} \equiv \tilde{\theta} \setminus \{ \Omega_j \}$, and that the same conditions hold on the prior. We can therefore write  
\begin{align}
P(d | \mathcal{M}_\Lambda) &= \int d\bar{\theta}\,  P\big(d | \bar{\theta}, \mathcal{M}_{\Lambda}\big) P\big(\bar{\theta}|\mathcal{M}_{\Lambda} \big) \nn\\
&= \int d\bar{\theta} \, P\big(d | \bar{\theta}_{X_i}, \Omega_{X_i}=0, \mathcal{M}_{\Lambda + i}\big) P\big( \bar{\theta}_{X_i}, \Omega_{X_i}=0 |  \mathcal{M}_{\Lambda + i}\big)  \nn \\
&= P \big( d | \Omega_{X_i}=0 , \mathcal{M}_{\Lambda + i}\big) \, . 
\end{align}
Applying now Bayes theorem to the third equality, 
\begin{align}
P \big( d | \Omega_{X_i}=0, \mathcal{M}_{\Lambda + i}  \big) = \frac{P \big( \Omega_{X_i}=0 | d, \mathcal{M}_{\Lambda + i} \big) P \big( d| \mathcal{M}_{\Lambda + i}\big) }{P \big( \Omega_{X_i}=0 | \mathcal{M}_{\Lambda + i} \big)} \, ,
\end{align}
leads to the SDDR, 
\begin{align}
 B_{\Lambda ( \Lambda + i)}\equiv 
 \frac{P(d| \mathcal{M}_\Lambda) }{P \big( d| \mathcal{M}_{\Lambda + i}\big) } =   \frac{P \big( \Omega_{X_i}=0 | d, \mathcal{M}_{\Lambda + i} \big) }{P \big( \Omega_{X_i}=0 | \mathcal{M}_{\Lambda + i} \big)}  \, ,
\end{align}
which tells us that the Bayes factor of the comparison between the nested model $\mathcal{M}_{\Lambda + i}$ and its sub-model $\mathcal{M}_\Lambda$ is equal to the ratio of the marginalized one-dimensional posterior distribution and marginalized one-dimensional prior of $\Omega_{X_i}$ obtained from the extended model, evaluated at the point where the simpler model is nested inside the extended model (i.e.\ at $\Omega_{X_i}=0$). Intuitively this can be understood as a measure of the amount by which the prior evolved into the posterior through knowledge update given new data. If the posterior grows at $\Omega_{X_i}=0$ this means that this value tends to be preferred by the data, i.e. $\mathcal{M}_\Lambda$ is preferred, and the Bayes factor increases. Conversely,  if the posterior decreases, a non-zero value of $\Omega_{X_i}$ is preferred. 

The procedure outlined above  allows us to compare $\mathcal{M}_\Lambda$ (or equivalently $\mathcal{M}_i$) with the nested model  $\mathcal{M}_{\Lambda + i}$.  However, we eventually want to compare directly $\mathcal{M}_\Lambda$ with $\mathcal{M}_i$. This can be done by comparing both to the extended model,
\bees
B_{\Lambda i} &\equiv& \frac{P(d | \mathcal{M}_{\Lambda})}{P(d | \mathcal{M}_{i})}
 = \frac{P(d | \mathcal{M}_{\Lambda})}{P(d | \mathcal{M}_{\Lambda + i})} 
\frac{P(d | \mathcal{M}_{\Lambda + i})}{P(d | \mathcal{M}_{i})}\nn\\
& =&  \frac{B_{\Lambda  (\Lambda + i) }}{B_{i  (\Lambda + i) }}
 = \frac{P(\Omega_{X_i} = 0 | d, \mathcal{M}_{\Lambda + i} ) }{  P(\Omega_{\Lambda} = 0 | d, \mathcal{M}_{\Lambda + i})} \, . \label{BF}
\ees
The expression involving the ratio of the Bayes factors always holds, but in
in the last identity we have assumed the equality of the marginalized one-dimensional prior, $P \big( \Omega_{X_i}=0 | \mathcal{M}_{\Lambda + i} \big) = P \big( \Omega_\Lambda=0 | \mathcal{M}_{\Lambda + i} \big)$. This shows how the SDDR method greatly simplifies the computation of the Bayes factor for the comparison of two models that can be nested together. Indeed, one only needs to know the final posterior distribution of the latter, for some set of data, which can be obtained by standard MCMC methods as the one we use (although occasionally multiple chains at different temperatures may be needed, see e.g.\ appendix A of\cite{Mukherjee:2010ve} for an example in a different context). Then one has just to take the ratio of two numbers, the marginalized 1d posterior of $\Omega_{\Lambda}$ evaluated at $\Omega_{\Lambda} = 0$, and the same for $\Omega_{X_i}$, within the extended model. If the priors do not cancel then one just computes the actual Bayes factors between the extended and the two nested models and uses their ratio.
% This shows the strength of Bayesian model selection using the SDDR method. %% already said :)

The last point that we need to discuss is the choice that we made for the parameters to vary for constructing the posterior distribution of the nested model, and their respective prior distribution. In Bayesian inference the choice of a (subjective) prior is part of the approach and should be chosen with care. In order to have sensible results, one needs to provide the least informative prior given by the current knowledge, before seeing the data. 
Since in the nested model we have two types of dark energies,  $\Omega_{X_i}$ and $\ola$, the flatness condition reads
$\Omega_{X_i}+\ola+\Omega_c=1-\ora$. 
In order to treat both type of dark energies on the same footing, we choose to vary both $\Omega_\Lambda$ and $\Omega_{X_i}$ and to take $\omega_c$ as derived. For the same reason,  we prefer to impose a flat prior on $\Omega_\Lambda$ and $\Omega_{X_i}$ rather than on, say, $\Lambda$ and $m^2$. 
It is natural to assume that the density fractions vary in the interval $[0,1]$. To avoid boundary effects  that could affect the determination of the SDDR,
we actually extend the allowed range of values to 
\begin{align}
\Omega_\Lambda \in [-0.2,1.2], \qquad \Omega_{X_i} \in [-0.2,1.2] \, ,
\end{align}
with a uniform prior and we then remove values that lie outside $[0,1]$ from the chain. The prior for both parameter being the same, the last equality in \eqref{BF} therefore make sense, and this formula can be applied in our case. 

We can now compute the likelihoods of the nested model,  perform the corresponding parameter estimation, and compute the  Bayes factors. The results are shown in Table~\ref{tab:nested}.
 
\begin{table}[t]
\begin{minipage}{1.\linewidth}
\centering
\resizebox{6cm}{!}{
\begin{tabular}{|l|c|c|c|} 
 \hline 
\multicolumn{1}{|l|}{ } & \multicolumn{2}{|c|}{BAO+Planck+JLA} \\ \hline
Param & ${\rm RT} + \Lambda$ &$ {\rm RR} + \Lambda$ \\ \hline 
$100~\omega_{b }$ &  $2.225_{-0.014}^{+0.014}$  & $2.225_{-0.015}^{+0.015}$ \\ 
$H_0$ & $68.18_{-0.68}^{+0.65}$  &  $68.28_{-1}^{+0.79}$\\ 
$\log (10^{10}A_{s })$ & $3.059_{-0.02}^{+0.017}$ & $3.061_{-0.025}^{+0.026}$ \\ 
$n_{s }$  &  $0.9646_{-0.0041}^{+0.004}$ & $0.9646_{-0.0044}^{+0.0044}$ \\ 
$\tau_{\rm re}$ &$0.06323_{-0.011}^{+0.0096}$  & $0.06379_{-0.013}^{+0.014}$ \\ 
$\Omega_{\Lambda }$ &$0.357_{-0.33}^{+0.35}$ & $0.538_{-0.17}^{+0.24}$ \\ 
$\Omega_{X}$ &$0.3368_{-0.35}^{+0.34}$ & $0.1566_{-0.24}^{+0.17}$ \\ 
$\Omega_{cdm}$ & $0.2569_{-0.023}^{+0.0025}$ & $0.2561_{-0.0072}^{+0.0076}$\\ 
$z_{\rm re}$ & $8.568_{-1}^{+1}$   & $8.604_{-1.2}^{+1.4}$  \\ 
$\sigma_8$ &$0.8217_{-0.0012}^{+0.033}$ & $0.8238_{-0.012}^{+0.011}$ \\ 
\hline 
$\chi^2_{\rm min}$ &$13631.0$ &$13630.8$\\
$B_{\Lambda i}$ &$1.02$ &$22.67$\\
\hline
\end{tabular}} \quad\quad
\resizebox{7.2cm}{!}{
\begin{tabular}{|l|c|c|c|} 
 \hline 
\multicolumn{1}{|l|}{ } & \multicolumn{2}{|c|}{BAO+Planck+JLA+($H_0=73.8 \pm 2.4$)} \\ \hline
Param & ${\rm RT} + \Lambda$ &${\rm RR} + \Lambda$ \\ \hline 
$100~\omega_{b }$ &   $2.23_{-0.014}^{+0.013}$  & $2.227_{-0.015}^{+0.015}$ \\ 
$H_0$ &  $68.62_{-0.58}^{+0.64}$  &  $69.08_{-1}^{+0.92}$ \\ 
$\log (10^{10}A_{s })$ & $3.066_{-0.017}^{+0.017}$ & $3.059_{-0.026}^{+0.028}$  \\ 
$n_{s }$  &  $0.9656_{-0.0039}^{+0.0039}$ & $0.9649_{-0.0045}^{+0.0045}$ \\ 
$\tau_{\rm re}$ &$0.06716_{-0.0092}^{+0.0092}$  &$0.06328_{-0.014}^{+0.015}$ \\ 
$\Omega_{\Lambda }$ &$0.2445_{-0.44}^{+0.13}$ &$0.3993_{-0.22}^{+0.21}$  \\ 
$\Omega_{X}$ &$0.454_{-0.17}^{+0.42}$  &  $0.3025_{-0.21}^{+0.22}$ \\ 
$\Omega_{cdm}$ &  $0.2527_{-0.022}^{+0.0035}$ & $0.2501_{-0.03}^{-0.00052}$ \\ 
$z_{\rm re}$ &  $8.935_{-0.86}^{+0.93}$    & $8.537_{-1.3}^{+1.5}$  \\ 
$\sigma_8$ &$0.8257_{-0.0017}^{+0.036}$  &$0.8306_{-0.035}^{+0.018}$ \\ 
\hline 
$\chi^2_{\rm min}$ &$13636.1$ &$13635.4$\\
$B_{\Lambda i}$ &$0.39$ &$2.38$\\
\hline
\end{tabular}}
\end{minipage}
\caption{Best fit values and Bayes factors for the nested models. Left: using BAO+Planck+JLA data.
Right: adding also $H_0=73.8 \pm 2.4$. \label{tab:nested}}
\end{table}

Not surprisingly, the mean values for the nested model $\mathcal{M}_{\Lambda+{\rm RT}}$ are always intermediate between the values obtained for $\mathcal{M}_{\Lambda}$ and those for 
$\mathcal{M}_{{\rm RT}}$, compare with Table~\ref{partable}, and the same for the nested model $\mathcal{M}_{\Lambda+{\rm RR}}$.
The conclusions drawn from the values  of the Bayes factors 
are fully consistent with that obtained in Section~\ref{sect:SDresults}
simply using $\Delta \chi^2_{ij}$. Namely, without a prior on $H_0$, the RT model and $\Lambda$CDM are statistically indistinguishable while, on the Jeffreys' scale, the evidence 
of  $\Lambda$CDM against the RR model is on the border between `definite' and `strong'. 
Adding a high prior on $H_0$ favors the non-local models, so that the RT models become slightly favored with respect to $\Lambda$CDM, with $B_{{\rm RT},\Lambda}\simeq 1/0.39\simeq 2.6$, and the RR model is now only slightly disfavored  with respect to $\Lambda$CDM,  by a similar amount, $B_{\Lambda,{\rm RR}}\simeq 2.4$, in both cases not very significant.

Also the evidence of the RT model with respect to the RR model can be computed easily, with
$B_{\rm RT,RR}=B_{{\rm RT},\Lambda}/B_{{\rm RR},\Lambda}=B_{\Lambda,{\rm RR}}/B_{\Lambda,{\rm RT}}$ ranging from a value
$22.67/1.02\simeq 22.2$ using BAO+Planck+JLA data, to a value $2.38/0.39\simeq 6.1$ when adding the high prior on the Hubble constant, $H_0=73.8 \pm 2.4$.

\section{Growth rate data and modified structure formation}\label{sect:growth}

\begin{figure}[t]
\centering
\includegraphics[width=0.65\columnwidth]{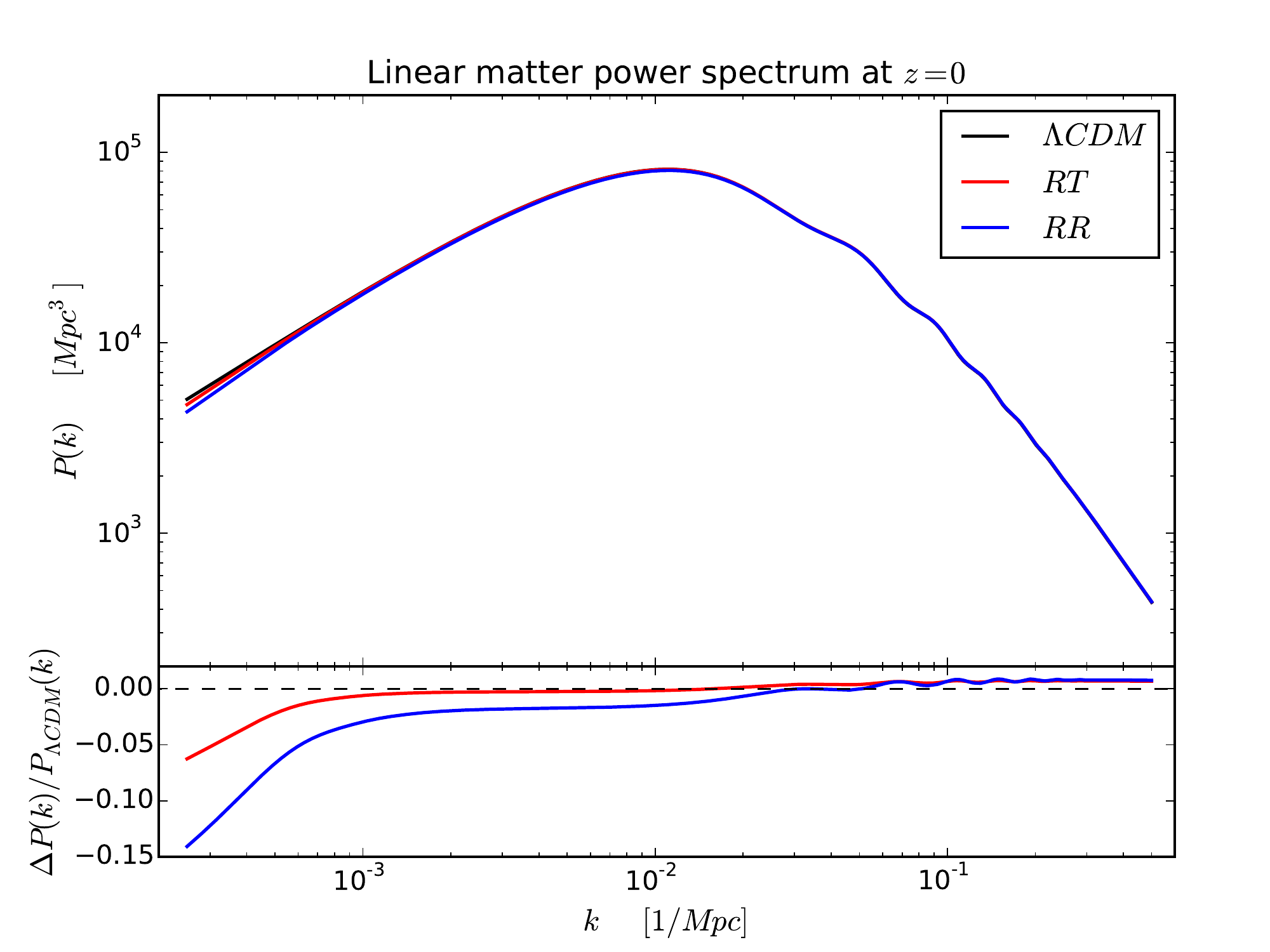} 
\caption{\label{pkz0} Upper: linear matter power spectrum for $\Lambda$CDM (black), RT (red) and RR (blue). Lower: relative difference of RT (red) and RR (blue), with respect to $\Lambda$CDM. These plots give  the linear power spectrum evaluated at redshift $z=0$.}
\end{figure}

\begin{figure}[t]
\centering
\includegraphics[width=0.65\columnwidth]{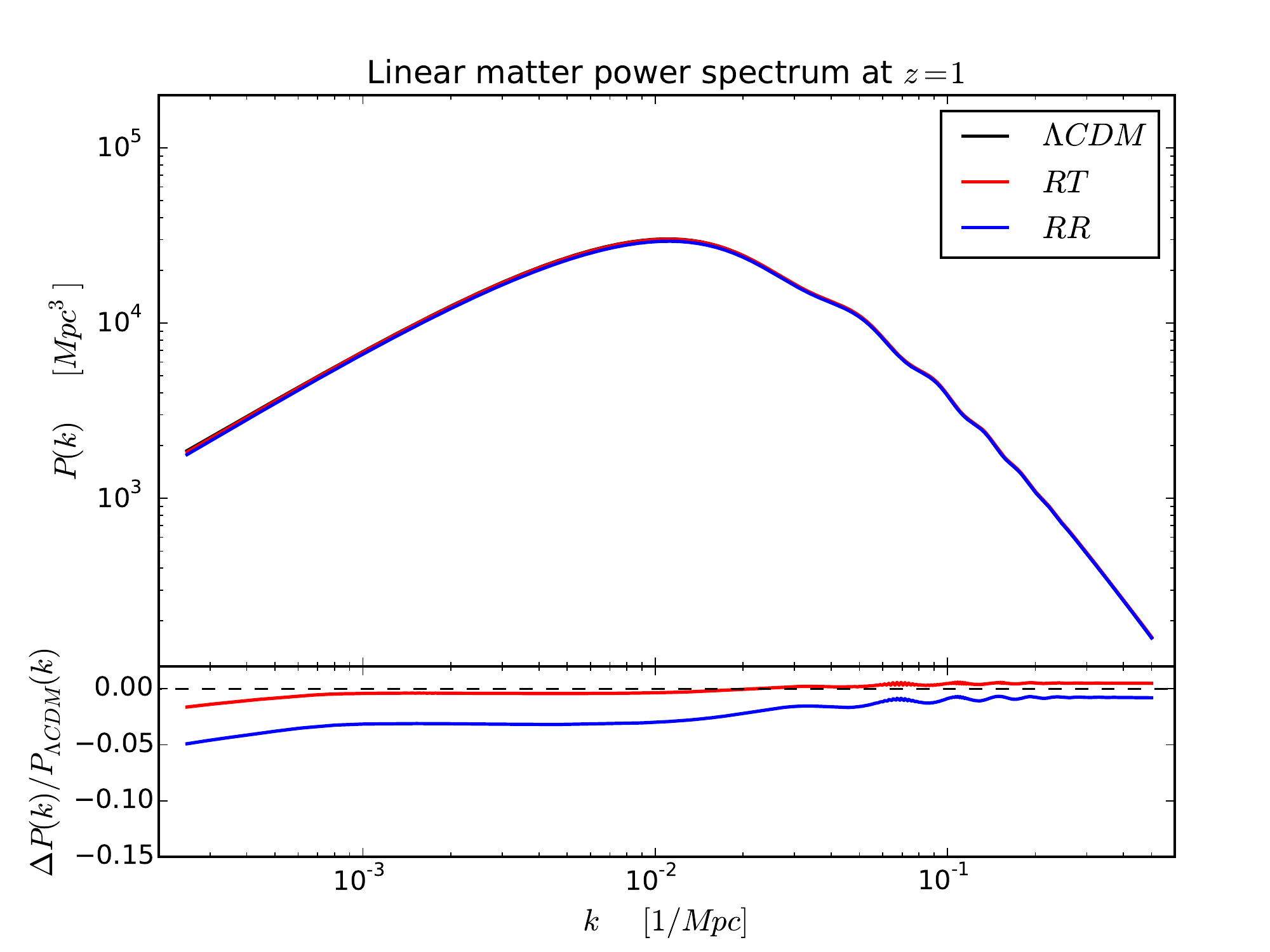}
\caption{\label{pkz1} As in Fig.~\ref{pkz0}, for redshift $z=1$.}
\end{figure}

\begin{figure}[t]
\centering
\includegraphics[scale=0.6]{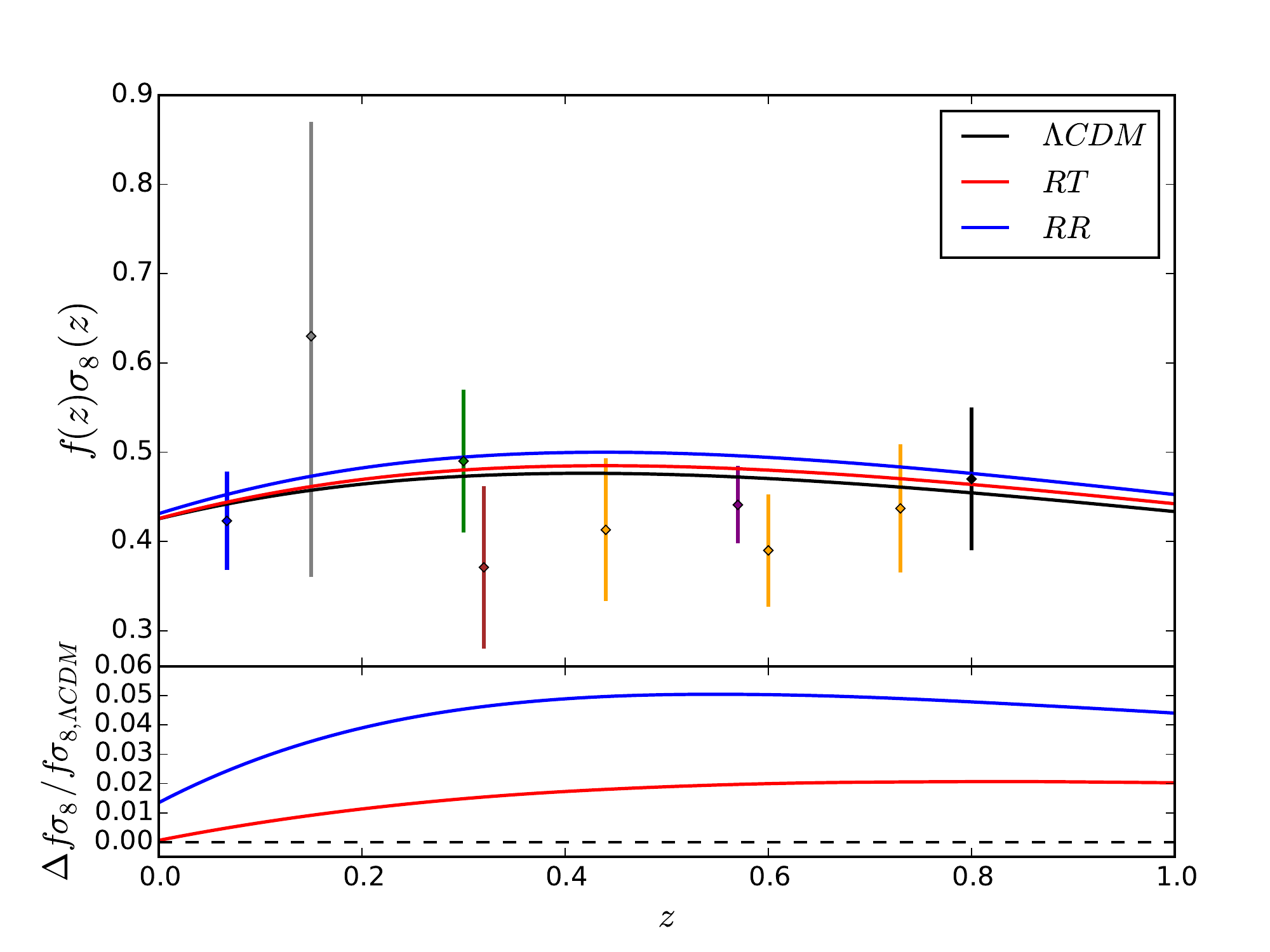}
\caption{\label{fsigma8} Upper panel: Growth rate computed in $\Lambda$CDM (black), RT (red) and RR (blue) constrained with various data points: 6dF GRS \cite{Beu_6dF_RSD_2012} (blue), SDSS LRG \cite{Oka_SDSS_2013} (grey), SDSS MGS \cite{How_SDSS_2014} (green), BOSS LOWZ \cite{Chuang_BOSS_2013} (brown), BOSS CMASS \cite{Samushia_BOSS_2013} (purple), WiggleZ \cite{Blake_WiggZ_2012} (orange), VIPERS \cite{delaTorre_VIPERS_2013} (black). Lower panel: Relative difference of the growth rate computed in RT (red) and in RR (blue) with respect to $\Lambda$CDM.}
\end{figure}

Redshift-space distortions (RSD) probe the velocity field, and are usually expressed as a constraint on the combination $f\sigma_8$, where $f \equiv d\ln \sigma_8 /d\ln a$ is the growth function \cite{Planck_2015_DE}, and $\sigma_8$ is the variance of the linear matter power spectrum in a radius of $8$ Mpc today. The translation to $f\sigma_8$ requires a fiducial model, which is taken to be $\Lambda$CDM. This could lead to problems when using the data for constraining  modified gravity models, especially if they exhibit a significant scale-dependence. However, the linear matter power spectrum of the RT and RR models look very much like that of $\Lambda$CDM on the relevant scales, as can be seen from Figs.~\ref{pkz0} and \ref{pkz1}, which show the linear matter power spectra at redshifts $z=0$ and $z=1$, respectively. For this reason we expect that we can use the RSD data to provide at least a rough test of our non-local models.
Furthermore, since RSD are mostly degenerated with the Alcock-Paczynski (AP) effect, measurements of $f\sigma_8$ are often combined with measurements of the angular diameter distance $D_A$ and the Hubble parameter $H$ at the corresponding effective redshift, or some combination of the latter. Whenever this is the case, we marginalize over these measurements and consider only the ones associated to $f\sigma_8$. From this point of view, the treatment which follows is only illustrative of the constraints partially imposed by RSD on the models under consideration. Of course, a more thorough study would consist in including the RSD datasets into a fit performed with MCMC techniques, similarly to what we performed in the previous section for CMB, SNe and BAO data.
For the purpose of our study, we fix the best-fit values of the models from the {\em Planck}+BAO+JLA data and check if  the predicted growth of perturbations agrees with the marginalized $f\sigma_8$ data. The result and the corresponding data used are shown in Fig.~\ref{fsigma8}.\footnote{Notice that the data point of SDSS LRG analyzed by \cite{Oka_SDSS_2013} is taken to be the resulting data from their ``full fit'' since other fits are explicitly based on $\Lambda$CDM-dependent priors such as the ones imposed on $\sigma_{8, {\rm nl}}$ (which is not applicable in our case, see $\Omega_m - \sigma_8$ contours in fig.~\ref{s8_Om}) or neglecting the AP effect which is motivated by the fit using the fiducial cosmology.}.
We find that the corresponding $\chi^2$ obtained from the data in Fig.~\ref{fsigma8} for $\Lambda$CDM, the RT and RR models are 
\be\label{chi2fsigma8}
\chi^2_{{\rm min}, \Lambda{\rm CDM}}= 3.9 \, ,\qquad
\chi^2_{\rm min, RT}= 4.7 \, ,\qquad 
\chi^2_{\rm min, RR}= 6.5 \, .
\ee
We see that the non-local models generically predict a larger growth rate, although for the RT model the difference with $\Lambda$CDM is again not statistically significant.  Once again, it should be stressed that this result is specific to our choice of neutrino masses. It is known (see e.g. \cite{Planck_2015_CP})   that increasing the  value of the neutrino masses has the effect of  decreasing both $\sigma_8$ and $H_0$. In $\Lambda$CDM a decrease in $H_0$  would increase the tension with local measurements of $H_0$. However, we have seen that the RR model predicts a higher value of $H_0$, so it can in principle be consistent with local measurement of $H_0$ even for larger neutrino masses. In turn such larger neutrino masses would have the effect of  decreasing   $\sigma_8$, which could render again the RR model fully competitive with $\Lambda$CDM, also from the point of view of fitting structure formation data. The corresponding analysis will be presented in \cite{DirainBarreira:inprep}.

\begin{figure}[th]
\centering
\includegraphics[width=0.65\columnwidth]{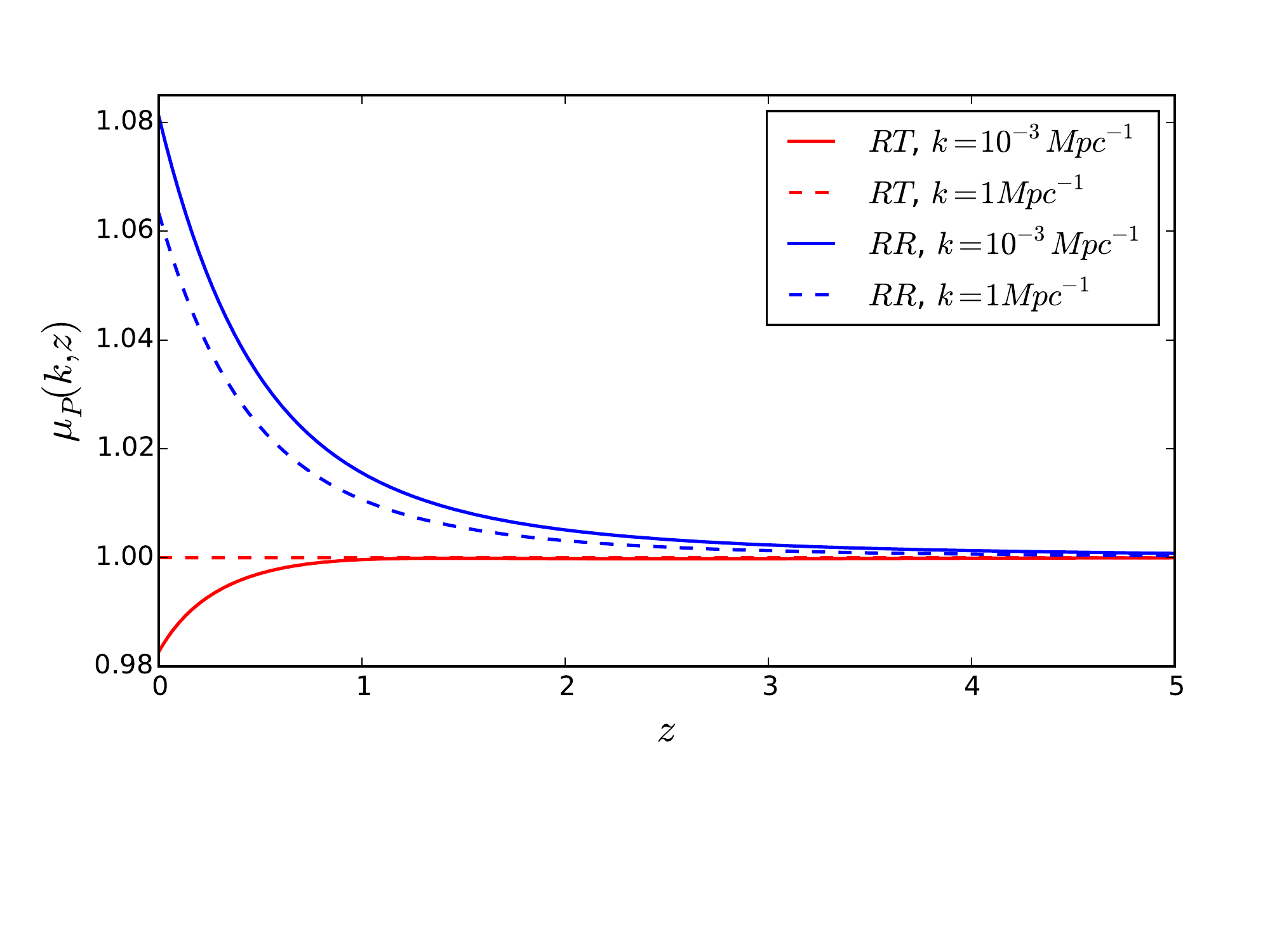}
\caption{\label{MuPlanck} Deviations of the non-local models from $\Lambda$CDM for the gravitational potential parametrized by $\mu_P$. }
\end{figure}

\begin{figure}[th]
\centering
\includegraphics[width=0.65\columnwidth]{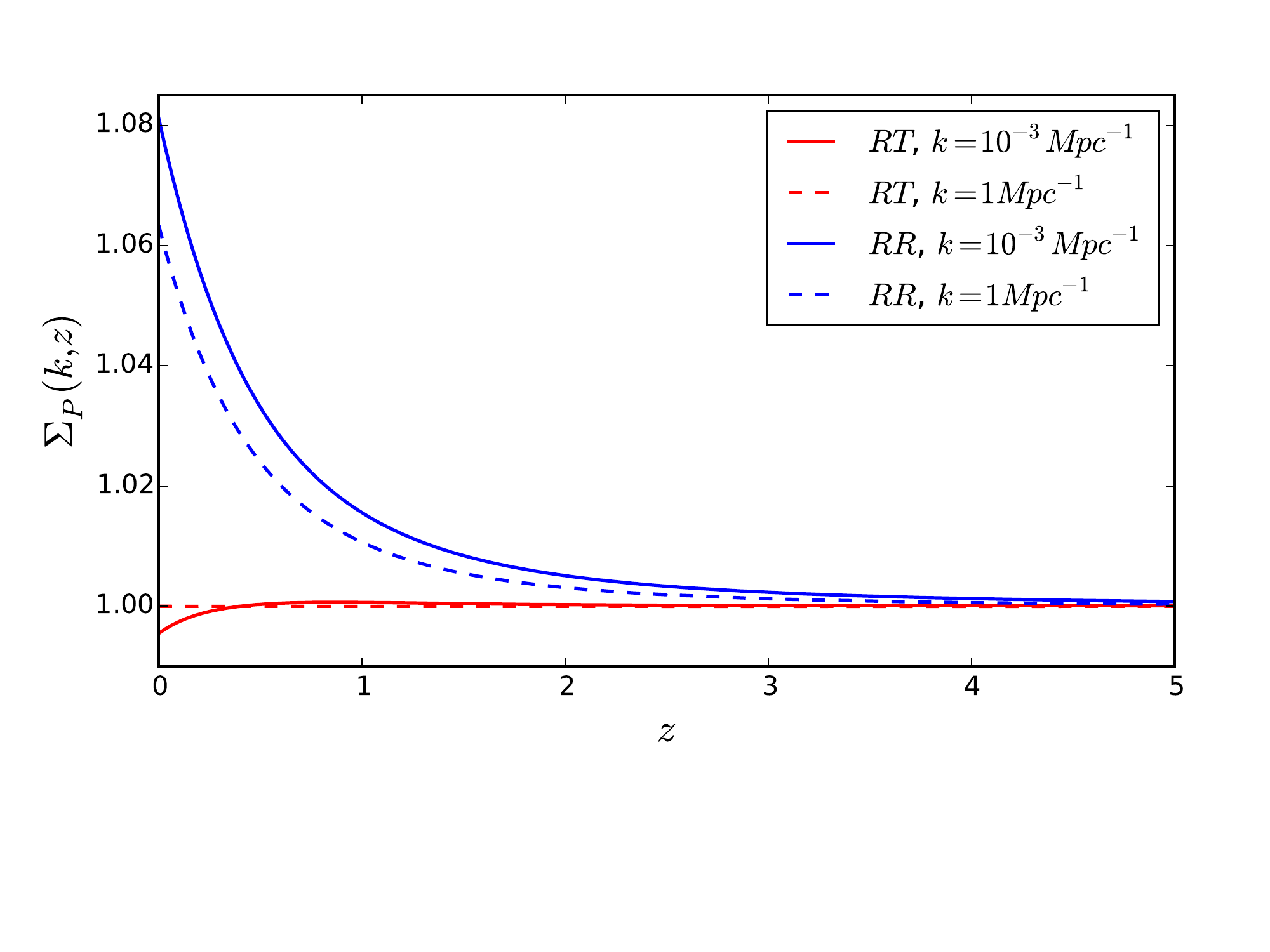}
\caption{\label{SigmaPlanck} Deviations of the non-local models from $\Lambda$CDM for the gravitational potential parametrized by $\Sigma_P$. }
\end{figure}

We have  also computed several other quantities which are relevant for the comparison with structure formation (in particular for assessing the possibility of discriminating the three  models with future observations), using  for each model the respective mean values of the parameters obtained from {\em Planck}+BAO+JLA data. 
When studying structure formation and lensing,
several indicators are used in the literature for parametrizing deviations from $\Lambda$CDM. We define the
Bardeen potentials $\Psi$ and $\Phi$ according to the sign convention 
\be\label{defPhiPsi}
ds^2 =  -(1+2 \Psi) dt^2 + a^2(t) (1 + 2 \Phi) \delta_{ij} dx^i dx^j\, .
\ee
Then, two useful quantities for parametrizing deviations from GR are the functions $\mu(k,z)$ \cite{Daniel:2010ky} and $\Sigma(k,z)$
\cite{Amendola:2007rr}, which are defined through
\bees
\Psi&=&\mu(k,z)\Psi_{\rm GR}\, ,\label{defmu}\\
\Psi-\Phi&=&\Sigma(k,z) (\Psi-\Phi)_{\rm GR}\label{defSigma}\, ,
\ees
where the subscript denotes the  quantities computed in GR, assuming a $\Lambda$CDM model.
The advantage of this  parametrization is that it neatly separates  the modifications to the motion of non-relativistic particles, which is described by $\mu$, from the modification to light propagation, which is encoded in  $\Sigma$. The functions $\mu$ and $\Sigma$ depend on the comoving momentum $k$ and on red-shift $z$.  We compute the time-evolution of $\Psi$ using the mean values for the non-local models and $\Psi_{\rm GR}$ using the mean values for $\Lambda$CDM.\footnote{Observe that in
\cite{Dirian:2014ara}, where a similar analysis (based however on $\Lambda$CDM mean values for the evolution of both $\Psi$ and $\Psi_{\rm GR}$) had been carried out, we used a different convention for the definition of $\mu$ and $\Sigma$, writing 
$\Psi =(1+\mu)\Psi_{\rm GR}$ and $(\Psi-\Phi)=(1+\Sigma) (\Psi-\Phi)_{\rm GR}$, respectively. With the definitions (\ref{defmu}) and (\ref{defSigma}) the GR limit corresponds to $\mu=\Sigma=1$ rather than $\mu=\Sigma=0$.} Often in the literature, e.g.\ in the Planck DE paper \cite{Planck_2015_DE}, a slightly different convention is used:
\bees
- k^2 \Psi&=&4 \pi G a^2 \mu_P (k,z) \rho \Delta \, ,\label{defmup}\\
-k^2 (\Psi-\Phi) &=& 8 \pi G a^2 \Sigma_P(k,z) \rho \Delta \label{defSigmap}\, .
\ees
(The subscript $P$ in $\mu_P$ and $\Sigma_P$ indicates that these definitions agree with those of the Planck analysis \cite{Planck_2015_DE}.)
In these definitions we compare the actual value of the gravitational potentials $\Phi$ and $\Psi$ to the value expected in GR due to the matter and radiation perturbations $\rho\Delta = \rho_m \Delta_m + \rho_r\Delta_r$, where $\Delta$ is the comoving density perturbation.  
In Fig.~\ref{MuPlanck} and \ref{SigmaPlanck} we plot  $\mu_P(k,z)$ and $\Sigma_P(k,z)$ for two representative momenta, as a function of $z$, for the three models. We see that, for the RT model,  the deviations from  $\Lambda$CDM  are tiny, below one percent, while for the RR model they reach a maximum value up to   $3-4\%$ at the typical redshifts, say, $z\,\gsim\,0.5$ relevant for observation of structure formation.

\begin{figure}[th]
\centering
\includegraphics[width=0.65\columnwidth]{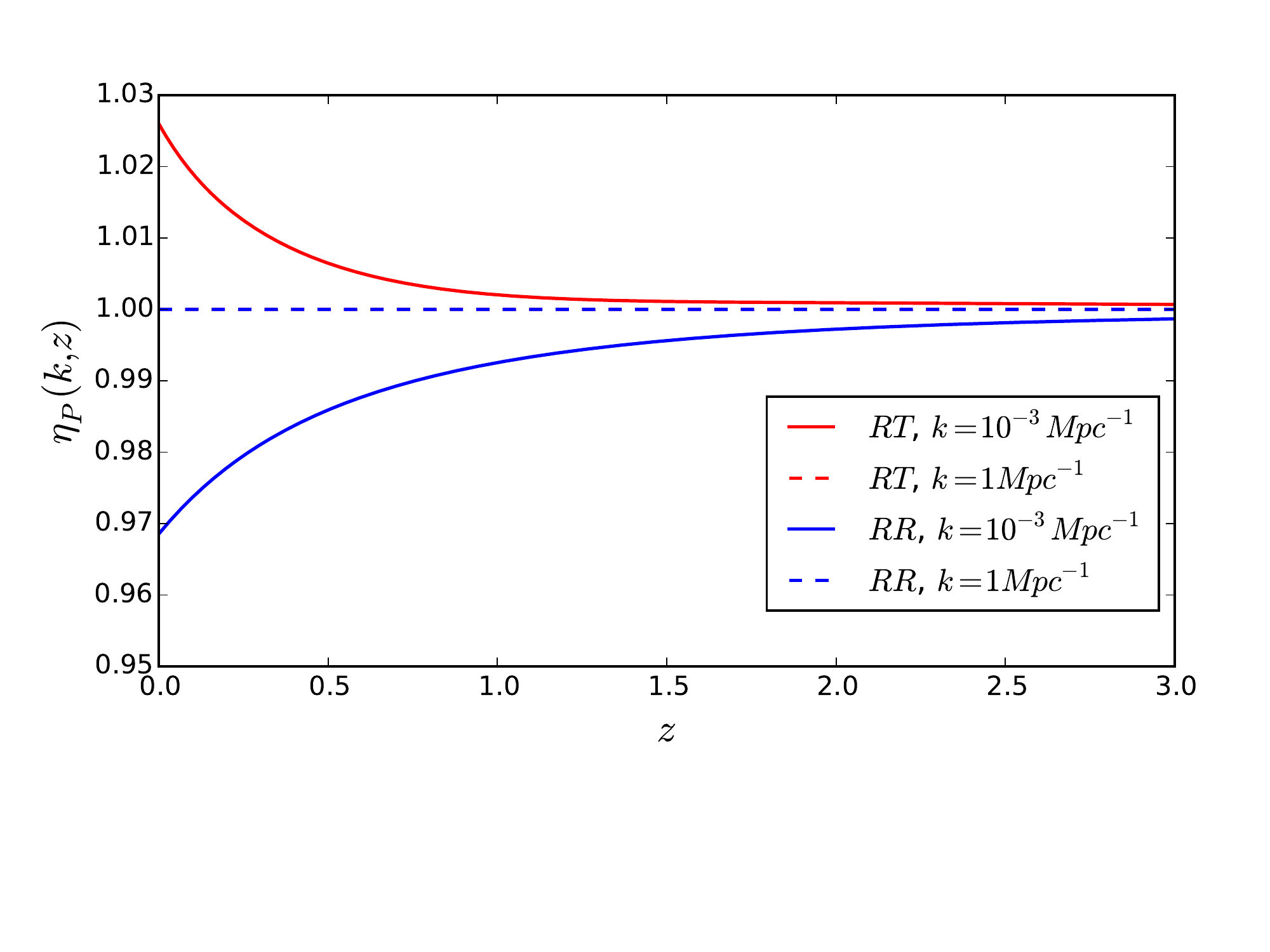}
\caption{\label{eta} The gravitational slip $\eta_P$ for the non-local models.}
\end{figure}

\begin{figure}[th]
\centering
\includegraphics[width=0.65\columnwidth]{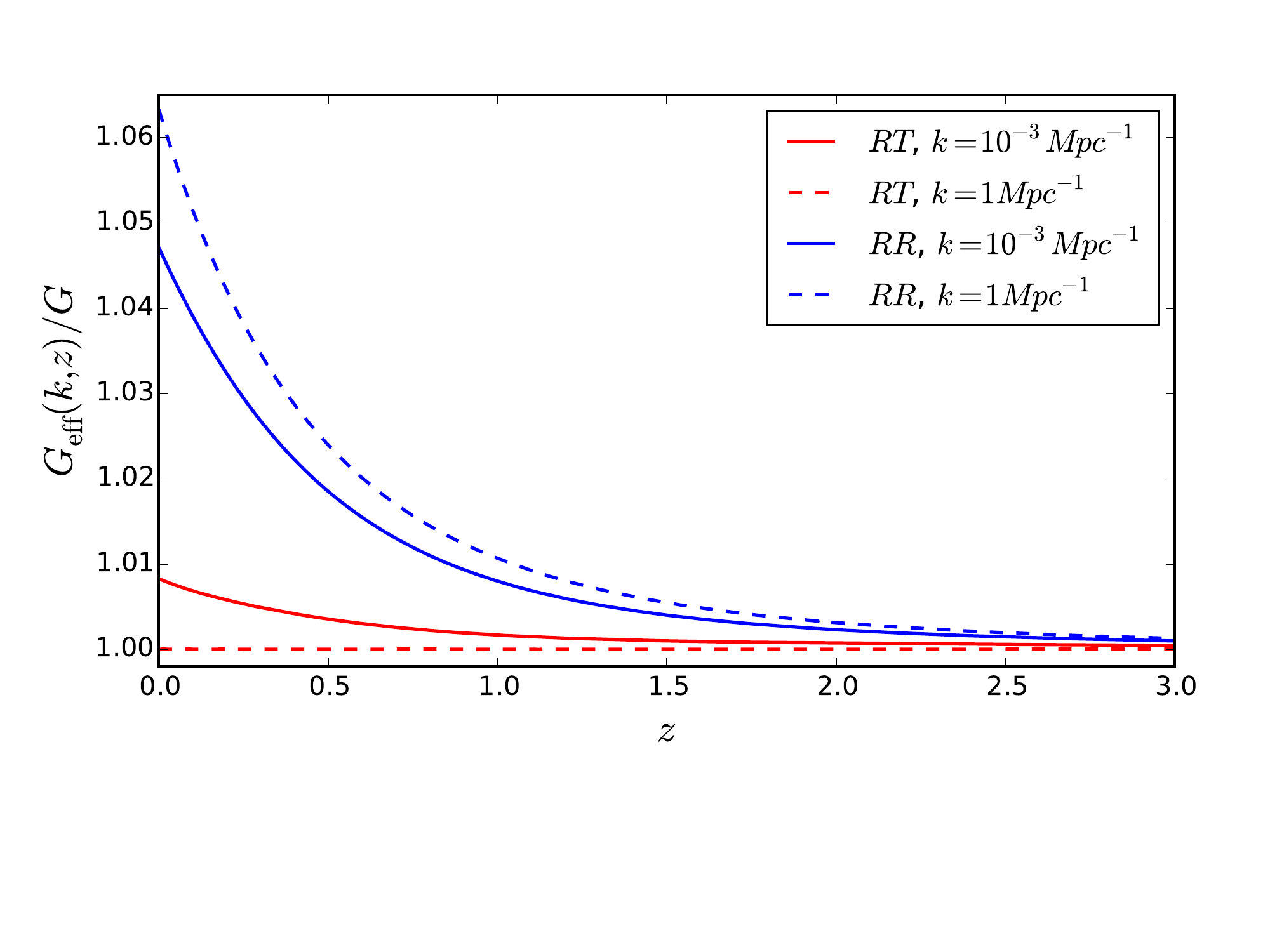}
\caption{\label{Geff} The effective Newton constant
$G_{\rm eff}(k,z)/G$.}
\end{figure}

\begin{figure}[th]
\centering
\includegraphics[scale=0.65]{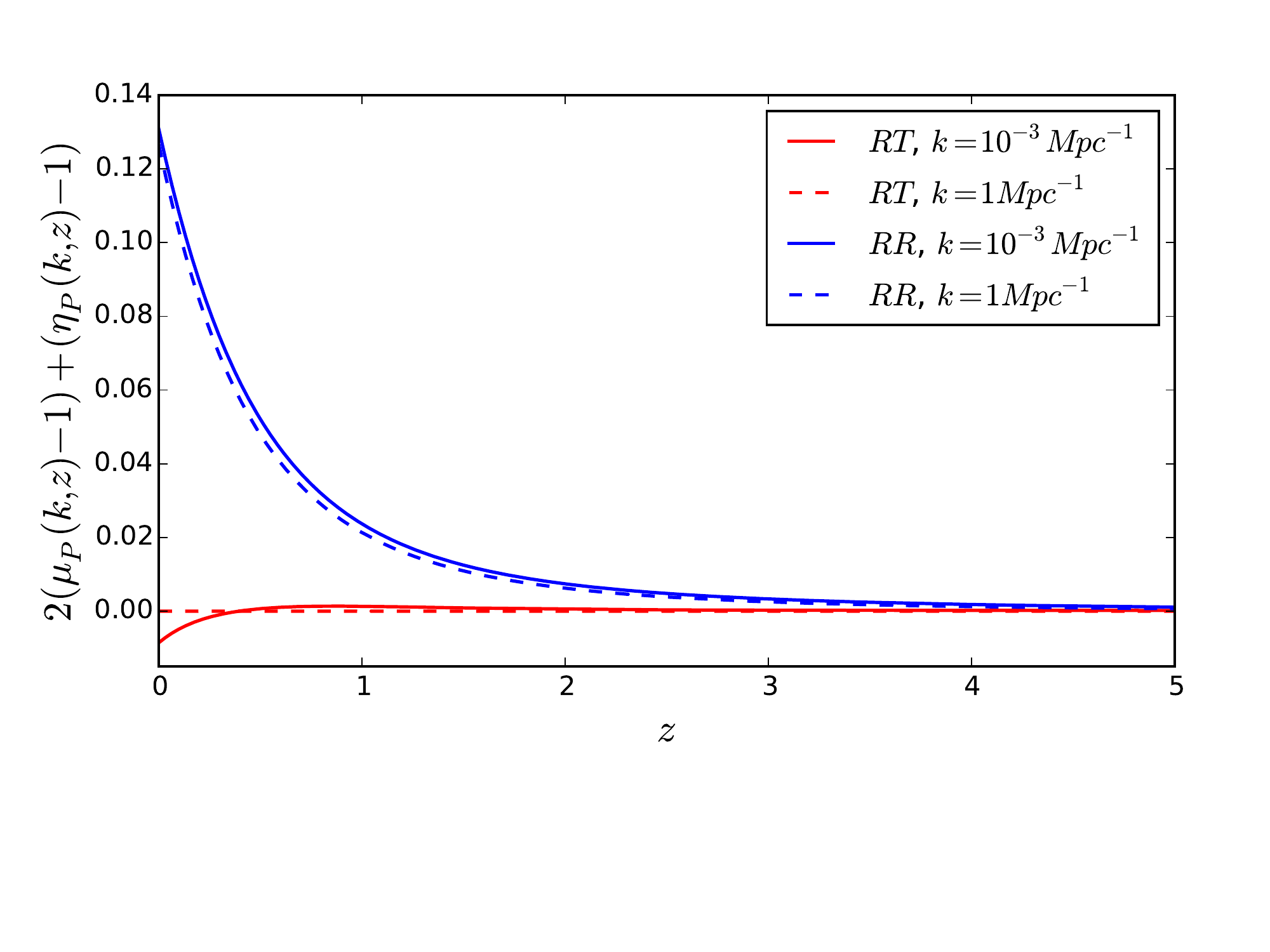}
\caption{\label{comb} The combination used in Planck 2015 dark energy paper \cite{Planck_2015_DE}.}
\end{figure}

\begin{figure}[th]
\centering
\includegraphics[width=0.65\columnwidth]{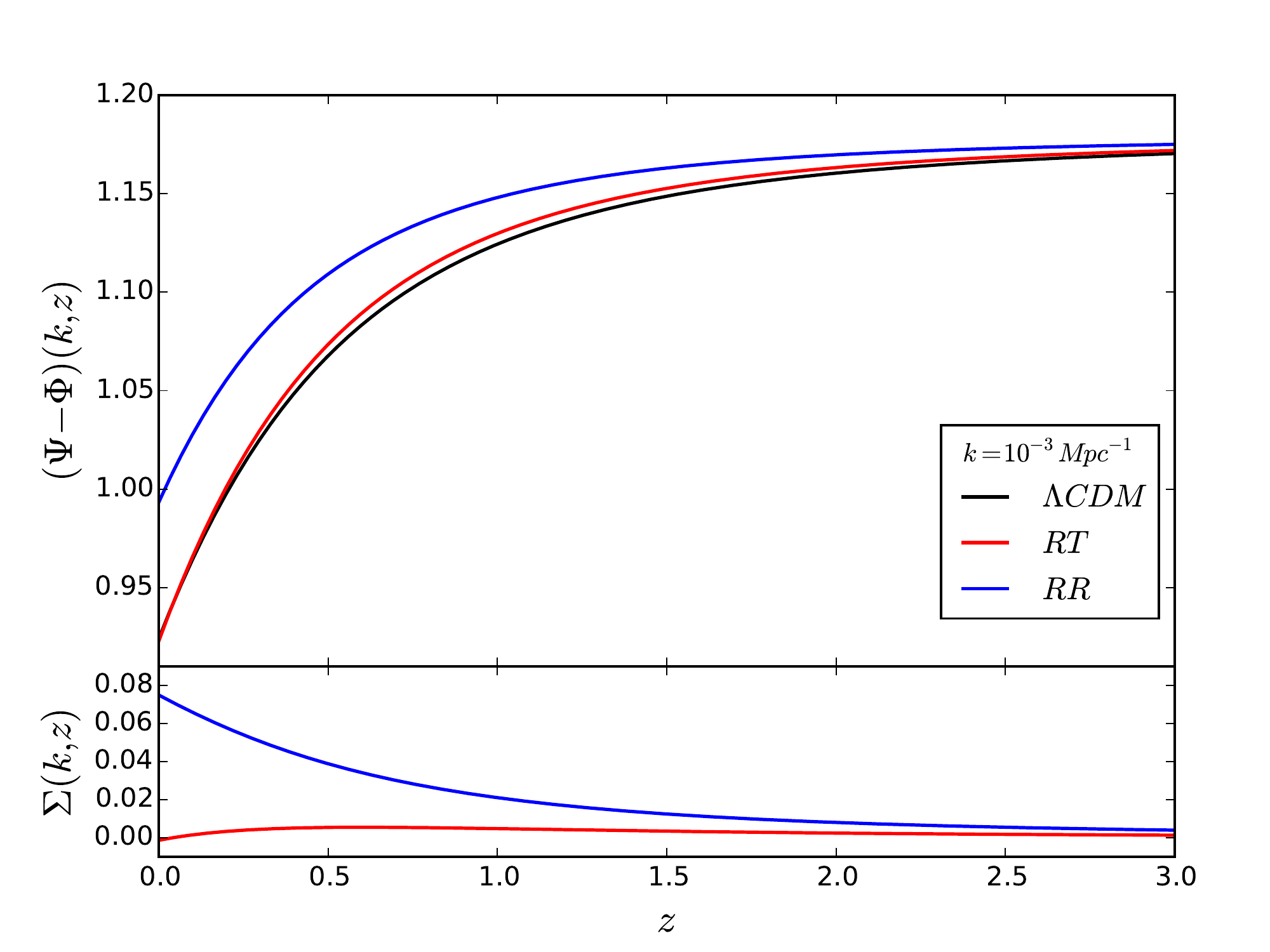}\\
\includegraphics[width=0.65\columnwidth]{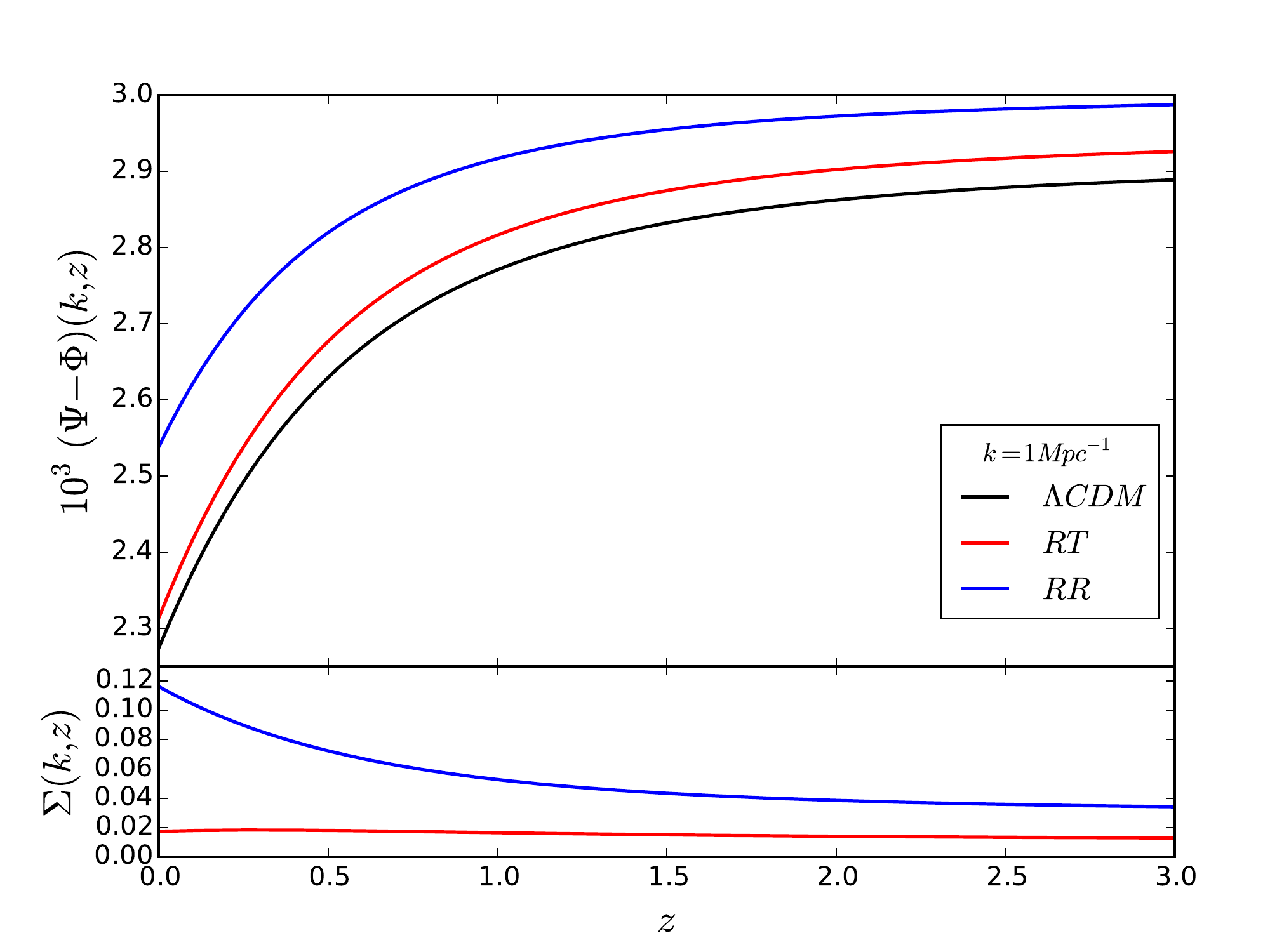}
\caption{\label{phippsi_kbig_and_ksmall} {The difference of the potentials $\Psi - \Phi$ at large scales (upper) and at small scales (lower), and their relative ratio to $\Lambda$CDM.}}
\end{figure}

An alternative   useful pair of indicators is provided by 
the slip function and the effective Newton constant~\cite{Amendola:2007rr,Park:2012cp,Dodelson:2013sma}.
The slip function is given by\footnote{Again the subscript P means that we use the Planck conventions. Another common definition, that we used in \cite{Dirian:2014ara}, is $\eta =(\Phi+\Psi)/\Phi$.}
\be\label{defeta}
\eta_P(k,z) = - \frac{\Phi}{\Psi}\, ,
\ee
This quantity
is equal to one in $\Lambda$CDM when the anisotropic stress is negligible. The effective Newton constant $G_{\rm eff}$  is defined as the constant that enters the Poisson equation for $\Phi$ (see sect.~4.3 and app.~A of 
\cite{Dirian:2014ara} for its explicit expression in the RR and RT models). Observe that 
$G_{\rm eff}(k,z)/G$ is the same as the function $Q(k,z)$ of \cite{Planck_2015_DE}. We plot these quantities in
Figs.~\ref{eta} and \ref{Geff}, while in  Fig.~\ref{comb} we show the combination
$2[\mu_P(k,z)-1]+[\eta_P(k,z)-1]$, which is the one used in the Planck 2015 dark energy paper~\cite{Planck_2015_DE}.
Finally, we show the difference $\Psi - \Phi$ both at large and small scales in 
Fig.~\ref{phippsi_kbig_and_ksmall}. 

\begin{figure}[th]
\centering
\includegraphics[width=0.65\columnwidth]{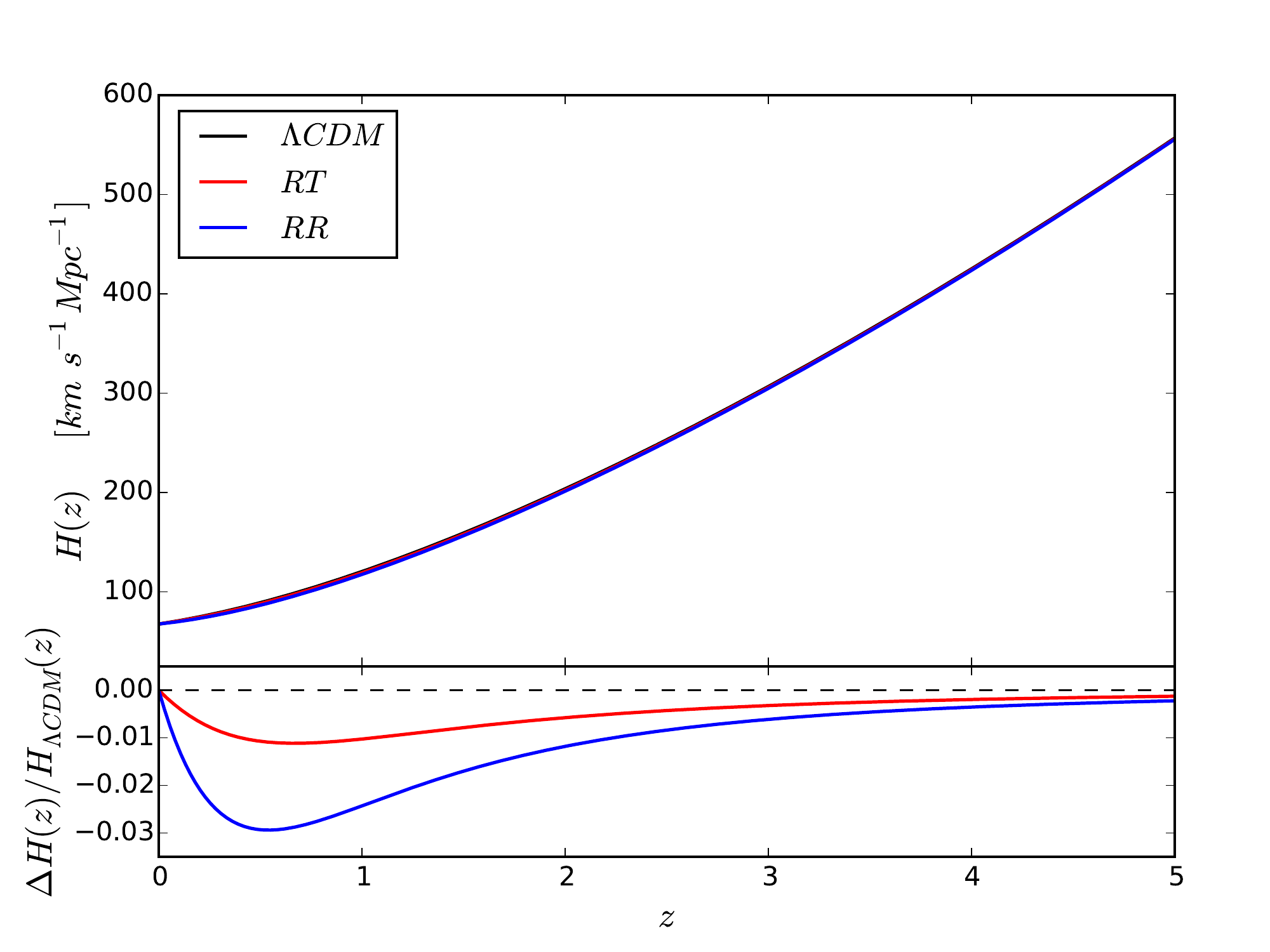}
\caption{\label{Hz} The Hubble parameter as a function of redshift $z$ for the three models, using in all three cases the same values of the parameters, chosen to be the  best-fit value for $\Lambda$CDM. In the upper panel the three curves are indistinguishable. In the lower panel we show the difference of the RT and RR models with respect to $\Lambda$CDM.}
\end{figure}

What we learn from these figures is that, at the typical redshifts of interest for the comparison with observations of structure formation, say $z\, \gsim\, 0.5$,  as far as structure formation is concerned
the RT model differs from $\Lambda$CDM at a level of at most $1\%$, which is  quite small compared to existing experimental uncertainties, but could be observable  with future missions such as Euclid 
\cite{Amendola:2012ys}. In contrast, the RR model shows differences which, at $z\simeq 0.5$, can be as large as  $6\%$, as for instance for the combination shown in 
Fig.~\ref{comb}.\footnote{By comparison, the Deser-Woodard model has been ruled out because $\mu$, near $z=0.5$, differs from the GR value by about 60\% and $\Sigma$ by about $20\%$, see Fig.~1 of \cite{Dodelson:2013sma}.}
As we see from Fig.~\ref{Geff}, these deviations go in the direction of producing a stronger effective Newton constant, and therefore more structures. This is the main reason why the growth displayed in Fig.~\ref{fsigma8} is stronger for the non-local models than for $\Lambda$CDM. A further reason is that, for the same values of the cosmological parameters,  the expansion rate in the RT and RR model is generically lower than in $\Lambda$CDM as can be seen in Fig.~\ref{Hz}, and this contributes (although to a smaller level) also to the enhancement of 
clustering.\footnote{This fact was also noticed in \cite{Barreira:2014kra} where N-body simulations revealed deeper gravitational potential wells and massive haloes slightly more abundant and concentrated.} 

\clearpage

\section{Tensor perturbations}\label{sect:tensor}

Finally, we briefly discuss tensor perturbations in the RR and the RT model, to check for their stability. The stability of perturbations in the tensor sector is a priori non-trivial. We have mentioned for instance in the introduction that in bigravity a potentially interesting self-accelerating branch is in fact plagued by an instability in the tensor sector~\cite{Lagos:2014lca,Cusin:2014psa}. For the RR and RT model, however, no such instability takes place.

\begin{figure}[H]
\centering
\includegraphics[scale=0.65]{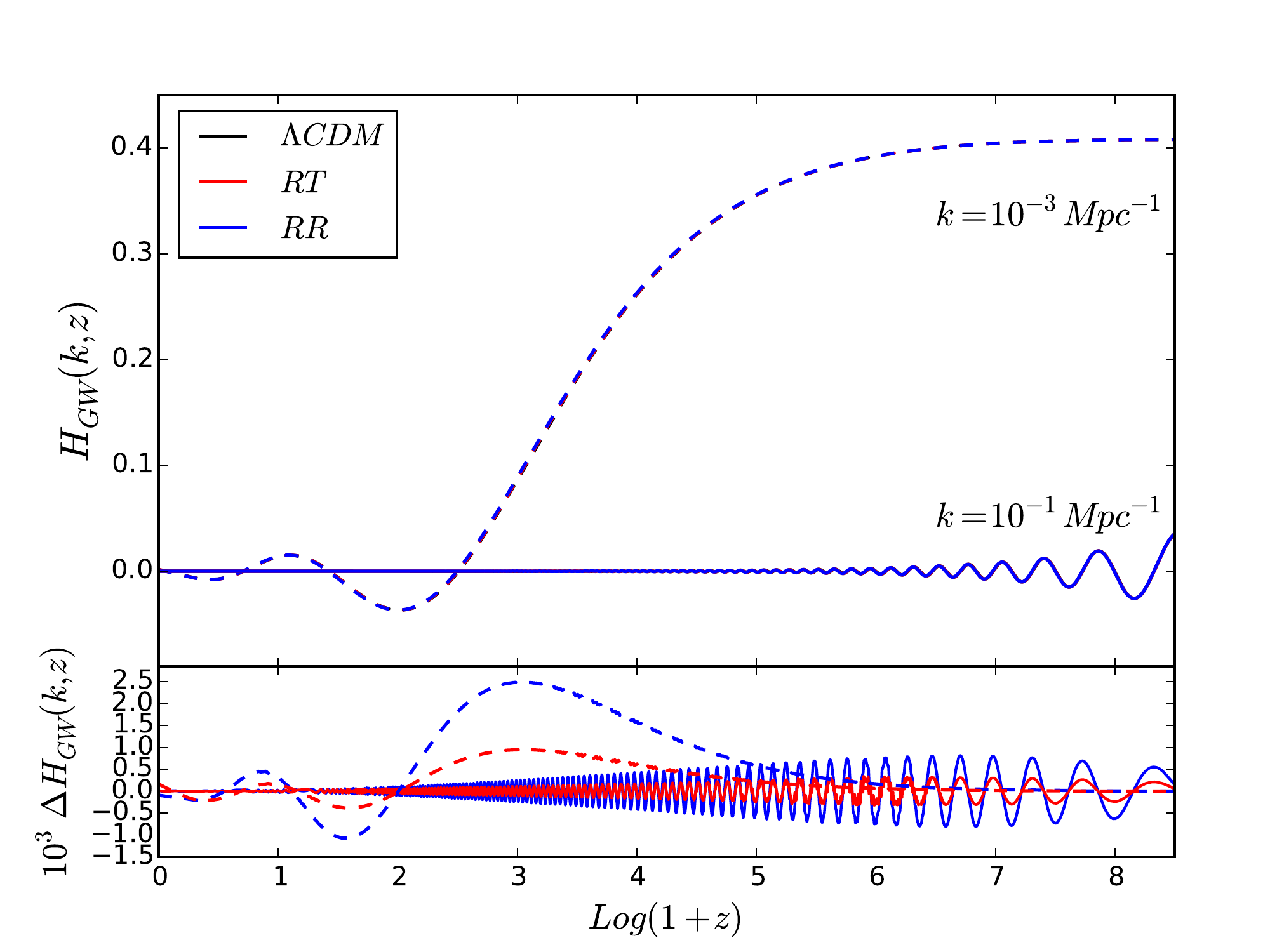}
\caption{\label{GW} Upper panel: Amplitude of the gravitational waves (see \eqref{GWRT}, \eqref{GWRR}) in $\Lambda$CDM (black), RT (red) and RR (blue)  for $k=10^{-1} ~ \mathrm{Mpc}^{-1}$ (solid), $k=10^{-3} ~ \mathrm{Mpc}^{-1}$ (dashed), fixing $r = 0.05$. For each $k$, the three curves corresponding to the three models are indistinguishable from each other on the scale of this plot. Lower panel:  Difference between the amplitude of the gravitational waves between RT and $\Lambda$CDM (red) and between RR and $\Lambda$CDM (blue) for $k=10^{-1} ~ \mathrm{Mpc}^{-1}$ (solid) and $k=10^{-3} ~ \mathrm{Mpc}^{-1}$ (dashed). }
\end{figure}

For the RT model, in the tensor sector the equations governing the dynamics of the gravitational waves are given by
\begin{align}
& \partial_\eta^2 h_{ij}^{TT} + \bigg( 2 \hc - 3 \gamma \bar{V} H_0^2 \bigg) \partial_\eta h_{ij}^{TT} + k^2 h_{ij}^{TT} = 8 \pi G a^2 \Sigma_{ij}^{TT}.  
\end{align}
Here $\bar{V}$ is the background value of the auxiliary field $V$ (see \cite{Dirian:2014ara} for notation) and $\gamma=m^2/(9H_0^2)$, as in  \eq{defgammamH0}, while
the source term $\Sigma_{ij}^{TT}$ is the traceless-transverse part of the anisotropic stress tensor. The latter is non-vanishing whenever relativistic free streaming particles are considered into the evolution \cite{DT_CMBGW, Wei_DampGW}. Decomposing the fields on the tensor eigenfunctions of the Laplacian, e.g. $h_{ij}^{TT}= H^{(T)} Q_{ij}^{(T)}$ \cite{Dur_CMB}, we can write, 
\begin{align}
& \partial_\eta^2  H^{(T)} + \bigg( 2 \hc - 3 \gamma \bar{V} H_0^2 \bigg) \partial_\eta  H^{(T)} + k^2  H^{(T)} = 8 \pi G a^2 \Sigma^{(T)}. \label{GWRT}  
\end{align}
Similarly, for the RR model we get
\begin{align}
& \bigg( 1 - 3\gamma \bar{V} \bigg) \bigg[ \partial_\eta^2 H^{(T)} + 2 \hc \partial_\eta H^{(T)} + k^2 H^{(T)} \bigg]  - 3 \gamma \partial_\eta \bar{V} \partial_\eta H^{(T)}  = 8 \pi G a^2 ~  \Sigma^{(T)} \, . \label{GWRR}
\end{align} 
These equations have no unstable mode.
 Fig.~\ref{GW} shows the result of the numerical integration of these equations, assuming an initial spectrum of perturbations corresponding to the one produced by single-field slow-roll inflation, where we indeed see that there is no instability in the tensor sector. These result have also been recently confirmed and extended in \cite{Cusin:2015rex}, where it has been found that, in contrast, the inclusion in the action of a term proportional to
 $C_{\mu\nu\rho\sigma} \Box^{-2}C^{\mu\nu\rho\sigma}$, where $C^{\mu\nu\rho\sigma}$ is the Weyl tensor, induces instabilities in the tensor sector.

\section{Conclusions}

In this paper we have performed a detailed comparison of the RT and RR non-local models with cosmological data. We have implemented  the cosmological perturbations of the model in a modified version of the  CLASS Boltzmann code  and we have compared these models to a rather complete set of data, including Planck 2015 data for temperature, polarization and lensing, isotropic and anisotropic BAO data, JLA supernovae, local measurement of $H_0$, as well as structure formation data.   The fact that these non-local models have reached the stage where such a detailed comparison is possible, is already by itself a rather non-trivial result. As we discussed in the introduction, from the purely phenomenological point of view no other attempt at modifying gravity in the infrared has reached this stage. We have found that, by themselves, both the RR and RT non-local models provide good fits to the data. We have then compared  the performances of the two non-local models between them and with $\Lambda$CDM. The result is that, with present data, $\Lambda$CDM and the RT model are statistically indistinguishable. They both fit the data extremely well. This is remarkable, considering that they are genuinely different models. The non-local models are not extensions of $\Lambda$CDM with extra free parameters. Rather, they have the same number of  parameters as $\Lambda$CDM, with a mass scale replacing the cosmological constant, and there is no formal limit in which they reduce to $\Lambda$CDM. 
The differences between the predictions of the RT model and those of $\Lambda$CDM  are of order of a few percent, for instance in structure formation; this is a level which allows both models to be consistent with present data, although they could in principle be distinguishable with the data of future missions.
The RR model, in contrast, performs less well in the comparison with data, compared both to $\Lambda$CDM and to the RT model. 

We should however stress that the comparison between the models 
generically depends on the prior assumed and in particular on the values of the sum of the neutrino masses, which could potentially be relevant in our case. In this paper we have used  the sum of the neutrino masses of the \textit{Planck} baseline analysis \cite{Planck_2015_CP}, i.e. a normal mass hierarchy with
$\sum_{\nu} m_{\nu}=0.06$~eV. We make this choice in order to facilitate the comparison with the \textit{Planck} analysis of $\Lambda$CDM. However, oscillation experiments can be consistent with larger neutrino masses, including an inverted hierarchy with $\sum_{\nu} m_{\nu}\, \gsim\, 0.1$~eV. The effect of increasing the neutrino masses is to lower $H_0$, which in $\Lambda$CDM would increase the tension with local measurements. However, we have seen that the non-local models, and particularly the RR model, predict higher values of $H_0$, so they can accommodate larger neutrino masses, without entering in tension with local measurements. Furthermore, larger neutrino masses also have the effect of  lowering $\sigma_8$ (see \cite{Planck_2015_CP} and refs. therein). In view of the result shown in Fig.~\ref{fsigma8}, this would improve the agreement of the non-local models with structure formation data. In particular one can expect that, leaving the neutrino masses as free parameters, within the priors allowed by oscillation experiments, should improve the agreement with the data of the RR model, rendering it again competitive with $\Lambda$CDM. This possibility is currently being investigated in \cite{DirainBarreira:inprep}.

These results encourage further studies of this class of non-local models, and  give precious hints for the search of the mechanism that generates the non-localities from a fundamental local theory.

\vspace{5mm}
\noindent
{\bf Acknowledgments.}
We thank Benjamin Audren,  Alex Barreira,
Julien Lesgourgues,  Baojiu Li, and Ignacy Sawiki for useful discussions.
The work of YD, SF, MK and MM is supported by the Fonds National Suisse. VP acknowledges the DFG TransRegio
TRR33 grant on The Dark Universe. The work of MM is supported by the SwissMap NCCR.
Part of the numerical work presented in this publication used the Baobab cluster of the University of Geneva.

The development of Planck has been supported by: ESA; CNES and CNRS/INSU-IN2P3-INP (France); ASI, CNR, and INAF (Italy); NASA and DoE (USA); STFC and UKSA (UK); CSIC, MICINN and JA (Spain); Tekes, AoF and CSC (Finland); DLR and MPG (Germany); CSA (Canada); DTU Space (Denmark); SER/SSO (Switzerland); RCN (Norway); SFI (Ireland); FCT/MCTES (Portugal).
A description of the Planck Collaboration and a list of its members, including the technical or scientific activities in which they have been involved, can be found at 
http://www.cosmos.esa.int/web/planck/planck-collaboration.

\appendix

\section{Implementation of the nonlocal models in CLASS}\label{app:class}

The full set of cosmological equations of motion for the RT and RR models, at the background and scalar linear perturbations level, has already been given in \cite{Dirian:2014ara}. In this appendix we briefly comment on the global strategy used in \textsc{CLASS} for evolving them. For completeness we also review here the equations used in the code, in the corresponding format. The code itself will became publicly available soon \cite{git_nonlocal}. 

CLASS uses the following convention for scalar perturbations\footnote{At the time of writing the present paper, the equations for the nonlocal models have been implemented only in Newtonian gauge into \textsc{CLASS}, and not in the synchronous one.}
\begin{align}
ds^2 = a^2 \big[ -(1+2 \psi) d\eta^2 + \[ (1 - 2 \phi) \delta_{ij}\] dx^i dx^j \big], \label{ds2}
\end{align}
which is different from the one we adopted in \eq{defPhiPsi}.
In order to facilitate the comparison with the code, we will use CLASS convention in this appendix.  
Since the nonlocal models  both have the same overall structure, that is, a set of modified Einstein equations including auxiliary fields along with second order differential equations dictating the evolution of the latter, their implementation in \textsc{CLASS} is similar.

\subsection{Implementation of the RT model}

We start by writing down the relevant cosmological background equations corresponding to the RT model \cite{Maggiore:2013mea,Foffa:2013vma}.
In the code, the $(00)$ component of Einstein equations is used to infer algebraically the Hubble parameter $H\equiv a'/a^2$ in terms of the energy density of the matter component that one wishes to take into account. In our case it reads,
\begin{align}
H = \frac{\gamma \bar{V} H_0^2}{2 a}  + \bigg[ \bigg(\frac{\gamma \bar{V} H_0^2}{2 a}\bigg) + \gamma\big(\bar{U} - \bar{V}' / a^2 \big) H_0^2+ \frac{8 \pi G}{3} \bar{\rho}\bigg]^{1/2}, \label{HRT}
\end{align}
\noindent where $H_0$ is the present value of the Hubble parameter and
$\bar{U}$ and $\bar{V}$ denote the background values of the auxiliary fields $U$ and $V$.  The derivative of the Hubble parameter is not computed numerically, but rather obtained algebraically from the trace of the $(ij)$ component of Einstein equations, in terms of the pressure,
\begin{align}
H' = - \frac{3}{2} a \bigg[H^2 + \frac{8 \pi G}{3} \bar{p} + \gamma \bigg( \frac{H}{a} \bar{V} - \bar{U} \bigg) H_0^2 \bigg]. \label{HpRT}
\end{align}
As already mentioned above, in our case, the evolution of the auxiliary fields is dictated by a set of second order differential equations,
\begin{align}
\bar{U}'' + 2 a H \bar{U}' = 6 a\big(H' + 2 a H^2 \big) \quad, \quad \bar{V}'' - a \big(H' + 5 a H^2 \big) \bar{V} = a^2 \bar{U}'\,
\end{align}
which need to be integrated numerically directly within the code with initial conditions $\bar{U}=\bar{U}'=\bar{V}=\bar{V}'=0$ set deep into the radiation era~\cite{Maggiore:2013mea,Foffa:2013vma}. Once the matter components have been specified, such that their energy density and pressure can be written explicitly (for example by using their separate energy-momentum conservation  together with an equation of state, or their unperturbed phase-space distribution functions) the system closes and the background evolution can be integrated.

Concerning the  linear scalar perturbations, the set of equations needed includes the conservation equation of each fluid component, along with the evolution equations for the auxiliary fields perturbations and two independent components of the modified Einstein equations. For the latter, the code originally uses the divergence of the $(0i)$ component in order to extract $\phi'$ algebraically. In our case we 
have\footnote{Here $\hc \equiv a' / a$, as used in the perturbation module of \textsc{CLASS}.}
\begin{align}
&\phi' = - \mathcal{H} \psi + 4 \pi G \frac{a^2}{k^2} ( \bar{\rho} + \bar{p}) \theta + \frac{3}{2}\gamma \bigg[ \mathcal{H} \delta Z - \frac{1}{2} \delta Z' + \frac{1}{2} \psi \bar{V} - \frac{1}{2} \delta V \bigg] H_0^2 \, .  \label{phipRT}
\end{align}
The expression of $\psi$ is obtained in terms of $\phi$ from the longitudinal-traceless part of the $(ij)$ component of Einstein equation (the one sourced by the scalar anisotropic stress $\sigma$, see eq. (4.8)-(4.9) of \cite{Dirian:2014ara} for more details), 
\begin{align}
&\psi = \phi - 12 \pi G \frac{a^2}{k^2} (\bar{\rho} + \bar{p}) \sigma  + 3 \gamma \delta Z H_0^2 \, , \label{psiRT}
\end{align}
and $\phi$ is obtained from \eqref{phipRT} by numerical integration. The equations governing the dynamics of the auxiliary fields are,
\begin{align}
 &\delta U'' + k^2 \delta U + 2 \mathcal{H} \delta U'  = ( \psi' + 3 \phi') (\bar{U}'-6 \mathcal{H} )+ 2 k^2 (\psi - 2 \phi) - 6 \phi '' ,\\
& \delta Z '' -2 (\mathcal{H}' + 2 \mathcal{H}^2 - k^2) \delta Z  = 2 a^2 \delta U - 4 \mathcal{H} \delta V - \delta V' + 3 \bar{V}' \psi + \big( \psi'  +2 \phi' \big) \bar{V} \, , \\
 & \delta V'' - \bigg( \mathcal{H}' + 4 \mathcal{H}^2 - \frac{k^2}{2} \bigg) \delta V  =   \\
& \hspace{1cm} a^2 \delta U'  - \frac{1}{2} k^2 \delta Z' + 2 \mathcal{H} k^2 \delta Z+\bigg[ \frac{k^2}{2}  \psi - \mathcal{H} (\psi' + 9 \phi') \bigg] \bar{V} + ( \psi' + 3 \phi') \bar{V}' + a^2 \bar{U}' \psi  \, .\nn
\end{align}
We see that they contain $\psi'$ and $\phi''$, which can also be extracted algebraically from the equations. In particular, $\phi''$ is obtained from the trace of the $(ij)$ component of Einstein equations,
\begin{align}
&\phi'' = - \psi (\mathcal{H}^2 + 2\mathcal{H}') - \mathcal{H} (\psi' + 2\phi') + \frac{k^2}{3} (\psi - \phi) \nn  \\
& \hspace{3.5cm} - \frac{3}{2} \bigg[ \gamma\bigg(a^2 \delta U + \big( \phi' + \mathcal{H} \psi \big) \bar{V} - \mathcal{H} \delta V - \frac{k^2}{3} \delta Z \bigg) H_0^2- \frac{8 \pi G}{3}a^2 \delta p\bigg], \label{psipp}
\end{align}
whereas $\psi'$ is obtained by taking the derivatives of \eqref{psiRT},
\begin{align}
& \psi '= \phi' - \frac{24 \pi G}{k^2} \hc a^2 (\bar{\rho} + \bar{p}) \sigma - 12 \pi G \frac{a^2}{k^2} \big[(\bar{\rho} + \bar{p}) \sigma\big]' + 3 \gamma \d Z' H_0^2 \, . \label{psipRT}
\end{align}
\noindent Within the version of \textsc{CLASS} that we used, the source term $\big[(\bar{\rho} + \bar{p}) \sigma\big]$ decomposes as
\begin{align}
\big[(\bar{\rho} + \bar{p}) \sigma\big] = \frac{4}{3} \big( \rho_\gamma \sigma_\gamma + \rho_{ur} \sigma_{ur} \big) + (\rho + p)_{ncdm} \sigma_{ncdm} \, .
\end{align}
where $\gamma$, $ur$ and $ncdm$ denote the photon, ultra-relativistic particles and non-cold dark matter (NCDM) components respectively. Therefore to compute its corresponding time derivative one needs, 
\begin{align}
\big[(\bar{\rho} + \bar{p}) \sigma\big]' = \frac{4}{3} \big( \rho_\gamma' \sigma_\gamma + \rho_{ur}' \sigma_{ur} +  \rho_\gamma \sigma_\gamma' + \rho_{ur} \sigma_{ur}' \big) + \big[(\rho + p)_{ncdm} \sigma_{ncdm}\big]' \, ,
\end{align}
and, since at background level each of the ultra-relativistic particles and photons components is conserved, one can use the background conservation equation, $\bar{\rho}' = - 3 \hc (\bar{\rho} + \bar{p})$ together with the equation of state for ultra-relativistic particle $\bar{p}=(1/3)\bar{\rho}$, to write,
\begin{align}
\big[(\bar{\rho} + \bar{p}) \sigma\big]' = \frac{4}{3} \rho_\g \big( \sigma_\g' - 4 \hc \sigma_\g \big) + \frac{4}{3} \rho_{ur} \big( \sigma_{ur}' - 4 \hc \sigma_{ur} \big) + \big[(\rho + p)_{ncdm} \sigma_{ncdm}\big]' \, , 
\end{align}
while the (massive) NCDM components have to be treated separately in terms of their phase-space description. This form is quite convenient for being implemented into the code since the various quantities have already been built into the original version.

\subsection{Implementation of the RR model}

In the case of the RR  model the general structure is the same as above and we only display the relevant equations needed for the implementation in the code. At the background level, the $(00)$ component of the modified Einstein equations   \cite{Maggiore:2014sia} leads to 
\begin{align}
H = \bigg( 1 - 3 \gamma \bar{V} \bigg)^{-1} \bigg\{ \frac{3 \gamma}{2a} \bar{V}' + \bigg[ \bigg(\frac{3 \gamma }{2 a} \bar{V}'\bigg)^2 +  \bigg( 1 - 3 \gamma \bar{V} \bigg) \bigg( \frac{\gamma}{4} \bar{U}^2 H_0^2 - \frac{\gamma}{2 a^2} \bar{V}' \bar{U}' + \frac{8 \pi G}{3} \rho \bigg)  \bigg]^{1/2} \bigg\} \, .
\end{align}
From the $(ij)$ component one gets
\begin{align}
H' = - \frac{3}{2} a\bigg\{ H^2 + \bigg( 1 - 3 \gamma  \bar{V} \bigg)^{-1} \bigg[\frac{8 \pi G}{3}\bar{p} - \gamma\bigg( \frac{1}{4} \bar{U}^2 H_0^2 + \bar{U} H_0^2 - \frac{H}{a} \bar{V}' + \frac{1}{2 a^2} \bar{V}' \bar{U}' \bigg) \bigg] \bigg\} \, . \label{H'class}
\end{align}
The equation of the auxiliary fields are
\begin{align}
\bar{U}'' + 2 a H \bar{U}' = 6 a \big( H' + 2 a H^2 \big), \qquad \bar{V}'' + 2 a H \bar{V}' = a^2 \bar{U} H_0^2.
\end{align}
To linear order in the scalar perturbations, the $(0i)$ component of the modified Einstein equations leads to, 
\begin{align}
\phi' = - \hc \psi+\frac{3}{2}\bigg( 1 - 3 \gamma \bar{V} \bigg)^{-1} \bigg[\frac{8 \pi G a^2}{3 k^2} (\bar{\rho} + \bar{p}) \theta - \gamma \bigg(\d V'- \bar{V}' \psi - \hc \d V + \frac{1}{2}\big( \bar{U'}\d V + \bar{V}' \d U \big) \bigg) \bigg].
\end{align}
The equation  sourced by the anisotropic stress gives 
\begin{align}
\psi =\phi + \bigg( 1 - 3 \gamma \bar{V} \bigg)^{-1} \bigg[ - 12 \pi G \frac{a^2}{k^2} (\bar{\rho} + \bar{p}) \sigma + 3 \gamma \d V  \bigg] \,.
\end{align}
The  perturbation of the equations for the auxiliary field equations gives
\begin{align}
&\delta U '' + 2 \hc \delta U' + k^2 \delta U = \big( \psi' + 3 \phi' \big) (\bar{U}'-6 \hc) + 2 k^2 \big( \psi - 2 \phi \big)  - 6 \phi'' \, , \\
& \delta V '' + 2 \hc \delta V' + k^2 \delta V = \big(\psi' +  3 \phi' \big) \bar{V}' + 2 a^2 \psi \bar{U} H_0^2 + a^2 \delta U H_0^2\, . 
\end{align}
In order to solve the system, one has also to provide expressions of $\psi'$ and $\phi''$ obtained similarly to the case of the RT model. One finds
\begin{align}
\psi '= \phi' + \bigg( 1 - 3 \gamma \bar{V} \bigg)^{-1} \bigg[ 3\gamma (\psi - \phi) \bar{V}'  - \frac{24 \pi G}{k^2} \hc a^2 (\bar{\rho} + \bar{p}) \sigma - 12 \pi G \frac{a^2}{k^2} \big[(\bar{\rho} + \bar{p}) \sigma\big]' + 3 \gamma \d V'  \bigg] \, , \label{psipRR}
\end{align}
and
\begin{align}
& \phi'' = - \psi(\mathcal{H}^2 + 2 \mathcal{H}') - \mathcal{H} ( \psi' + 2 \phi') + \frac{k^2}{3}(\psi - \phi) \nn \\
& \hspace{1cm} - \frac{3}{2}\bigg( 1 - 3 \gamma \bar{V} \bigg)^{-1}  \bigg\{ \gamma \bigg[ \frac{1}{2} a^2 \bar{U} \delta U H_0^2- 2 a^2 \psi \bar{U} H_0^2- \big(  2 \phi' - 2 \hc \psi + \psi' + \psi \bar{U}' \big) \bar{V}'  \nn \\
& \hspace{1cm} + \delta V'' + \hc \delta V'+ \bigg( \hc^2 + 2 \hc' + \frac{2 k^2}{3}\bigg) \delta V +\frac12 \big( \bar{U}' \delta V'  + \bar{V}' \delta U' \big) \bigg] -  \frac{8 \pi G}{3}a^2 \delta p \bigg\} \, . \label{EE4co}
\end{align}

\section{Comparison with Lunar Laser Ranging}\label{app:solar}

Ref.~\cite{Barreira:2014kra} has raised the question of the consistency of the 
RR model with Lunar Laser Ranging experiments. The concern arises from the fact that, in the RR model, among the equations obtained perturbing over the cosmological solution,  we have a Poisson equation where the term $\n^2\Phi$ is multiplied by a factor
$[1-(m^2 \bar{S}/3)]$, where $\bar{S}$ is the background value of   the auxiliary field $S$ (see \cite{Dirian:2014ara} for definitions and notation), leading to an effective Newton constant 
\be\label{GeffA1}
G_{\rm eff}=G \[ 1-\frac{m^2 \bar{S}}{3}\]^{-1} \[ 1+{\cal O}\(\frac{H_0^2}{k^2}\)\]\, ,
\ee
see eqs.~(23,24) of
\cite{Barreira:2014kra} or eq.~(5.3) of \cite{Dirian:2014ara}. If  one now uses the time dependence of $\bar{S}(t)$ found from the cosmological background solution, and applies \eq{GeffA1}  to the Earth-Moon system, ones find a value $\dot{G}_{\rm eff}/G\simeq 92\times 10^{-13} \, {\rm yr}^{-1}$~\cite{Barreira:2014kra}, which exceeds the current bound  from Lunar Laser Ranging (LLR),
$\dot{G}_{\rm eff}/G\simeq (4\pm 9)\times 10^{-13} \, {\rm yr}^{-1}$ \cite{Williams:2004qba}.

As a first remark, we observe that this potential problem does not appear in the 
RT model. Indeed, for the RT model  inside the horizon $G_{\rm eff}/G=1+{\cal O}(H_0^2/k^2)$,
see \cite{Kehagias:2014sda,Nesseris:2014mea,Dirian:2014ara}, so for the RT model the issue raised
in \cite{Barreira:2014kra} does not apply, and there is no problem with  LLR.

For the RR model the issue is  more delicate. An obvious question, indeed mentioned in \cite{Barreira:2014kra}, is whether one is allowed to use the background expansion rate for the field
$\bar{S}(t)$,
computed at cosmological scales, inside a system such as the Earth-Moon system.
The Earth-Moon system, and more generally the solar system or even the Local Group, are not expanding with the Hubble flow,  so the numerical estimate suggesting a value  $\dot{G}_{\rm eff}/G\simeq 92\times 10^{-13} \, {\rm yr}^{-1}$ cannot be taken  literally. Nevertheless, there still remains the possibility that, even if the background geometry at local scales (i.e. at the scales of the Earth-Moon system, or of the solar system) no longer follows the cosmological expansion, a scalar field  could still have  a residual time dependence inherited from its behavior at cosmological scales. Indeed,  this can happen in scalar-tensor theories in which the scalar field $\varphi$ has a shift symmetry $\varphi\ra\varphi+ {\rm const.}$ \cite{Babichev:2011iz}. In this case the equations of motion admit a separation of variables of the form 
$\varphi(t,r)=\varphi(r)+\varphi_{\rm cosm}(t)$, with $\dot{\varphi}_{\rm cosm}(t)\sim H(t)$.
Since the LLR bound, expressed in terms of $H_0$, reads $\dot{G}/G= (0.6\pm 1.3)\times 10^{-2} H_0$,   if this scalar field enters in the effective Newton constant (as happens in most tensor-scalar theories, as well as in the RR model, but not in the RT model), this could induce, depending on the precise numerical factor, a  violation of the LLR bound.\footnote{Note however that 
it is possible  that this ansatz  leads to an differential equation for 
$\varphi(r)$ which has no regular solution~\cite{Babichev:2011iz}, in which case this would not  be the solution relevant to the physics of the problem, i.e. the  solution corresponding to a metric interpolating from a Schwarzschild-like behavior in the local region to the FRW behavior at cosmological distances.}

In our case,  in the RR model there is no shift symmetry $S\ra S+{\rm const.}$ for the  auxiliary field $S$ (see e.g. eqs.~(4)-(7) of \cite{Maggiore:2014sia}),  so the above argument does not  apply directly.   
Nevertheless, for the RR model, in the absence of an explicit  study of the equations including as sources both a cosmological fluid  and a  localized static source, one cannot exclude a priori the presence of a residual time dependence in the auxiliary field $S$, on local scales.
Similarly, it would be interesting to study the stability of such $r$- and $t$-dependent solution under time-dependent perturbations. We leave these issues for future work.

\bibliographystyle{utphys}						
%\bibliography{myrefs_massive,biblioYves,biblio}
\bibliography{myrefs_massive,biblioYves}

\end{document}

%% file: mydefs.tex
\newcommand{\hc}{{\cal H}}

\newcommand{\nn}{\nonumber}

\newcommand{\iBox}{\Box^{-1}}

\renewcommand\({\left(}
\renewcommand\){\right)}
\renewcommand\[{\left[}
\renewcommand\]{\right]}

\newcommand\n{{\mbox {\boldmath $\nabla$}}}
\newcommand{\ra}{\rightarrow}

\def\lsim{\raise 0.4ex\hbox{$<$}\kern -0.8em\lower 0.62
ex\hbox{$\sim$}}

\def\gsim{\raise 0.4ex\hbox{$>$}\kern -0.7em\lower 0.62
ex\hbox{$\sim$}}

\def\lbar{{\hbox{$\lambda$}\kern -0.7em\raise 0.6ex
\hbox{$-$}}}

\newcommand\eq[1]{eq.~(\ref{#1})}

\newcommand\p{\partial}

\newcommand\ee{\end{equation}}
\newcommand\be{\begin{equation}}
\def\bea{\begin{array}}
\def\eea{\end{array}}\def\ea{\end{array}}
\newcommand\ees{\end{eqnarray}}
\newcommand\bees{\begin{eqnarray}}
\def\nn{\nonumber}

\def\s{\sigma}
\def\g{\gamma}

\def\d{\delta}

\def\dslash{\hspace{-1mm}\not{\hbox{\kern-2pt $\partial$}}}
\def\Dslash{\not{\hbox{\kern-4pt $D$}}}
\def\pslash{\not{\hbox{\kern-2.1pt $p$}}}
\def\kslash{\not{\hbox{\kern-2.3pt $k$}}}
\def\qslash{\not{\hbox{\kern-2.3pt $q$}}}

\def\p1{{\bf p}_1}
\def\p2{{\bf p}_2}
\def\k1{{\bf k}_1}
\def\k2{{\bf k}_2}

\newcommand{\gmn}{g_{\mu\nu}}

\newcommand{\Rmn}{R_{\mu\nu}}
\newcommand{\Gmn}{G_{\mu\nu}}
\newcommand{\RMN}{R^{\mu\nu}}

\newcommand{\Tmn}{T_{\mu\nu}}

\newcommand{\dddM}{\kern 0.2em \raise 1.9ex\hbox{$...$}\kern -1.0em \hbox{$M$}}
\newcommand{\dddQ}{\kern 0.2em \raise 1.9ex\hbox{$...$}\kern -1.0em \hbox{$Q$}}
\newcommand{\dddI}{\kern 0.2em \raise 1.9ex\hbox{$...$}\kern -1.0em\hbox{$I$}}
\newcommand{\dddJ}{\kern 0.2em \raise 1.9ex\hbox{$...$}\kern-1.0em
\hbox{$J$}}
\newcommand{\dddcalJ}{\kern 0.2em \raise 1.9ex\hbox{$...$}\kern-1.0em
\hbox{${\cal J}$}}

\newcommand{\dddO}{\kern 0.2em \raise 1.9ex\hbox{$...$}\kern -1.0em
\hbox{${\cal O}$}}
\def\dddz{\raise 1.5ex\hbox{$...$}\kern -0.8em \hbox{$z$}}
\def\dddd{\raise 1.8ex\hbox{$...$}\kern -0.8em \hbox{$d$}}
\def\dddbd{\raise 1.8ex\hbox{$...$}\kern -0.8em \hbox{${\bf d}$}}
\def\ddbd{\raise 1.8ex\hbox{$..$}\kern -0.8em \hbox{${\bf d}$}}
\def\dddx{\raise 1.6ex\hbox{$...$}\kern -0.8em \hbox{$x$}}
\newcommand{\mplr}{m_{\rm Pl}}

\newcommand{\ora}{\Omega_{R}}

\newcommand{\ola}{\Omega_{\Lambda}}

%% file: nonlocal_long250416jcap.bbl
\providecommand{\href}[2]{#2}\begingroup\raggedright\begin{thebibliography}{100}

\bibitem{Bull:2015stt}
P.~Bull {\em et.~al.}, ``{Beyond $\Lambda$CDM: Problems, solutions, and the
  road ahead},'' \href{http://xxx.lanl.gov/abs/1512.05356}{{\tt 1512.05356}}.

\bibitem{Dvali:2000hr}
G.~Dvali, G.~Gabadadze, and M.~Porrati, ``{4-D gravity on a brane in 5-D
  Minkowski space},'' {\em Phys.Lett.} {\bf B485} (2000) 208--214,
  \href{http://xxx.lanl.gov/abs/hep-th/0005016}{{\tt hep-th/0005016}}.

\bibitem{deRham:2010ik}
C.~de~Rham and G.~Gabadadze, ``{Generalization of the Fierz-Pauli Action},''
  {\em Phys.Rev.} {\bf D82} (2010) 044020,
  \href{http://xxx.lanl.gov/abs/1007.0443}{{\tt 1007.0443}}.

\bibitem{deRham:2010kj}
C.~de~Rham, G.~Gabadadze, and A.~J. Tolley, ``{Resummation of Massive
  Gravity},'' {\em Phys.Rev.Lett.} {\bf 106} (2011) 231101,
  \href{http://xxx.lanl.gov/abs/1011.1232}{{\tt 1011.1232}}.

\bibitem{Hassan:2011hr}
S.~Hassan and R.~A. Rosen, ``{Resolving the Ghost Problem in non-Linear Massive
  Gravity},'' {\em Phys.Rev.Lett.} {\bf 108} (2012) 041101,
  \href{http://xxx.lanl.gov/abs/1106.3344}{{\tt 1106.3344}}.

\bibitem{Hinterbichler:2011tt}
K.~Hinterbichler, ``{Theoretical Aspects of Massive Gravity},'' {\em
  Rev.Mod.Phys.} {\bf 84} (2012) 671--710,
  \href{http://xxx.lanl.gov/abs/1105.3735}{{\tt 1105.3735}}.

\bibitem{deRham:2014zqa}
C.~de~Rham, ``{Massive Gravity},'' {\em Living Rev. Rel.} {\bf 17} (2014) 7,
  \href{http://xxx.lanl.gov/abs/1401.4173}{{\tt 1401.4173}}.

\bibitem{Hassan:2011zd}
S.~Hassan and R.~A. Rosen, ``{Bimetric Gravity from Ghost-free Massive
  Gravity},'' {\em JHEP} {\bf 1202} (2012) 126,
  \href{http://xxx.lanl.gov/abs/1109.3515}{{\tt 1109.3515}}.

\bibitem{Schmidt-May:2015vnx}
A.~Schmidt-May and M.~von Strauss, ``{Recent developments in bimetric
  theory},'' \href{http://xxx.lanl.gov/abs/1512.00021}{{\tt 1512.00021}}.

\bibitem{Deffayet:2000uy}
C.~Deffayet, ``{Cosmology on a brane in Minkowski bulk},'' {\em Phys.Lett.}
  {\bf B502} (2001) 199--208,
  \href{http://xxx.lanl.gov/abs/hep-th/0010186}{{\tt hep-th/0010186}}.

\bibitem{Deffayet:2001pu}
C.~Deffayet, G.~Dvali, and G.~Gabadadze, ``{Accelerated universe from gravity
  leaking to extra dimensions},'' {\em Phys.Rev.} {\bf D65} (2002) 044023,
  \href{http://xxx.lanl.gov/abs/astro-ph/0105068}{{\tt astro-ph/0105068}}.

\bibitem{Luty:2003vm}
M.~A. Luty, M.~Porrati, and R.~Rattazzi, ``{Strong interactions and stability
  in the DGP model},'' {\em JHEP} {\bf 0309} (2003) 029,
  \href{http://xxx.lanl.gov/abs/hep-th/0303116}{{\tt hep-th/0303116}}.

\bibitem{Nicolis:2004qq}
A.~Nicolis and R.~Rattazzi, ``{Classical and quantum consistency of the DGP
  model},'' {\em JHEP} {\bf 0406} (2004) 059,
  \href{http://xxx.lanl.gov/abs/hep-th/0404159}{{\tt hep-th/0404159}}.

\bibitem{Gorbunov:2005zk}
D.~Gorbunov, K.~Koyama, and S.~Sibiryakov, ``{More on ghosts in DGP model},''
  {\em Phys.Rev.} {\bf D73} (2006) 044016,
  \href{http://xxx.lanl.gov/abs/hep-th/0512097}{{\tt hep-th/0512097}}.

\bibitem{Charmousis:2006pn}
C.~Charmousis, R.~Gregory, N.~Kaloper, and A.~Padilla, ``{DGP Specteroscopy},''
  {\em JHEP} {\bf 0610} (2006) 066,
  \href{http://xxx.lanl.gov/abs/hep-th/0604086}{{\tt hep-th/0604086}}.

\bibitem{Izumi:2006ca}
K.~Izumi, K.~Koyama, and T.~Tanaka, ``{Unexorcized ghost in DGP brane world},''
  {\em JHEP} {\bf 0704} (2007) 053,
  \href{http://xxx.lanl.gov/abs/hep-th/0610282}{{\tt hep-th/0610282}}.

\bibitem{D'Amico:2011jj}
G.~D'Amico, C.~de~Rham, S.~Dubovsky, G.~Gabadadze, D.~Pirtskhalava, and A.~J.
  Tolley, ``{Massive Cosmologies},'' {\em Phys.Rev.} {\bf D84} (2011) 124046,
  \href{http://xxx.lanl.gov/abs/1108.5231}{{\tt 1108.5231}}.

\bibitem{DeFelice:2013awa}
A.~De~Felice, A.~E. Gumrukcuoglu, C.~Lin, and S.~Mukohyama, ``{Nonlinear
  stability of cosmological solutions in massive gravity},'' {\em JCAP} {\bf
  1305} (2013) 035, \href{http://xxx.lanl.gov/abs/1303.4154}{{\tt 1303.4154}}.

\bibitem{Volkov:2011an}
M.~S. Volkov, ``{Cosmological solutions with massive gravitons in the bigravity
  theory},'' {\em JHEP} {\bf 01} (2012) 035,
  \href{http://xxx.lanl.gov/abs/1110.6153}{{\tt 1110.6153}}.

\bibitem{Comelli:2011zm}
D.~Comelli, M.~Crisostomi, F.~Nesti, and L.~Pilo, ``{FRW Cosmology in Ghost
  Free Massive Gravity},'' {\em JHEP} {\bf 03} (2012) 067,
  \href{http://xxx.lanl.gov/abs/1111.1983}{{\tt 1111.1983}}. [Erratum:
  JHEP06,020(2012)].

\bibitem{vonStrauss:2011mq}
M.~von Strauss, A.~Schmidt-May, J.~Enander, E.~Mortsell, and S.~Hassan,
  ``{Cosmological Solutions in Bimetric Gravity and their Observational
  Tests},'' {\em JCAP} {\bf 1203} (2012) 042,
  \href{http://xxx.lanl.gov/abs/1111.1655}{{\tt 1111.1655}}.

\bibitem{Akrami:2012vf}
Y.~Akrami, T.~S. Koivisto, and M.~Sandstad, ``{Accelerated expansion from
  ghost-free bigravity: a statistical analysis with improved generality},''
  {\em JHEP} {\bf 03} (2013) 099, \href{http://xxx.lanl.gov/abs/1209.0457}{{\tt
  1209.0457}}.

\bibitem{Tamanini:2013xia}
N.~Tamanini, E.~N. Saridakis, and T.~S. Koivisto, ``{The Cosmology of
  Interacting Spin-2 Fields},'' {\em JCAP} {\bf 1402} (2014) 015,
  \href{http://xxx.lanl.gov/abs/1307.5984}{{\tt 1307.5984}}.

\bibitem{Fasiello:2013woa}
M.~Fasiello and A.~J. Tolley, ``{Cosmological Stability Bound in Massive
  Gravity and Bigravity},'' {\em JCAP} {\bf 1312} (2013) 002,
  \href{http://xxx.lanl.gov/abs/1308.1647}{{\tt 1308.1647}}.

\bibitem{Akrami:2013ffa}
Y.~Akrami, T.~S. Koivisto, D.~F. Mota, and M.~Sandstad, ``{Bimetric gravity
  doubly coupled to matter: theory and cosmological implications},'' {\em JCAP}
  {\bf 1310} (2013) 046, \href{http://xxx.lanl.gov/abs/1306.0004}{{\tt
  1306.0004}}.

\bibitem{Konnig:2013gxa}
F.~Koennig, A.~Patil, and L.~Amendola, ``{Viable cosmological solutions in
  massive bimetric gravity},'' {\em JCAP} {\bf 1403} (2014) 029,
  \href{http://xxx.lanl.gov/abs/1312.3208}{{\tt 1312.3208}}.

\bibitem{Comelli:2014bqa}
D.~Comelli, M.~Crisostomi, and L.~Pilo, ``{FRW Cosmological Perturbations in
  Massive Bigravity},'' \href{http://xxx.lanl.gov/abs/1403.5679}{{\tt
  1403.5679}}.

\bibitem{Solomon:2014dua}
A.~R. Solomon, Y.~Akrami, and T.~S. Koivisto, ``{Cosmological perturbations in
  massive bigravity: I. Linear growth of structures},''
  \href{http://xxx.lanl.gov/abs/1404.4061}{{\tt 1404.4061}}.

\bibitem{DeFelice:2014nja}
A.~De~Felice, A.~E. Gumrukcuoglu, S.~Mukohyama, N.~Tanahashi, and T.~Tanaka,
  ``{Viable cosmology in bimetric theory},'' {\em JCAP} {\bf 1406} (2014) 037,
  \href{http://xxx.lanl.gov/abs/1404.0008}{{\tt 1404.0008}}.

\bibitem{Konnig:2014dna}
F.~Koennig and L.~Amendola, ``{A minimal bimetric gravity model that fits
  cosmological observations},'' \href{http://xxx.lanl.gov/abs/1402.1988}{{\tt
  1402.1988}}.

\bibitem{Lagos:2014lca}
M.~Lagos and P.~G. Ferreira, ``{Cosmological perturbations in massive
  bigravity},'' {\em JCAP} {\bf 1412} (2014) 026,
  \href{http://xxx.lanl.gov/abs/1410.0207}{{\tt 1410.0207}}.

\bibitem{Cusin:2014psa}
G.~Cusin, R.~Durrer, P.~Guarato, and M.~Motta, ``{Gravitational waves in
  bigravity cosmology},'' {\em JCAP} {\bf 1505} (2015) 030,
  \href{http://xxx.lanl.gov/abs/1412.5979}{{\tt 1412.5979}}.

\bibitem{Konnig:2015lfa}
F.~Koennig, ``{Higuchi Ghosts and Gradient Instabilities in Bimetric
  Gravity},'' {\em Phys. Rev.} {\bf D91} (2015) 104019,
  \href{http://xxx.lanl.gov/abs/1503.07436}{{\tt 1503.07436}}.

\bibitem{Comelli:2012db}
D.~Comelli, M.~Crisostomi, and L.~Pilo, ``{Perturbations in Massive Gravity
  Cosmology},'' {\em JHEP} {\bf 1206} (2012) 085,
  \href{http://xxx.lanl.gov/abs/1202.1986}{{\tt 1202.1986}}.

\bibitem{Konnig:2014xva}
F.~Koennig, Y.~Akrami, L.~Amendola, M.~Motta, and A.~R. Solomon, ``{Stable and
  unstable cosmological models in bimetric massive gravity},'' {\em Phys. Rev.}
  {\bf D90} (2014) 124014, \href{http://xxx.lanl.gov/abs/1407.4331}{{\tt
  1407.4331}}.

\bibitem{Akrami:2015qga}
Y.~Akrami, S.~F. Hassan, F.~K{\"o}nnig, A.~Schmidt-May, and A.~R. Solomon,
  ``{Bimetric gravity is cosmologically viable},'' {\em Phys. Lett.} {\bf B748}
  (2015) 37--44, \href{http://xxx.lanl.gov/abs/1503.07521}{{\tt 1503.07521}}.

\bibitem{Deser:2007jk}
S.~Deser and R.~Woodard, ``{Nonlocal Cosmology},'' {\em Phys.Rev.Lett.} {\bf
  99} (2007) 111301, \href{http://xxx.lanl.gov/abs/0706.2151}{{\tt 0706.2151}}.

\bibitem{Deser:2013uya}
S.~Deser and R.~Woodard, ``{Observational Viability and Stability of Nonlocal
  Cosmology},'' {\em JCAP} {\bf 1311} (2013) 036,
  \href{http://xxx.lanl.gov/abs/1307.6639}{{\tt 1307.6639}}.

\bibitem{Woodard:2014iga}
R.~Woodard, ``{Nonlocal Models of Cosmic Acceleration},'' {\em Found.Phys.}
  {\bf 44} (2014) 213--233, \href{http://xxx.lanl.gov/abs/1401.0254}{{\tt
  1401.0254}}.

\bibitem{Barvinsky:2003kg}
A.~Barvinsky, ``{Nonlocal action for long distance modifications of gravity
  theory},'' {\em Phys.Lett.} {\bf B572} (2003) 109,
  \href{http://xxx.lanl.gov/abs/hep-th/0304229}{{\tt hep-th/0304229}}.

\bibitem{Barvinsky:2011hd}
A.~Barvinsky, ``{Dark energy and dark matter from nonlocal ghost-free gravity
  theory},'' {\em Phys.Lett.} {\bf B710} (2012) 12--16,
  \href{http://xxx.lanl.gov/abs/1107.1463}{{\tt 1107.1463}}.

\bibitem{Barvinsky:2011rk}
A.~O. Barvinsky, ``{Serendipitous discoveries in nonlocal gravity theory},''
  {\em Phys.Rev.} {\bf D85} (2012) 104018,
  \href{http://xxx.lanl.gov/abs/1112.4340}{{\tt 1112.4340}}.

\bibitem{Koivisto:2008xfa}
T.~Koivisto, ``{Dynamics of Nonlocal Cosmology},'' {\em Phys.Rev.} {\bf D77}
  (2008) 123513, \href{http://xxx.lanl.gov/abs/0803.3399}{{\tt 0803.3399}}.

\bibitem{Koivisto:2008dh}
T.~Koivisto, ``{Newtonian limit of nonlocal cosmology},'' {\em Phys.Rev.} {\bf
  D78} (2008) 123505, \href{http://xxx.lanl.gov/abs/0807.3778}{{\tt
  0807.3778}}.

\bibitem{Capozziello:2008gu}
S.~Capozziello, E.~Elizalde, S.~Nojiri, and S.~D. Odintsov, ``{Accelerating
  cosmologies from non-local higher-derivative gravity},'' {\em Phys.Lett.}
  {\bf B671} (2009) 193--198, \href{http://xxx.lanl.gov/abs/0809.1535}{{\tt
  0809.1535}}.

\bibitem{Elizalde:2011su}
E.~Elizalde, E.~Pozdeeva, and S.~Y. Vernov, ``{De Sitter Universe in Non-local
  Gravity},'' {\em Phys.Rev.} {\bf D85} (2012) 044002,
  \href{http://xxx.lanl.gov/abs/1110.5806}{{\tt 1110.5806}}.

\bibitem{Zhang:2011uv}
Y.~Zhang and M.~Sasaki, ``{Screening of cosmological constant in non-local
  cosmology},'' {\em Int.J.Mod.Phys.} {\bf D21} (2012) 1250006,
  \href{http://xxx.lanl.gov/abs/1108.2112}{{\tt 1108.2112}}.

\bibitem{Elizalde:2012ja}
E.~Elizalde, E.~Pozdeeva, and S.~Y. Vernov, ``{Reconstruction Procedure in
  Nonlocal Models},'' {\em Class.Quant.Grav.} {\bf 30} (2013) 035002,
  \href{http://xxx.lanl.gov/abs/1209.5957}{{\tt 1209.5957}}.

\bibitem{Park:2012cp}
S.~Park and S.~Dodelson, ``{Structure formation in a nonlocally modified
  gravity model},'' {\em Phys.Rev.} {\bf D87} (2013) 024003,
  \href{http://xxx.lanl.gov/abs/1209.0836}{{\tt 1209.0836}}.

\bibitem{Bamba:2012ky}
K.~Bamba, S.~Nojiri, S.~D. Odintsov, and M.~Sasaki, ``{Screening of
  cosmological constant for De Sitter Universe in non-local gravity,
  phantom-divide crossing and finite-time future singularities},'' {\em
  Gen.Rel.Grav.} {\bf 44} (2012) 1321--1356,
  \href{http://xxx.lanl.gov/abs/1104.2692}{{\tt 1104.2692}}.

\bibitem{Dodelson:2013sma}
S.~Dodelson and S.~Park, ``{Nonlocal Gravity and Structure in the Universe},''
  {\em Phys.Rev.} {\bf D90} (2014) 043535,
  \href{http://xxx.lanl.gov/abs/1310.4329}{{\tt 1310.4329}}.

\bibitem{Maggiore:2013mea}
M.~Maggiore, ``{Phantom dark energy from nonlocal infrared modifications of
  general relativity},'' {\em Phys.Rev.} {\bf D89} (2014) 043008,
  \href{http://xxx.lanl.gov/abs/1307.3898}{{\tt 1307.3898}}.

\bibitem{ArkaniHamed:2002fu}
N.~Arkani-Hamed, S.~Dimopoulos, G.~Dvali, and G.~Gabadadze, ``{Nonlocal
  modification of gravity and the cosmological constant problem},''
  \href{http://xxx.lanl.gov/abs/hep-th/0209227}{{\tt hep-th/0209227}}.

\bibitem{Dvali:2006su}
G.~Dvali, ``{Predictive Power of Strong Coupling in Theories with Large
  Distance Modified Gravity},'' {\em New J.Phys.} {\bf 8} (2006) 326,
  \href{http://xxx.lanl.gov/abs/hep-th/0610013}{{\tt hep-th/0610013}}.

\bibitem{Dvali:2007kt}
G.~Dvali, S.~Hofmann, and J.~Khoury, ``{Degravitation of the cosmological
  constant and graviton width},'' {\em Phys.Rev.} {\bf D76} (2007) 084006,
  \href{http://xxx.lanl.gov/abs/hep-th/0703027}{{\tt hep-th/0703027}}.

\bibitem{Porrati:2002cp}
M.~Porrati, ``{Fully covariant van Dam-Veltman-Zakharov discontinuity, and
  absence thereof},'' {\em Phys.Lett.} {\bf B534} (2002) 209,
  \href{http://xxx.lanl.gov/abs/hep-th/0203014}{{\tt hep-th/0203014}}.

\bibitem{Jaccard:2013gla}
M.~Jaccard, M.~Maggiore, and E.~Mitsou, ``{A non-local theory of massive
  gravity},'' {\em Phys.Rev.} {\bf D88} (2013) 044033,
  \href{http://xxx.lanl.gov/abs/1305.3034}{{\tt 1305.3034}}.

\bibitem{Maggiore:2014sia}
M.~Maggiore and M.~Mancarella, ``{Non-local gravity and dark energy},'' {\em
  Phys.Rev.} {\bf D90} (2014) 023005,
  \href{http://xxx.lanl.gov/abs/1402.0448}{{\tt 1402.0448}}.

\bibitem{Cusin:2015rex}
G.~Cusin, S.~Foffa, M.~Maggiore, and M.~Mancarella, ``{Nonlocal gravity with a
  Weyl-square term},'' {\em Phys. Rev.} {\bf D93} (2016) 043006,
  \href{http://xxx.lanl.gov/abs/1512.06373}{{\tt 1512.06373}}.

\bibitem{Ferreira:2013tqn}
P.~G. Ferreira and A.~L. Maroto, ``{A few cosmological implications of tensor
  nonlocalities},'' {\em Phys.Rev.} {\bf D88} (2013) 123502,
  \href{http://xxx.lanl.gov/abs/1310.1238}{{\tt 1310.1238}}.

\bibitem{Maggiore:2015rma}
M.~Maggiore, ``{Dark energy and dimensional transmutation in $R^2$ gravity},''
  \href{http://xxx.lanl.gov/abs/1506.06217}{{\tt 1506.06217}}.

\bibitem{Maggiore:2016fbn}
M.~Maggiore, ``{Perturbative loop corrections and nonlocal gravity},'' {\em
  Phys. Rev.} {\bf D93} (2016) 063008,
  \href{http://xxx.lanl.gov/abs/1603.01515}{{\tt 1603.01515}}.

\bibitem{Kehagias:2014sda}
A.~Kehagias and M.~Maggiore, ``{Spherically symmetric static solutions in a
  non-local infrared modification of General Relativity},'' {\em JHEP} {\bf
  1408} (2014) 029, \href{http://xxx.lanl.gov/abs/1401.8289}{{\tt 1401.8289}}.

\bibitem{Barreira:2014kra}
A.~Barreira, B.~Li, W.~A. Hellwing, C.~M. Baugh, and S.~Pascoli, ``{Nonlinear
  structure formation in Nonlocal Gravity},'' {\em JCAP} {\bf 1409} (2014) 031,
  \href{http://xxx.lanl.gov/abs/1408.1084}{{\tt 1408.1084}}.

\bibitem{Foffa:2013vma}
S.~Foffa, M.~Maggiore, and E.~Mitsou, ``{Cosmological dynamics and dark energy
  from non-local infrared modifications of gravity},'' {\em Int.J.Mod.Phys.}
  {\bf A29} (2014) 1450116, \href{http://xxx.lanl.gov/abs/1311.3435}{{\tt
  1311.3435}}.

\bibitem{Dirian:2014ara}
Y.~Dirian, S.~Foffa, N.~Khosravi, M.~Kunz, and M.~Maggiore, ``{Cosmological
  perturbations and structure formation in nonlocal infrared modifications of
  general relativity},'' {\em JCAP} {\bf 1406} (2014) 033,
  \href{http://xxx.lanl.gov/abs/1403.6068}{{\tt 1403.6068}}.

\bibitem{Nesseris:2014mea}
S.~Nesseris and S.~Tsujikawa, ``{Cosmological perturbations and observational
  constraints on nonlocal massive gravity},'' {\em Phys.Rev.} {\bf D90} (2014)
  024070, \href{http://xxx.lanl.gov/abs/1402.4613}{{\tt 1402.4613}}.

\bibitem{Modesto:2013jea}
L.~Modesto and S.~Tsujikawa, ``{Non-local massive gravity},'' {\em Phys.Lett.}
  {\bf B727} (2013) 48--56, \href{http://xxx.lanl.gov/abs/1307.6968}{{\tt
  1307.6968}}.

\bibitem{Foffa:2013sma}
S.~Foffa, M.~Maggiore, and E.~Mitsou, ``{Apparent ghosts and spurious degrees
  of freedom in non-local theories},'' {\em Phys.Lett.} {\bf B733} (2014)
  76--83, \href{http://xxx.lanl.gov/abs/1311.3421}{{\tt 1311.3421}}.

\bibitem{Conroy:2014eja}
A.~Conroy, T.~Koivisto, A.~Mazumdar, and A.~Teimouri, ``{Generalized quadratic
  curvature, non-local infrared modifications of gravity and Newtonian
  potentials},'' {\em Class. Quant. Grav.} {\bf 32} (2015) 015024,
  \href{http://xxx.lanl.gov/abs/1406.4998}{{\tt 1406.4998}}.

\bibitem{Cusin:2014zoa}
G.~Cusin, J.~Fumagalli, and M.~Maggiore, ``{Non-local formulation of ghost-free
  bigravity theory},'' {\em JHEP} {\bf 1409} (2014) 181,
  \href{http://xxx.lanl.gov/abs/1407.5580}{{\tt 1407.5580}}.

\bibitem{Dirian:2014xoa}
Y.~Dirian and E.~Mitsou, ``{Stability analysis and future singularity of the
  $m^2 R \Box^{-2} R$ model of non-local gravity},'' {\em JCAP} {\bf 10} (2014)
  065, \href{http://xxx.lanl.gov/abs/1408.5058}{{\tt 1408.5058}}.

\bibitem{Mitsou:2015yfa}
E.~Mitsou, {\em {Aspects of Infrared Non-local Modifications of General
  Relativity}}.
\newblock PhD thesis, Geneva U., 2015.
\newblock \href{http://xxx.lanl.gov/abs/1504.04050}{{\tt 1504.04050}}.

\bibitem{Barreira:2015fpa}
A.~Barreira, B.~Li, E.~Jennings, J.~Merten, L.~King, {\em et.~al.}, ``{Galaxy
  cluster lensing masses in modified lensing potentials},'' {\em Mon. Not. Roy.
  Astron. Soc.} {\bf 454} (2015) 4085,
  \href{http://xxx.lanl.gov/abs/1505.03468}{{\tt 1505.03468}}.

\bibitem{Barreira:2015vra}
A.~Barreira, M.~Cautun, B.~Li, C.~Baugh, and S.~Pascoli, ``{Weak lensing by
  voids in modified lensing potentials},'' {\em JCAP} {\bf 1508} (2015) 028,
  \href{http://xxx.lanl.gov/abs/1505.05809}{{\tt 1505.05809}}.

\bibitem{Codello:2015pga}
A.~Codello and R.~K. Jain, ``{Covariant Effective Field Theory of Gravity II:
  Cosmological Implications},'' \href{http://xxx.lanl.gov/abs/1507.07829}{{\tt
  1507.07829}}.

\bibitem{Cusin:2016nzi}
G.~Cusin, S.~Foffa, M.~Maggiore, and M.~Mancarella, ``{Conformal symmetry and
  nonlinear extensions of nonlocal gravity},'' {\em Phys. Rev.} {\bf D93}
  (2016) 083008, \href{http://xxx.lanl.gov/abs/1602.01078}{{\tt 1602.01078}}.

\bibitem{Codello:2016neo}
A.~Codello and R.~K. Jain, ``{A Unified Evolution of the Universe},''
  \href{http://xxx.lanl.gov/abs/1603.00028}{{\tt 1603.00028}}.

\bibitem{Dirian:2014bma}
Y.~Dirian, S.~Foffa, M.~Kunz, M.~Maggiore, and V.~Pettorino, ``{Non-local
  gravity and comparison with observational datasets},'' {\em JCAP} {\bf 1504}
  (2015) 044, \href{http://xxx.lanl.gov/abs/1411.7692}{{\tt 1411.7692}}.

\bibitem{git_nonlocal}
\url{https://github.com/dirian/class_public/tree/nonlocal}.

\bibitem{Class}
D.~{Blas}, J.~{Lesgourgues}, and T.~{Tram}, ``{The Cosmic Linear Anisotropy
  Solving System (CLASS). Part II: Approximation schemes},'' {\em JCAP} {\bf 7}
  (July, 2011) 34, \href{http://xxx.lanl.gov/abs/1104.2933}{{\tt 1104.2933}}.

\bibitem{MP}
B.~{Audren}, J.~{Lesgourgues}, K.~{Benabed}, and S.~{Prunet}, ``{Conservative
  constraints on early cosmology with MONTE PYTHON},'' {\em JCAP} {\bf 2}
  (Feb., 2013) 1, \href{http://xxx.lanl.gov/abs/1210.7183}{{\tt 1210.7183}}.

\bibitem{Planck_2015_CP}
{\bf Planck} Collaboration, P.~Ade {\em et.~al.}, ``{Planck 2015 results. XIII.
  Cosmological parameters},'' \href{http://xxx.lanl.gov/abs/1502.01589}{{\tt
  1502.01589}}.

\bibitem{Planck_2015_DE}
{\bf Planck} Collaboration, P.~A.~R. Ade {\em et.~al.}, ``{Planck 2015 results.
  XIV. Dark energy and modified gravity},''
  \href{http://xxx.lanl.gov/abs/1502.01590}{{\tt 1502.01590}}.

\bibitem{Planck_2015_1}
{\bf Planck} Collaboration, R.~Adam {\em et.~al.}, ``{Planck 2015 results. I.
  Overview of products and scientific results},''
  \href{http://xxx.lanl.gov/abs/1502.01582}{{\tt 1502.01582}}.

\bibitem{Planck_2013_1}
{\bf Planck} Collaboration, P.~A.~R. Ade {\em et.~al.}, ``{Planck 2013 results.
  I. Overview of products and scientific results},'' {\em Astron. Astrophys.}
  {\bf 571} (2014) A1, \href{http://xxx.lanl.gov/abs/1303.5062}{{\tt
  1303.5062}}.

\bibitem{Planck_2015_wiki}
{\bf Planck} Collaboration, ``{Planck 2015 release explanatory supplement},''
  \href{http://xxx.lanl.gov/abs/http://wiki.cosmos.esa.int/planckpla2015/index.php/CMB\_spectrum\_\%26\_Likelihood\_Code}{{\tt
  http://wiki.cosmos.esa.int/planckpla2015/index.php/CMB\_spectrum\_\%26\_Likelihood\_Code}}.

\bibitem{Planck_2015_Lkl}
{\bf Planck} Collaboration, N.~Aghanim {\em et.~al.}, ``{Planck 2015 results.
  XI. CMB power spectra, likelihoods, and robustness of parameters},'' {\em
  Submitted to: Astron. Astrophys.} (2015)
  \href{http://xxx.lanl.gov/abs/1507.02704}{{\tt 1507.02704}}.

\bibitem{Planck_2015_lens}
{\bf Planck} Collaboration, P.~A.~R. Ade {\em et.~al.}, ``{Planck 2015 results.
  XV. Gravitational lensing},'' \href{http://xxx.lanl.gov/abs/1502.01591}{{\tt
  1502.01591}}.

\bibitem{JLA_2014}
{\bf SDSS} Collaboration, M.~Betoule {\em et.~al.}, ``{Improved cosmological
  constraints from a joint analysis of the SDSS-II and SNLS supernova
  samples},'' {\em Astron.Astrophys.} {\bf 568} (2014) A22,
  \href{http://xxx.lanl.gov/abs/1401.4064}{{\tt 1401.4064}}.

\bibitem{Beutler_6dF_BAO_2012}
F.~{Beutler}, C.~{Blake}, M.~{Colless}, D.~H. {Jones}, L.~{Staveley-Smith},
  L.~{Campbell}, Q.~{Parker}, W.~{Saunders}, and F.~{Watson}, ``{The 6dF Galaxy
  Survey: baryon acoustic oscillations and the local Hubble constant},'' {\em
  Mon.Not.Roy.Astron.Soc.} {\bf 416} (Oct., 2011) 3017--3032,
  \href{http://xxx.lanl.gov/abs/1106.3366}{{\tt 1106.3366}}.

\bibitem{Ross_SDSS_2014}
A.~J. Ross, L.~Samushia, C.~Howlett, W.~J. Percival, A.~Burden, and M.~Manera,
  ``{The clustering of the SDSS DR7 main Galaxy sample - I. A 4 per cent
  distance measure at $z = 0.15$},'' {\em Mon. Not. Roy. Astron. Soc.} {\bf
  449} (2015), no.~1 835--847, \href{http://xxx.lanl.gov/abs/1409.3242}{{\tt
  1409.3242}}.

\bibitem{And_BOSS_2013}
{\bf BOSS} Collaboration, L.~Anderson {\em et.~al.}, ``{The clustering of
  galaxies in the SDSS-III Baryon Oscillation Spectroscopic Survey: Baryon
  Acoustic Oscillations in the Data Release 10 and 11 galaxy samples},'' {\em
  Mon.Not.Roy.Astron.Soc.} {\bf 441} (2014), no.~1 24--62,
  \href{http://xxx.lanl.gov/abs/1312.4877}{{\tt 1312.4877}}.

\bibitem{Efsta_H0_2013}
G.~Efstathiou, ``{$H_0$ Revisited},'' {\em Mon.Not.Roy.Astron.Soc.} {\bf 440}
  (2014), no.~2 1138--1152, \href{http://xxx.lanl.gov/abs/1311.3461}{{\tt
  1311.3461}}.

\bibitem{Riess:2011yx}
A.~G. Riess {\em et.~al.}, ``{A 3\% Solution: Determination of the Hubble
  Constant with the Hubble Space Telescope and Wide Field Camera 3},'' {\em
  Astrophys.J.} {\bf 730} (2011) 119,
  \href{http://xxx.lanl.gov/abs/1103.2976}{{\tt 1103.2976}}.

\bibitem{Riess:2016jrr}
A.~G. Riess {\em et.~al.}, ``{A 2.4\% Determination of the Local Value of the
  Hubble Constant},'' \href{http://xxx.lanl.gov/abs/1604.01424}{{\tt
  1604.01424}}.

\bibitem{Planck_2013_CP}
{\bf Planck} Collaboration, P.~Ade {\em et.~al.}, ``{Planck 2013 results. XVI.
  Cosmological parameters},'' {\em Astron.Astrophys.} {\bf 571} (2014) A16,
  \href{http://xxx.lanl.gov/abs/1303.5076}{{\tt 1303.5076}}.

\bibitem{DirainBarreira:inprep}
A.~Barreira and Y.~Dirian {\em in preparation}.

\bibitem{Trotta_BITS}
R.~Trotta, ``{Bayes in the sky: Bayesian inference and model selection in
  cosmology},'' {\em Contemp. Phys.} {\bf 49} (2008) 71--104,
  \href{http://xxx.lanl.gov/abs/0803.4089}{{\tt 0803.4089}}.

\bibitem{Ade:2015tva}
{\bf BICEP2, Planck} Collaboration, P.~Ade {\em et.~al.}, ``{Joint Analysis of
  BICEP2/$Keck\, Array$ and $Planck$ Data},'' {\em Phys. Rev. Lett.} {\bf 114}
  (2015) 101301, \href{http://xxx.lanl.gov/abs/1502.00612}{{\tt 1502.00612}}.

\bibitem{Trotta_ABS}
R.~Trotta, ``{Applications of Bayesian model selection to cosmological
  parameters},'' {\em Mon. Not. Roy. Astron. Soc.} {\bf 378} (2007) 72--82,
  \href{http://xxx.lanl.gov/abs/astro-ph/0504022}{{\tt astro-ph/0504022}}.

\bibitem{Nesseris:2012cq}
S.~Nesseris and J.~Garcia-Bellido, ``{Is the Jeffreys' scale a reliable tool
  for Bayesian model comparison in cosmology?},'' {\em JCAP} {\bf 1308} (2013)
  036, \href{http://xxx.lanl.gov/abs/1210.7652}{{\tt 1210.7652}}.

\bibitem{Wagen_SDDR}
E.-J. Wagenmakers, T.~Lodewyckx, H.~Kuriyal, and G.~Raoul, ``{Bayesian
  hypothesis testing for psychologists: a tutorial on the Savage-Dickey
  method.},'' {\em Cogn Psychol.} {\bf 60(3)} (2010) 158?189.

\bibitem{Mukherjee:2010ve}
P.~Mukherjee, J.~Urrestilla, M.~Kunz, A.~R. Liddle, N.~Bevis, and M.~Hindmarsh,
  ``{Detecting and distinguishing topological defects in future data from the
  CMBPol satellite},'' {\em Phys. Rev.} {\bf D83} (2011) 043003,
  \href{http://xxx.lanl.gov/abs/1010.5662}{{\tt 1010.5662}}.

\bibitem{Oka_SDSS_2013}
A.~Oka, S.~Saito, T.~Nishimichi, A.~Taruya, and K.~Yamamoto, ``{Simultaneous
  constraints on the growth of structure and cosmic expansion from the
  multipole power spectra of the SDSS DR7 LRG sample},'' {\em
  Mon.Not.Roy.Astron.Soc.} {\bf 439} (2014) 2515--2530,
  \href{http://xxx.lanl.gov/abs/1310.2820}{{\tt 1310.2820}}.

\bibitem{Beu_6dF_RSD_2012}
F.~{Beutler}, C.~{Blake}, M.~{Colless}, D.~H. {Jones}, L.~{Staveley-Smith},
  G.~B. {Poole}, L.~{Campbell}, Q.~{Parker}, W.~{Saunders}, and F.~{Watson},
  ``{The 6dF Galaxy Survey: $z \approx 0$ measurements of the growth rate and
  {$\sigma$}$_{8}$},'' {\em Mon.Not.Roy.Astron.Soc.} {\bf 423} (July, 2012)
  3430--3444, \href{http://xxx.lanl.gov/abs/1204.4725}{{\tt 1204.4725}}.

\bibitem{How_SDSS_2014}
C.~Howlett, A.~Ross, L.~Samushia, W.~Percival, and M.~Manera, ``{The Clustering
  of the SDSS Main Galaxy Sample II: Mock galaxy catalogues and a measurement
  of the growth of structure from Redshift Space Distortions at $z=0.15$},''
  \href{http://xxx.lanl.gov/abs/1409.3238}{{\tt 1409.3238}}.

\bibitem{Chuang_BOSS_2013}
C.-H. Chuang, F.~Prada, F.~Beutler, D.~J. Eisenstein, S.~Escoffier, {\em
  et.~al.}, ``{The clustering of galaxies in the SDSS-III Baryon Oscillation
  Spectroscopic Survey: single-probe measurements from CMASS and LOWZ
  anisotropic galaxy clustering},''
  \href{http://xxx.lanl.gov/abs/1312.4889}{{\tt 1312.4889}}.

\bibitem{Samushia_BOSS_2013}
L.~Samushia, B.~A. Reid, M.~White, W.~J. Percival, A.~J. Cuesta, {\em et.~al.},
  ``{The Clustering of Galaxies in the SDSS-III Baryon Oscillation
  Spectroscopic Survey (BOSS): measuring growth rate and geometry with
  anisotropic clustering},'' {\em Mon.Not.Roy.Astron.Soc.} {\bf 439} (2014)
  3504--3519, \href{http://xxx.lanl.gov/abs/1312.4899}{{\tt 1312.4899}}.

\bibitem{Blake_WiggZ_2012}
C.~Blake, S.~Brough, M.~Colless, C.~Contreras, W.~Couch, {\em et.~al.}, ``{The
  WiggleZ Dark Energy Survey: Joint measurements of the expansion and growth
  history at $z < 1$},'' {\em Mon.Not.Roy.Astron.Soc.} {\bf 425} (2012)
  405--414, \href{http://xxx.lanl.gov/abs/1204.3674}{{\tt 1204.3674}}.

\bibitem{delaTorre_VIPERS_2013}
S.~de~la Torre, L.~Guzzo, J.~Peacock, E.~Branchini, A.~Iovino, {\em et.~al.},
  ``{The VIMOS Public Extragalactic Redshift Survey (VIPERS). Galaxy clustering
  and redshift-space distortions at $z=0.8$ in the first data release},'' {\em
  Astron.Astrophys.} {\bf 557} (2013) A54,
  \href{http://xxx.lanl.gov/abs/1303.2622}{{\tt 1303.2622}}.

\bibitem{Daniel:2010ky}
S.~F. Daniel {\em et.~al.}, ``{Testing General Relativity with Current
  Cosmological Data},'' {\em Phys.Rev.} {\bf D81} (2010) 123508,
  \href{http://xxx.lanl.gov/abs/1002.1962}{{\tt 1002.1962}}.

\bibitem{Amendola:2007rr}
L.~Amendola, M.~Kunz, and D.~Sapone, ``{Measuring the dark side (with weak
  lensing)},'' {\em JCAP} {\bf 0804} (2008) 013,
  \href{http://xxx.lanl.gov/abs/0704.2421}{{\tt 0704.2421}}.

\bibitem{Amendola:2012ys}
{\bf Euclid Theory Working Group} Collaboration, L.~Amendola {\em et.~al.},
  ``{Cosmology and fundamental physics with the Euclid satellite},'' {\em
  Living Rev.Rel.} {\bf 16} (2013) 6,
  \href{http://xxx.lanl.gov/abs/1206.1225}{{\tt 1206.1225}}.

\bibitem{DT_CMBGW}
R.~Durrer and T.~Kahniashvili, ``{CMB anisotropies caused by gravitational
  waves: A Parameter study},'' {\em Helv.Phys.Acta} {\bf 71} (1998) 445--457,
  \href{http://xxx.lanl.gov/abs/astro-ph/9702226}{{\tt astro-ph/9702226}}.

\bibitem{Wei_DampGW}
S.~Weinberg, ``{Damping of tensor modes in cosmology},'' {\em Phys.Rev.} {\bf
  D69} (2004) 023503, \href{http://xxx.lanl.gov/abs/astro-ph/0306304}{{\tt
  astro-ph/0306304}}.

\bibitem{Dur_CMB}
R.~Durrer, {\em {The Cosmic Microwave Background}}.
\newblock Cambridge Univ. Press, Cambridge, 2008.

\bibitem{Williams:2004qba}
J.~G. Williams, S.~G. Turyshev, and D.~H. Boggs, ``{Progress in lunar laser
  ranging tests of relativistic gravity},'' {\em Phys. Rev. Lett.} {\bf 93}
  (2004) 261101, \href{http://xxx.lanl.gov/abs/gr-qc/0411113}{{\tt
  gr-qc/0411113}}.

\bibitem{Babichev:2011iz}
E.~Babichev, C.~Deffayet, and G.~Esposito-Farese, ``{Constraints on
  Shift-Symmetric Scalar-Tensor Theories with a Vainshtein Mechanism from
  Bounds on the Time Variation of G},'' {\em Phys. Rev. Lett.} {\bf 107} (2011)
  251102, \href{http://xxx.lanl.gov/abs/1107.1569}{{\tt 1107.1569}}.

\end{thebibliography}\endgroup
